\documentclass[twocolumn]{aastex7}
\usepackage{graphicx}
\usepackage{amsmath}	
\usepackage{amssymb}	
\usepackage{enumitem}
\usepackage{verbatim} 
\usepackage[caption=false]{subfig}
\usepackage{array}
\usepackage{listings}
\usepackage{hyperref}
\usepackage{braket}
\usepackage{tabularx}
\usepackage{capt-of}
\usepackage{mathrsfs}
\usepackage{mathtools}
\usepackage{multirow}
\newcommand{\code}[1]{\texttt{#1}}
\usepackage{rotating}

\usepackage[mathlines]{lineno}
\linenumbers

\newcommand{\citett}[1]{\textsuperscript{#1}}
\begin{document}
\bibliographystyle{aasjournalv7}

\title[]{Planet-Induced Stellar Flare Candidates from the TESS Mission}

\makeatletter\let\frontmatter@title@above=\relax

\newcommand{\todo}[1]{\textcolor{red}{ToDo: #1}}

\author[0000-0002-7960-8064]{Nathan Whitsett}
\email{whitsett.n@wustl.edu}
\affiliation{Department of Physics, Washington University, St. Louis, MO 63130, USA}

\author[0000-0002-6939-9211]{Tansu Daylan}
\email{tansu@wustl.edu}
\affiliation{Department of Physics, Washington University, St. Louis, MO 63130, USA}
\affiliation{McDonnell Center for the Space Sciences, Washington University, St. Louis, MO 63130, USA}


\begin{abstract}
The empirical underabundance of close-in planets with radii 1.5-2.0 times that of the Earth, referred to as the radius irradiation valley, may be linked to the inability of gas dwarfs under certain conditions to retain their volatile envelopes due to photoevaporation or core-powered mass loss. In either case, the extent to which a planet can preserve its atmosphere critically depends on poorly understood planetary magnetism. An effective probe of planetary magnetic fields is the interaction between a star and its close-in planet, where the planet magnetically interacts with its host, inducing flares as it moves near its periastron within the Alfvén surface of its host star. We construct a pipeline, \textsc{ardor}, to detect and characterize potentially planet-induced flares in time-series photometric data using a physically motivated forward model of star-planet interactions, with a focus on recovering flares located near the noise floor. We perform extensive injection-recovery simulations to determine our sensitivity to flares correlating with the planetary phase over a range of stellar types and orbital architectures. We identify one close-in, eccentric ($e=0.18$) system, TOI-1062\,b, which exhibits flaring during periastron consistent with being induced with $p_{KS}=2.2\times10^{-5}$ and a $5.1\sigma$ detection through unbinned likelihood analysis and goodness-of-fit tests. We also identify an additional eccentric ($e=0.363$) candidate, Gliese 49\,b, which exhibits moderately significant ($2.5\sigma$) flare clustering at periastron, requiring additional photometric observations to confirm its significance. TOI-1062\,b and Gliese 49\,b are promising candidates for induced flares, underscoring the need for radio and UV follow-up observations.
\end{abstract}


\section{Introduction}
\label{Sect:Introduction}
The Transiting Exoplanet Survey Satellite (TESS) \citep{Ricker2015} has recently accelerated the discovery of small exoplanets amenable to comparative atmospheric characterization \citep{Daylan2021a, Guenther2019a, Badenas-Agusti2020a} in multiplanetary systems that harbor small planets of various sizes subject to similar irradiation histories. While photoevaporation and core-powered mass loss can generally explain radius-irradiation demographics of these small exoplanets\citep*{Van2018, Hardegree2020, Bean2021}, planetary magnetism remains as an outstanding knowledge gap that determines the ability of a gas dwarf to retain its volatile envelope. Furthermore, recent discoveries of hot Neptunes, such as LTT 9779\,b, TOI-849\,b, and TOI-332\,b, challenge the generality of photoevaporation and core-powered mass loss \citep{Jenkins2020, Armstrong2020, Osborn2023}. Exoplanetary magnetic fields likely play a significant role in atmospheric loss; however, observational efforts, both direct and indirect, have been inconclusive \citep{Shkolnik2008, Vidotto2013, Khodachenko2015, Ilin2022, Gupta2023}.

The impact of frequently flaring hosts on atmospheric compositions, in tandem with bulk loss, is also a poorly recorded effect but has been identified in several systems \citep{Bisikalo2018, Odert2017, Chadney2017}. Consequently, understanding the connection between exoplanetary magnetic fields, space weather, and atmospheric loss is crucial for developing a comprehensive theory of planetary formation.

\begin{figure}[!ht]
    \centering
    \includegraphics{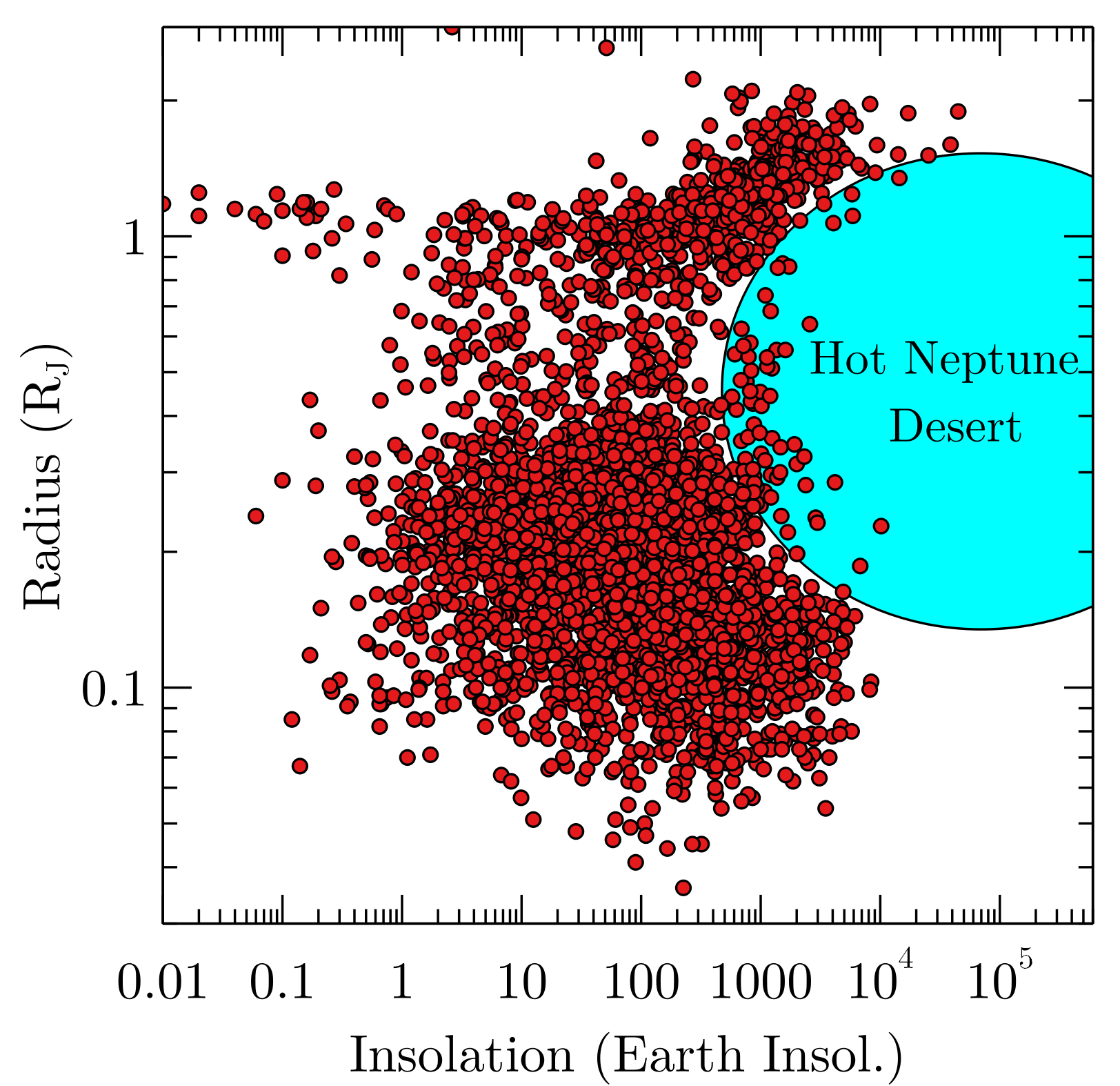}
    \caption{Scatterplot of known exoplanet radii plotted against insolation flux. The hot Neptune Valley is indicated in cyan, showing an under-population of a close-in, Neptune-sized planet.}
    \label{Fig: Hot-Neptune Desert}
\end{figure}

Observing planetary magnetic fields directly is a challenging task. The most promising detection signatures are auroral cyclotron-maser emission and synchrotron radiation in low-frequency radio emission, the exact mechanisms that drive the decametric/decimetric emission of the Jupiter-Io system, respectively \citep{Carr1983, Belcher1987}. Although searches for direct planetary radio emission have so far yielded no candidates, the detection of brown dwarf auroras and radiation belts strongly suggests that facility limitations \citep{Hallinan2008, Narang2022, Kao2023} may be a factor. 

Alternative probes for planetary magnetic fields include phase-correlated magnetic star-planet interactions, with proposed observables through three primary mechanisms: chromospheric activity via the calcium infrared triplet (IRT); coherent radio emission through Alfvén wave propagation; and induced stellar flares \citep{Shkolnik2008, Lanza2012, Vidotto2014, Lanza2018, Route2019, Feinstein2021, Ilin2022, Ilin2024}. For reviews of these mechanisms and the basis of magnetic SPIs, see \citet{Strugarek2021, Strugarek2025}. 

\begin{figure}[!ht]
\label{Figure 1: SPI_Visual}
\includegraphics{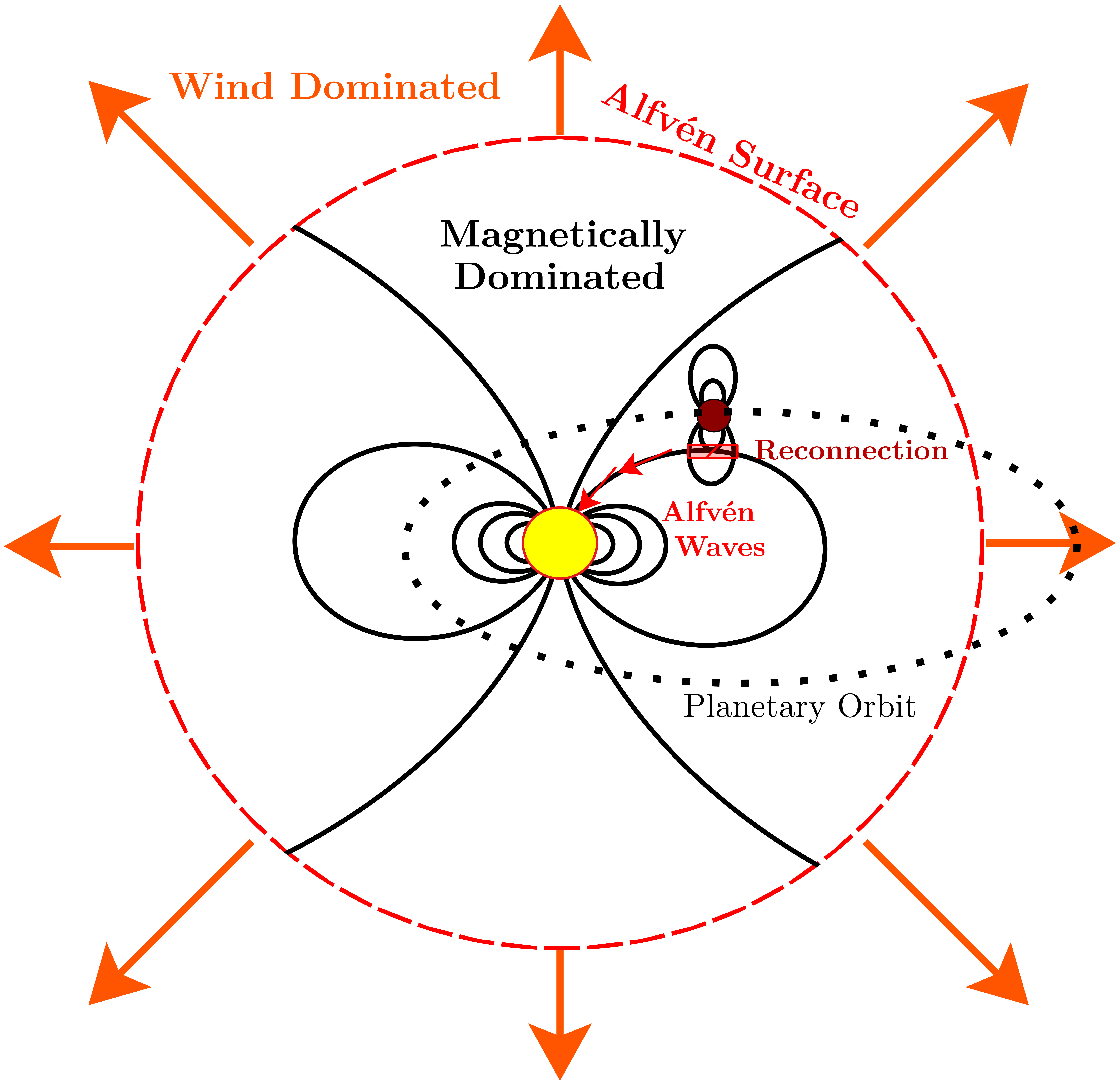}
\caption{Proposed induced-flare mechanisms. A close-orbiting, magnetically active planet (red circle) with an elliptic orbit passes below its host's Alfvén surface (black circle). As it does so, it can magnetically interact with its host, causing induced stellar flares preferentially at periastron. This diagram illustrates two possible mechanisms: reconnection ("bend-and-snap") and Alfvén waves. The red lines show simplified dipole fields. The obliquity between the planetary orbit and the field lines is chosen for illustrative purposes.}
\end{figure}

The Kepler Space Telescope \citep{Borucki2010} and TESS missions provide short-cadence photometric data allowing for white-light flare surveys with year-long baselines for thousands of confirmed planet hosts and TESS planet candidate hosts \citep{Davenport2014, Gunther2020, Pietras2022}. Therefore, identifying magnetic SPI through induced flaring should be feasible, but requires a close examination of flare-detection paradigms.

Many state-of-the-art flare detection pipelines, e.g., \citet{Davenport2014, ilin2021, Pietras2022}, identify flares by flagging three sequential data points above the median flux by a detection threshold, typically $2.5-3\sigma$. This procedure successfully circumvents false positives. The drawback is the inherent loss of sensitivity to flares with short durations and/or short amplitudes. For 2-minute cadence TESS data, flares returning to the quiescent flux within $\sim$6 minutes will yield low recovery rates. A study by \citet{Brasseur2019} found that 95\% of UV flares ($N = 1904$) around solar-like stars had durations of less than five minutes and had white-light analogs in Kepler light curves. As the induced-flare amplitude-FWHM space is not well-constrained, ignoring small flares may lower our sensitivity to the signal, despite the cost of false positives. 

Section~\ref{section: SPMI Models} presents a model by \citet{Lanza2018} predicting magnetic SPI interactions dependent on the instantaneous separation of a sub-Alfvénic planet and its host. In Section~\ref{subsection: Estimating Flare Parameters}, we use empirical relationships of stellar age, mass-loss, and magnetic field strengths to estimate the exoplanet host star's Alfvén surface, with limitations and uncertainty discussed in Section~\ref{subsubsection: Alfven Error}. We present a flare pipeline, \texttt{ardor} \footnote{https://github.com/AstroMusers/ardor}, that detects, vets, and characterizes stellar flares in time-series photometry. \texttt{ardor} operates in successive "tiers" that follow established flare pipeline paradigms, which refine flare candidates to increasingly strict detection criteria. We present pipeline diagnostics, performance, and completion metrics based on injection-recovery and precision-recall tests in Section~\ref{subsection: Completeness}. Analysis methods used in this paper, including three goodness-of-fit tests and unbinned likelihood analysis, are presented in Section~\ref{subsection: Stat Tests}. 

To test the detectability of induced flares, we construct two toy probability density functions (PDFs) in Section~\ref{subsection: Flare Probability Dist.}, motivated by the physical model in Section~\ref{section: SPMI Models}. We apply these distributions to simulated light curves based on empirical flare rates and typical TESS data volumes. We permute over various SPI strengths to determine when SPIs become distinguishable over quiescent flaring. We report on the results of both simulations in Section~\ref{subsection: Simulation Analysis} and in Figure \ref{figure: All_Simulations}. Limitations of both models are discussed in Section \ref{subsection: Limitations Future Work} with suggestions for improvement. Additionally, we develop a metric, $\beta_{\mathrm{SPI}}$, that quantifies the agreement between the flare sample and periastron-dependent SPIs by using the scaling of the instantaneous separation and expected flare power described in Section~\ref{section: SPMI Models}.

In Section~\ref{section: Data}, we present target criteria for the TESS photometry to analyze using \texttt{ardor}. This divides our targets into three sub-samples: exoplanet hosts with a constraint on their argument of periapsis, $\omega$; exoplanet hosts without a constraint on $\omega$; and TESS Planet Candidate (PC) hosts. In Section~\ref{section: Results}, we discuss notable flares, common false-positives, construct flare frequency distributions of all targets (FFDs), and present the results of the statistical tests presented in Section \ref{subsection: Stat Tests}. Additionally, we investigate flare correlation with stellar rotation periods ($P_{\star, rot}$) and orbital-rotational synodic periods ($P_{syn}$) to identify signal contamination with induced flare candidates. During this process, we also constructed a stellar rotation catalog of planet and planet candidate hosts using the \texttt{SpinSpotter} pipeline. In Section~\ref{section: Discussion}, we discuss two induced flare candidates, TOI-1062\,b and Gliese 49\,b, as well as detections arising from correlations between $P_{\star, rot}$ and $P_{syn}$. Section~\ref{section: Results} is continued by discussing the orbital parameters of each system and the number and quality of the flares observed. In Section \ref{subsection: Impacts Methodology}, the impacts of our refined flare detection paradigms, the stability of our results to additional flare detections, and the importance of model-agnostic flare metrics are discussed. Lastly, in Section~\ref{section: Conclusion}, we propose future work and observations that could strengthen the induced flare candidates presented in this work.

\section{Flaring Star-Planet Magnetic Interactions}
\label{section: SPMI Models}
The underlying dynamics of magnetic SPIs have been widely studied by, e.g., \citet*{Lanza2012, Strugarek2017, Fischer2022}. This section highlights a particular model of interest, focusing on the sub-Alfvénic dipole interaction of a magnetized planet with the stellar field. Most known exoplanet hosts are weakly or moderately active main-sequence, solar-like stars. This differs from early or pre-main-sequence stars, which can have highly extended, closed field lines, similar to those found in T Tauri stars. The magnetic field present at the position of the planet is described by:
\begin{equation}
    \mathbf{B} = \mathbf{B}_{\star} + \mathbf{B}_{MP}(C_{mp} - 1) + C_{D}\mathbf{B_{P}},
\end{equation}
where $\mathbf{B}_{\star}$ is the unperturbed stellar field, $\mathbf{B}_{MP}$ is the field produced by the magnetopause currents, and $\mathbf{B}_{P}$ is the field produced by the assumed planetary dynamo, ionospheric currents, and magnetotail currents \citep{Lanza2018}. The coefficients $C_{mp}$ and $C_{D}$ represent the reconnection efficiency between the solar field and the planet's magnetosphere. Due to the unknown properties of the planetary field, ionosphere, and magnetotail currents, $C_{D}=0$, which assumes that the interior field lines are confined within the magnetopause. Conversely, $C_{D}\neq 0$ assumes flux of the stellar field across the field associated with the magnetopause currents. Both coefficients vary over time as the planet encounters different magnetic environments while orbiting its host, e.g., trans-Alfvénic orbits or coronal structures. Nevertheless, as long as the interaction time of the magnetic SPI is assumed to be sufficiently shorter than the timescale of the parameter change, this equation holds. The energy associated with this field can be written as
\begin{equation}
\label{equation: B integral}
    2\mu E_{e} = \int_{V_{e}} \mathbf{B}^{2} dV =  \int_{V_{e}} [ \mathbf{B} - (1-C_{mp})\nabla u_{mp}]^2 dV
\end{equation}
where $u_{mp}$ is the potential generated by the magnetopause current such that $B_{mp}=-\nabla_{mp}$ \citep{Lanza2018}. For this to be evaluated, a magnetospheric geometry is assumed, constituting a hemispherical head of radius $R_{m}$ and a cylindrical tail of the same radius, as outlined in \citet{Griessmeier2004}. This radius is defined by
\begin{equation}
    R_{m} = 2(2f_{0})^{1/3} \left[\frac{B_{Planet}}{B(\mathbf{r_{P}})}\right]^{1/3} R_{Planet}.
\end{equation}
Here, $f_{0}$ is the shape factor, $B_{Planet}$ is the polar magnetic field strength of the planet, and $R_{Planet}$ is the planetary radius. The energy decrease of the stellar field due to the presence of the close-orbiting planet is then written as:
\begin{multline}
\label{equation: Field Energy Difference}
2\mu \Delta E = -\pi R^{3}_{m}B^{2}( \mathbf{r}_{p}) \left(  \left(1-\frac{1}{3}C^{2}_{mp}\right) +\right. \\
\left. \left[1+(1-C^{2}_{mp})\cos^{2}\xi \right]\frac{1}{3}\left(\frac{r_{p}}{R_{m}} \right)  \right).
\end{multline}

The variable $\xi$ is the angle between the north pole of the planet and the stellar field vector at the planet's position. Thus, if the stellar field points toward the planet, $\xi=\pi / 2$, causing $\cos^{2}\xi = 0$. The maximum energy difference in this scenario occurs when there is no reconnection between the magnetopause field and the stellar field, such that $C_{mp} = 0$. 

Equation~\ref{equation: Field Energy Difference} has several important observational implications. First, $\Delta E \propto{\frac{1}{r_p^3}}$, showing the energy difference is maximal when the planet is at periastron. The timescale of the interaction at periastron is given by
\begin{multline}
\label{eq: timescale}
    \tau \geq \frac{2}{\pi} P_{orb}(2f_{0})^{1/3} \left( \frac{a}{R}\right) \\
    \frac{(1-e)^{7/6}}{(1+e)^{1/2}}  \left( \frac{B_{Planet}}{B_{0}}\right)^{1/3} \left(\frac{R_{Planet}}{R}\right).
\end{multline}

This mechanism only ensures an energy drop in the stellar field. While the interaction power is similar to flares, the perturbation may not necessarily drive a flare event.

\subsection{Estimating Induced Flare Parameters}
\label{subsection: Estimating Flare Parameters}

This section develops a framework to estimate the relevant parameters that dictate the magnetic star-planet interaction described in Section~\ref{section: SPMI Models}. We generate estimates for solar-like, main-sequence stars and M dwarfs. We use empirical trends to create these estimates, covering as many stellar hosts as possible. We use this to estimate stellar Alfvén surfaces, derived from the magnetic condiment parameter, $\eta_{\star}$, which depends on the stellar radius, magnetic field strength, mass-loss rate, and terminal wind speed.

\subsubsection{Stellar Surface Magnetic Fields}
\label{subsubsection: Estimating Stellar Fields}

Magnetic SPIs rely on the magnetic environment of the host star, commonly inferred through Zeeman Doppler Imaging (ZDI) reconstruction. \citet{Vidotto2014} analyzed a large sample of main-sequence stars with known field strengths and constructed empirical relationships between the unsigned surface field $\braket{B_{V}}$, stellar age $t_{age}$, and stellar rotation period $P_{rot}$. These are:
\begin{equation}
    \left\langle B_{V} \right\rangle \propto t^{-0.655\pm 0.045}; \left\langle B_{V}\right\rangle \propto P_{rot}^{-0.132\pm 0.14}.
\end{equation}
M dwarfs with known ages and rotations in this sample showed a similar trend, but with $B_{V}$ strengths an order of magnitude higher. Estimating saturated M dwarfs and highly active stars, which have $\left\langle B_{V}\right\rangle \approx 2 \; \mathrm{kG}$, is out of scope for this empirical analysis. Any $B_V$ estimate for either unsaturated or saturated M dwarfs can then be seen as a lower bound.

\subsubsection{Mass-Loss Rates of Main Sequence Stars}
\label{subsubsection: Mass loss rate}

Main-sequence mass-loss rates are challenging since the wind outflow does not radiate strongly compared to stars on the asymptotic giant branch. One method developed by \citet{Wood2002} identified hot HI gas between the heliopause and bow shock, which leads to red-shifted absorption in the stellar $\mathrm{Ly}\mathrm{\alpha}$ lines, and structures in the stellar atmosphere lead to blue-shifted absorption. The absorption depends on the density, size, and velocity of the stellar wind, which can be used to deduce mass loss rates \citep{Holzwarth2007}. In another paper by \cite{Wood2005}, they derive an empirical relationship between mass-loss rates of main-sequence stars and stellar age:
\begin{equation}
\label{equation: mass loss}
    \dot{M} \propto t^{-2.33\pm0.55}
\end{equation}
In a similar fashion to the other estimated stellar parameters, we use the solar value of $3\times10^{-14}\; \mathrm{M_{\odot}/yr}$ to estimate other stellar mass loss rates.

\subsubsection{Terminal Wind Speed}
\label{subsubsection: Terminal Wind Speed}
The last required stellar parameter is terminal wind speed, $v_{\infty}$. For main-sequence stars, we estimate this as the escape velocity $v_{\infty} \approx v_{esc}$:
\begin{equation}
\label{equation: Terminal Wind}
    v_{\infty} \approx v_{esc} = \sqrt{\frac{2GM_{\star}}{R_{\star}}}
\end{equation}

This approximation breaks down in the case of highly luminous giant stars and young T Tauri-type stars, which are excluded from our model approximations.

\subsubsection{The Alfvén Surface}
\label{subsubsection: Alfvén Surface}

The Alfvén surface is the "interaction surface" for magnetic perturbations, as it defines the surface where an Alfvén wave can return to the stellar surface \citep{Deforest2014}. Estimates of the Alfvén surface are difficult due to the complexity of stellar field structures and their temporal variability over short and long timescales. Exact location(s) of the solar Alfvén surface were recently confirmed by the Parker Solar Probe (PSP) \emph{in situ}, with a typical distance of $10-12\;\mathrm{R_{\odot}}$, but extended to $30\;\mathrm{R_{\odot}}$ for regions with sub-Alfvénic winds associated with low-mach boundary layers \citep{Cranmer2023, Jiao2023}. This description of a 'bumpy' surface is consistent with simulations, for example, from \citet{Vidotto2014}, based on ZDI reconstructions. We simplify the model by assuming a spherical Alfvén surface.

We introduce the \emph{magnetic confinement parameter}, $\eta_{\star}$, defined in \citet{Ud2009} and \citet{Owocki2009}:
\begin{equation}
\label{equation: Magnetic confinement}
    \eta_{\star} \equiv \frac{B_{\star}^{2}R_{\star}^{2}}{\dot{M} v_{\infty}}
\end{equation}
This parameter can be thought of as a star's capacity to maintain distant, closed field lines, where stellar size $R_{\star}$ and surface field strength $B_{\star}$ generate large-scale magnetic structures, whereas mass loss $ \dot{M}$ and terminal wind speed $v_{\infty}$ drive a stellar wind which competes against these structures. Following the prescription of \citet{Vidotto2013}, we approximate that the field derived from the ZDI technique satisfies the definition for $\eta_{\star}$ such that $B_{\star} = \left\langle B_{V} \right\rangle$, as both refer to typical surface field strengths.

To estimate the Alfvén surface, we follow the relation set by \citet{Owocki2009}
\begin{equation}
\label{equation: Alfvén 1}
    \frac{R_{A}}{R_{\star}} \approx 1+(\eta_{\star}+1/4)^{1/(2q-2)}-(1/4)^{1/(2q-2)},
\end{equation}
where $R_{A}$ is the estimated Alfvén radius, $R_{\star}$ is the stellar radius, and $q$ refers to the radial field scaling. We will assume the stellar field is dipolar, which implies $q=3$, though we acknowledge that stellar fields have higher-order contributions. With this, Equation~\ref{equation: Alfvén 1} reduces to:
\begin{equation}
\label{equation: Alfvén 2}
\frac{R_{A}}{R_{\star}} \approx 0.29 + (\eta_{\star} + 0.25)^{
1/4} \eta_{\star} >> 1.
\end{equation}

Applying Equation \ref{equation: Magnetic confinement} to known solar values yields:
\begin{equation}
\label{equation: Solar Alfvén}
    \eta_{\star, \odot} = \frac{(2.9\; \mathrm{G})^{2}(6.96 \times 10^{10} \; \mathrm{cm})^{2}}{(4\times 10^{7} \; \mathrm{cm/s})(2.5 \times 10^{-14}\; M_{\odot}/\mathrm{yr})} \approx 650.
\end{equation}
Thus, solar-like stars seem to satisfy the condition of $\eta_{\star} >> 1$, which is in the "strong confinement" regime. Using this value gives us an estimated Alfvén surface of $R_{\mathrm{A}, \odot} \sim 5.3 R_{\odot}$. This discrepancy from the \emph{in situ} measurements from the PSP likely stems from the dipole approximation. Assuming the scaling of the confinement parameter holds, we normalize equation~\ref{equation: Alfvén 2} to the solar value:
\begin{equation}
\label{equation: Alfvén 3}
     \frac{R_{A}}{R_{\star}} \approx 0.61 + 2.1(\eta_{\star} + 0.25)^{1/4} .
\end{equation}

We use this relation to derive the expected average Alfvén radii for the exoplanet hosts analyzed in this work.

\begin{figure*}[!ht]
    \centering
    \includegraphics[width=\textwidth]{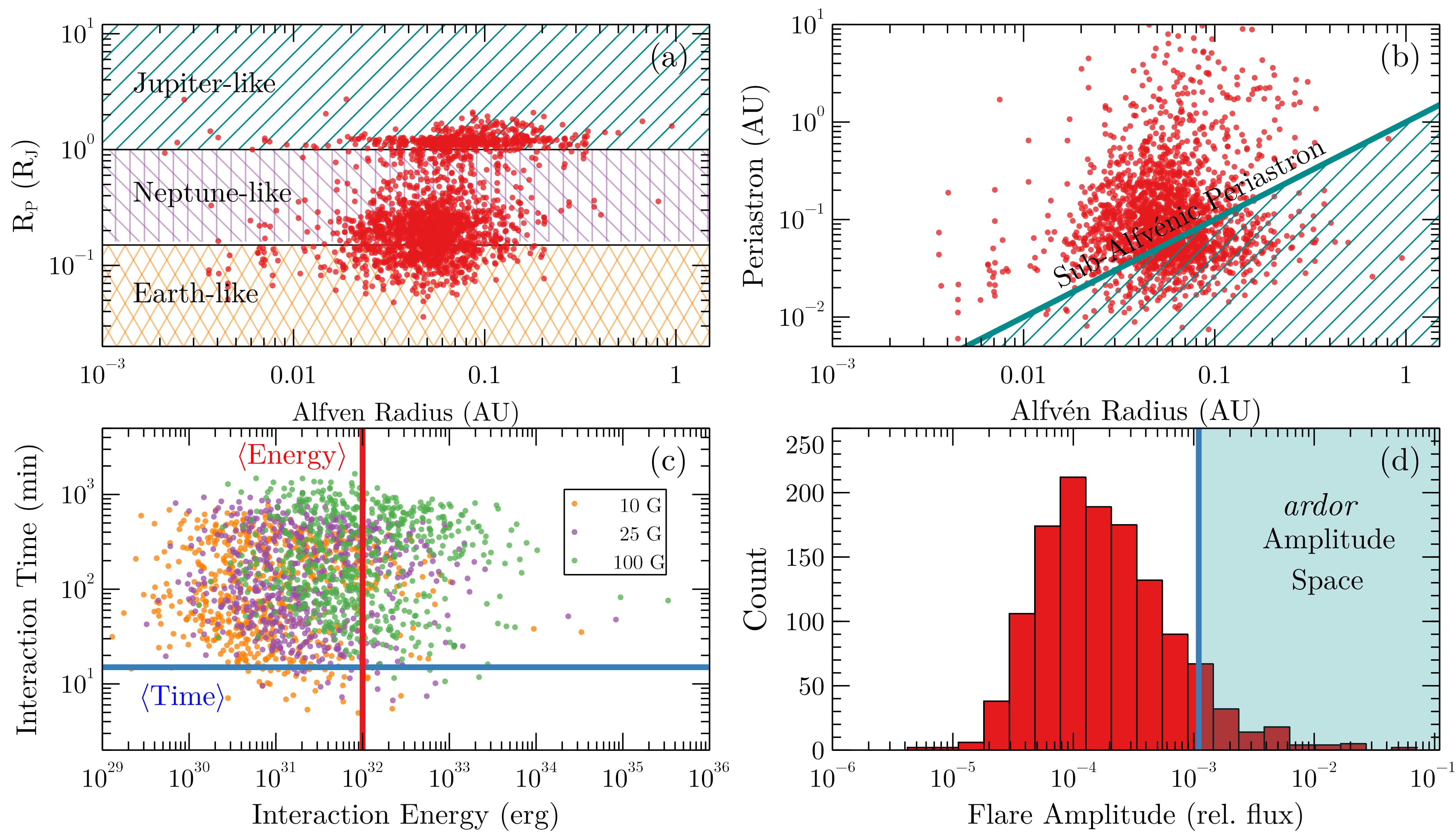}
    \caption{Estimated induced-flare parameters derived from quantities defined throughout Section~\ref{section: SPMI Models} for known exoplanet systems. (a) Scatterplot of host Alfvén radius vs closest orbiting planet radius. (b) Scatterplot of host Alfvén radius vs. periastron distance of the closest approaching planet. (c) Scatterplot of induced-flare amplitudes using the timescale and energy estimates predicted by equation~\ref{equation: Field Energy Difference} and equation~\ref{eq: timescale}. The vertical and horizontal lines indicate the average energies and full-width at half-maximum parameters derived from the flare catalog presented in this work. (d) Histogram of predicted induced-flare amplitudes at periastron distance for a planet with a polar field strength of 100 G.}
    \label{Fig: Estimated Results}
\end{figure*}

\subsubsection{Uncertainties in Alfvén Surface Estimations}
\label{subsubsection: Alfven Error}

Section~\ref{subsection: Estimating Flare Parameters} develops plausibility that many known planetary hosts likely have sub or trans-Alfvénic planetary systems and that induced flare energies are detectable in TESS photometry. To assess the error in the Alfvén surface values, we use Monte Carlo (MC) methods, randomly sampling each relevant parameter from assumed normal distributions. We then construct $1\sigma$ confidence intervals from the resulting samples of Alfvén radii. The error accumulation in the measured parameters and the power-law relationships result in $\mathbf{1\sigma}$ confidence intervals with typical relative uncertainties $\sim\mathbf{200-300}\%$. (a) and (b) in Figure~\ref{Fig: Estimated Results} are maximum likelihood estimates of host Alfvén surfaces and should be interpreted cautiously.

Improved estimates of stellar Alfvén surfaces require characterization of the stellar magnetic environment using ZDI in tandem with MHD simulations (e.g., \citet{Vidotto2014}). Even in this situation, the Alfvén surface is radially inhomogeneous and time-dependent on short (i.e., rotation of active regions) and long (i.e., stellar cycles) time scales. Additionally, estimates are informed only by the large-scale structures inferred from ZDI imaging. Small-scale structures (e.g., helmet streamers and coronal loops) play a pivotal role in shaping the Alfvén surface. This was observed \emph{in situ} by the Parker Solar Probe's first entrance of the surface at $\sim0.09\, \mathrm{AU}$, particularly extended from its typical value of $\sim0.07 \, \mathrm{AU}$ due to the presence of a helmet streamer \citep{Kasper2021, Chhiber2024}.

\section{Analysis Methods}
\label{section: Analysis Methods}
We present our flare detection pipeline, \texttt{ardor}, and the diagnostic tests used to evaluate its efficacy in both the injection-recovery and precision-recall space. We then present two statistical tools, goodness-of-fit tests and unbinned likelihood analysis, which we use to identify candidate induced flares.
\subsection{\textsc{ardor}: A Multi-Tiered Bayesian Flare Detection Pipeline}
\subsubsection{TESS 2-Minute Photometry}
\label{subsection: TESS Data}
Since 2018, the TESS mission has produced time-series photometry for 200,000 stars across 96\% of the celestial sphere \citep{Ricker2010}. It uses four CCD cameras with a bandpass of $600-1000\,\mathrm{nm}$ and a total field of view of $24 \times 96$ square degrees. Each sector contains \~27 days of photometric data. The Science Processing Operation Center (SPOC) pipeline generates time-series photometry for $\approx20000$ stars at a 2-minute cadence, with $\approx1000$ at a 20-second cadence \citep{Jenkins2016}. Time-series light curves are publicly available with each TESS data release and can be accessed via the Mikulski Archive for Space Telescopes (MAST) database. Each TESS light curve contains Simple Aperture Photometry (SAP) flux and Pre-search Data Conditioning Simple Aperture Photometry (PDCSAP) flux data. PDCSAP flux is pre-processed to remove long-term systematic baseline trends and scattered light from when the Earth or Moon comes into view. PDCSAP flux is the data product used in this analysis, as removing systematic baseline trends has little effect on the timescales of flares and is generally easier to work with.

\subsubsection{Tier 0: Detrending}
\label{subsection: Tier 0}
Photometric time-series data can exhibit long-term trends, such as sinusoidal baselines resulting from stellar activity. These must be removed in flare searches without significantly altering the underlying signal. For \textsc{ardor}, this is implemented by applying a Savitzky-Golay (SG) filter, defined by:

\begin{equation}
Y_{j} = \Sigma^{m / 2 - 1 / 2}_{i=1 / 2 - m / 2} C_{i} y_{j+i}
\end{equation}

where $Y_{j}$ is the smoothed data, $m$ is the polynomial order, $C_{i}$ is the $i\mathrm{th}$ convolution coefficient, and $y_{i+j}$ is the raw data \citep{Savitzky1964}. \textsc{ardor} wraps the \code{flatten} method in the \emph{LightKurve} package, which implements SG filtering while simultaneously excluding specified outliers in a window. This work utilized a window size of 401, $m=3$, and excluded outliers up to $3\sigma$ deviation from the local median flux. This method is computationally efficient but can encounter difficulties when the baseline frequency is high (see Section~\ref{subsection: False-Positives}).

\subsubsection{Tier 1: Flare Candidate Identification}
\label{subsection: Tier 1}
After detrending, flare candidates are identified using a $3\sigma$ cutoff above the median baseline flux. Since the baseline noise can vary significantly within individual light curves, the standard deviation value used in this cutoff is based on a local window of 100 data points, symmetrically centered around the analyzed data point, and excluding outliers at $3\sigma$. To reduce computation time, the window shifts every 10 points, except at the edges of the light curve, where the window remains static for the first and last 100 data points. Each candidate is vetted further depending on which cadence data is used. For the two-minute cadence, we require a single data point to be above the $3\sigma$ threshold, with the following two points above at least $1\sigma$ above the baseline to be considered a candidate. For the 20-second cadence, we require one point to be above $3\sigma$ and five subsequent data points to be above $1\sigma$. This ensures that individual, spurious peaks are not considered candidates. This approach differs marginally from other flare pipelines \citep[e.g.][]{Gunther2020, Pietras2022, Ilin2024}, which require three consecutive points to be above $3\sigma$. Consequently, \textsc{ardor} is more sensitive to shorter time scale flares at the cost of precision, which is addressed in Section~\ref{subsection: Completeness}.

\subsubsection{Tier 2: Coarse Model Fitting}
\label{subsection: Tier 2}
Once the flare candidates are identified, the time, flux, and flux error values associated with the candidate are isolated. 50 data points on either side of the peak are kept as a buffer to establish the relative baseline flux further along in the pipeline. Using the duration of the flare given by Tier 1, the horizontal asymptote of the flare is removed by subtracting off the value of the data point after the duration given by Tier 1, under the condition that it is lower in value than the previous point to avoid domain issues with the $\log$ function. If not, then the subsequent value is used, and so forth. A $\log$ transform is applied to the candidate flare beginning at the peak value, as shown here: 
\begin{equation}
\label{equation: log-linear transform}
    y = a\exp[-bt] \rightarrow\ln(y)=\ln{a}-bt.
\end{equation}

Here, $\ln{a}$ is used to find the amplitude, and $b$ is the decay constant used to estimate the FWHM. Error is propagated for the relative flux data to account for the log transform and asymptote removal:
\begin{equation}
    \label{equation: tier 2 error prop}
    \delta{\,\ln{y}} = \frac{\sqrt{\delta{y}^{2}+\delta{y_{n+1}^2}}}{y}.
\end{equation}
$y_{n+1}$ refers to the data point used to remove the horizontal asymptote, while $y$ is an arbitrary flux point in the decay profile of the flare. During flare vetting, \textsc{ardor} requires $a,b>0$, ensuring a positive amplitude and exponential decay. The tentative flare epoch is then defined as the time when the maximum amplitude in the data occurs. To assess goodness-of-fit, we compute the reduced chi-square statistic, $\chi^{2}_{\nu}$, on the model parameters derived by using the \texttt{scipy.optimize.curve\_fit} method on the log-transformed data. This is defined by:

\begin{equation}
    \chi^{2}_{\nu} = \frac{\chi^{2}}{\nu}; \; \chi^2 = \sum_{i} \left(\frac{(x_{i}- y_{i})}{\sigma_{i}} \right)^{2}.
\end{equation}

$\nu = n-m$ is the degrees of freedom, where $n$ is the number of data points in the flare, and $m=2$ is the number of fitted parameters. Then, $x_{i}$ is the logarithmic transformed flux, $y_{x}$ is the linear model, and $\sigma_{i}$ is the uncertainty of the respective data point. This test filters spurious noise signals from shorter timescale flares and large structures that may fool the initial Tier 1 flare detection. Importantly, flares do not always exhibit exponential decay profiles, and secondary outbursts are common after the initial peak. In order not to overly constrain candidates to an exponential decay profile (see \citet{Davenport2014} for discussions of flares that deviate from this prescription), we set $\chi^{2}_{\nu} = 20$. In most applications, this would be considered too high a threshold; however, following precision-recall analysis in \ref{subsubsection: PR}, this threshold sufficiently encompasses the variation seen in the decay phases of real flares.

\begin{figure*}[!ht]
    \centering
    \label{ardor: Tier 1/Tier2}
    \includegraphics[width=\textwidth]{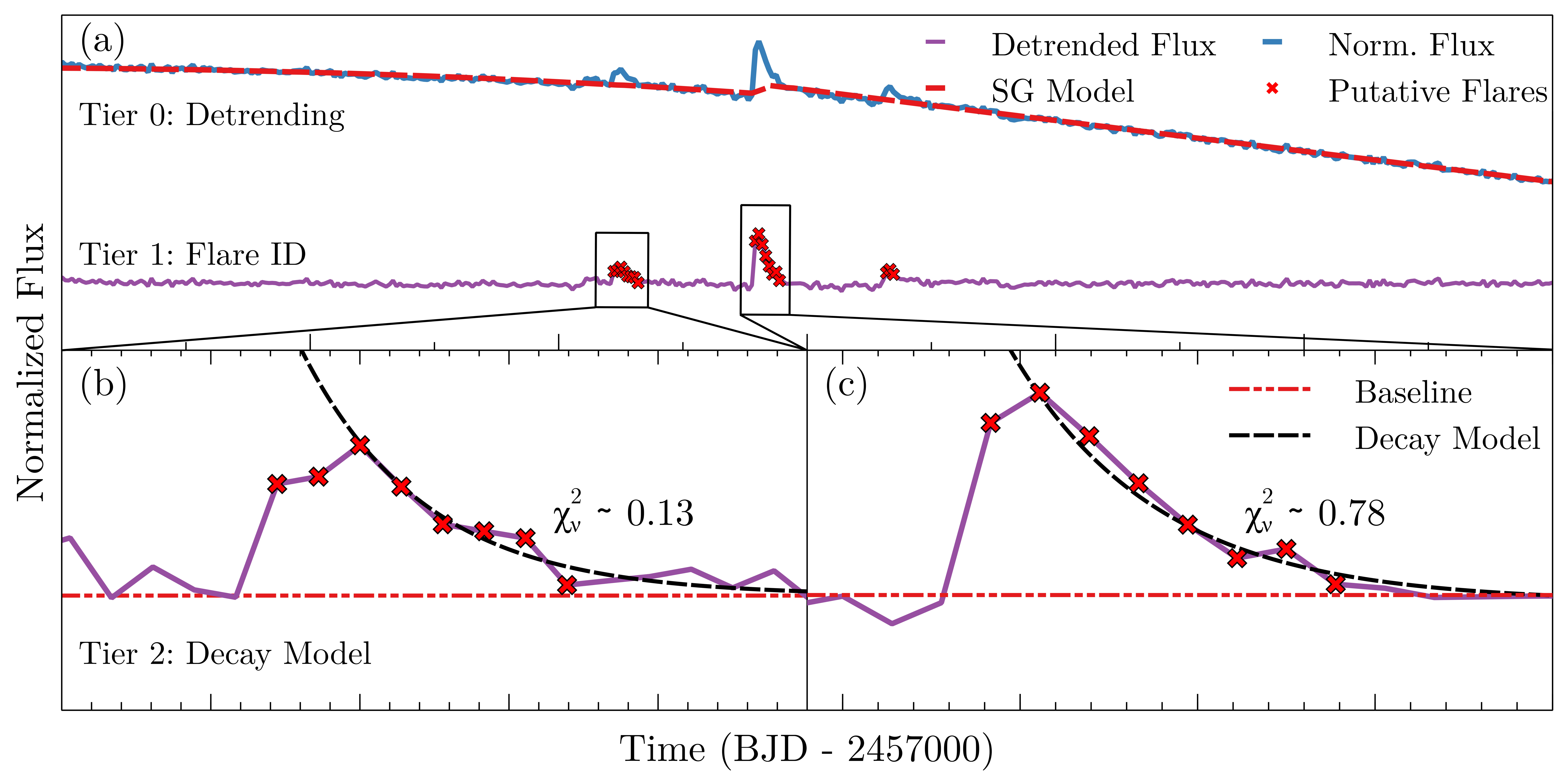}
    \caption{A diagram of the first two tiers of \textsc{ardor}. (a) Tiers 0 \& 1 of \textsc{ardor}. The normalized PDCSAP flux (blue) is shown, along with the underlying trend (dashed red), determined by the Savitzky-Golay filter. The purple shows the detrended normalized flux, with the red crosses representing putative flare detections from Tier 1 of \textsc{ardor}. (b-c) Candidate flares which pass the $\chi^{2}$ test from Section~\ref{subsection: Tier 2}. The dashed lines represent the best-fit exponential decay profiles, along with their corresponding $\chi_{\nu}^{2}$ values. NOTE: As described in Section \ref{subsection: Tier 2}, the putative flares are log-transformed with a linear fit used to compute $\chi^{2}_{\nu}$. The curves are shown to visualize decay profiles in the natural parameter space.}
\end{figure*}

\subsubsection{Tier 3: Model Comparison}
\label{subsection: Tier 3}

To ensure that the outputs of Tier 2 are consistent with accepted flare morphologies, the single peak flare model developed by \citet{Davenport2014} is fitted to each flare using Dynamic Nested Sampling, performed using \textsc{allesfitter} \citep{allesfitter-code, allesfitter-paper}. Additionally, a hybrid spline is fitted to model locally variable noise structures. Following the analysis of \citet{Gunther2020}, the Bayes factor, $\log{Z}$, is calculated for both models using Dynamic Nested Sampling \citep{Speagle2020dynesty}. To pass tier 3, the flare model must increase the Bayes factor by at least two, as shown in equation~\ref{equation: Bayes factor}. This was done to recover as many low-energy flares as possible, thereby covering as much of the flare parameter space as possible, consistent with the estimated flare energies in Figure~\ref{Fig: Estimated Results}. Although we include all flares with $\Delta \log{Z}>2$ in our search, it is a critical metric to evaluate the quality of the induced flare detection. 
\begin{equation}
\label{equation: Bayes factor}
\log{Z}_{Flare} - \log{Z}_{Noise} = \Delta \log{Z} \geq 2
\end{equation}

\begin{figure}
    \includegraphics[scale=0.575]{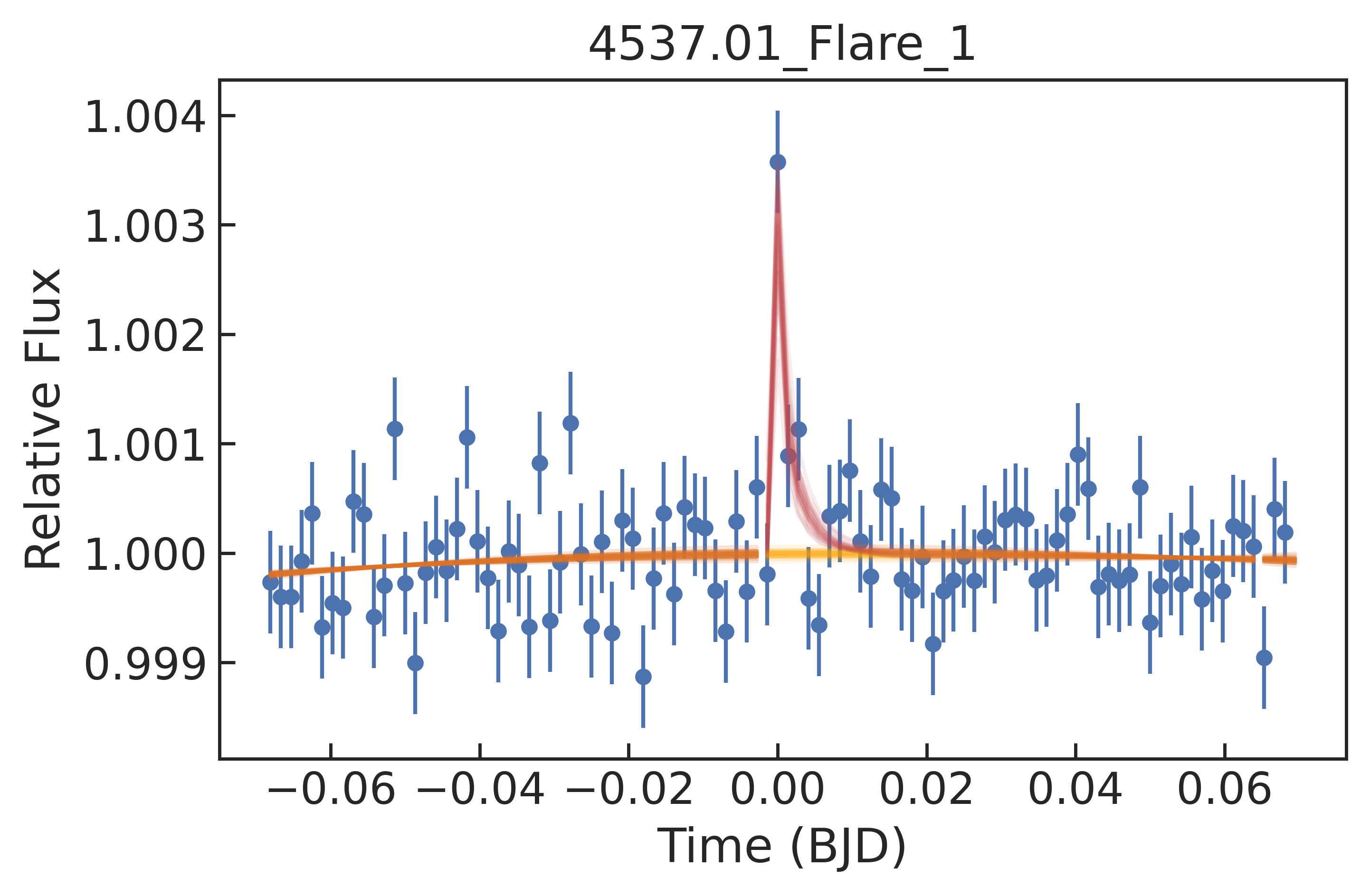}
    \caption{Two example outputs of the model comparison described in Section \ref{subsection: Tier 3} using dynamic Nested Sampling. The top graph shows a distinct flare, with the red line corresponding to the flare model derived from Dynamic Nested Sampling. The $\Delta \log{Z}$ from this fit was 1305, suggesting the flare model fits the data significantly better than the baseline model. This flare is consequently accepted. Comparatively, the bottom graph shows a candidate flare with a $\Delta \log{Z} = 0.045$, suggesting that this candidate is likely a noise structure. This candidate is rejected.}
    \label{fig:Model-Compare}
\end{figure}

Once a flare has passed model comparison, the posterior distributions of the parameters of the single-peak flare model (flare epoch ($t_{0}$), amplitude ($A$), and full width at half maximum (FWHM, $\tau$) are sampled via Markov-Chain Monte Carlo (MCMC) \citep{allesfitter-code, allesfitter-paper}. We use a hybrid spline to estimate the local baseline flux around the flare, with the peak data point (i.e., the best guess for the amplitude from the decay fit) centered at $t=0$. The prior distributions are created from the output of Section~\ref{subsection: Tier 2}. The best-guess values of each parameter are derived from \ref{equation: log-linear transform}, with amplitude derived from $a$, FWHM is estimated analytically from $b$, and the flare epoch is calculated by $t_{0}$. The MCMC is run for 1000 steps, 400 burn-in steps, and 10 walkers. 

We only consider 'flare morphologies with a single peak; however, as suggested in other flare analyses \citep[e.g.,][]{Davenport2014, Gunther2020}, flares commonly occur in 'outbursts,' with multiple sub-peaks and independent decay phases. In searching for magnetic SPIs, we are interested in the epochs of individual events, so outbursts with multiple sub-peaks are treated as separate events. However, models with sub-peaks are expected to be incorporated as \textsc{ardor} is updated.

\subsubsection{Estimating Bolometric Flare Energies}
\label{subsection: Flare energy}

To estimate the bolometric flare energy, we follow \citet{Shibayama2013}. Each star's luminosity is determined either by reported luminosity values in the NASA Exoplanet Archive or estimated using the luminosity-radius-mass equation:
\begin{equation}
\label{equation: RTL relation}
    L = 4\pi R^{2} \sigma_{B} T^{4}
\end{equation}
To relate the star's luminosity to the flare's luminosity, we treat the flare as a blackbody with $T = 9000 \pm 500 \mathrm{K}$, which has been used as a lower limit in several other flare studies \citep[e.g.]{Kowalski2009, Davenport2014, Davenport2019, Gunther2020}. We can express the star and flare luminosity in the TESS bandpass as
\begin{equation}
\label{equation: star Lumin}
    L_{\mathrm{\star}} = \pi R_{\star} \int_{600 \mathrm{nm}} ^{1000 \mathrm{nm}} R_{\lambda} B_{\lambda} (T_{\mathrm{eff}}) d\lambda
\end{equation}
\begin{equation}
\label{equation: flare Lumin}
    L_{\mathrm{flare}} = \pi A_{\mathrm{flare}} \int_{600 \mathrm{nm}} ^{1000 \mathrm{nm}} R_{\lambda} B_{\lambda} (T_{\mathrm{flare}}) d\lambda
\end{equation}

Here, $R_{\lambda}$ corresponds to the TESS response function, $B_{\lambda}$ is the Planck function evaluated at either the effective stellar temperature for the star or those above $9000 \mathrm{K}$ for the flare. $A_{\mathrm{flare}}$ is the integral of the flare over the normalized baseline flux of the star. The normalized baseline flux derived from the TESS data gives the relative flare amplitude:
\begin{equation}
\label{equation: Relative flux}
    \frac{\Delta F_{\mathrm{flare}}}{F_{\mathrm{\star}}} (t) = \frac{L_{\mathrm{flare}}}{L_{\star}}
\end{equation}
We can substitute (\ref{equation: star Lumin}) and (\ref{equation: flare Lumin}) to solve for $A_{\mathrm{flare}}$:
\begin{equation}
\label{equation: Flare area}
    A_{\mathrm{flare}} = \frac{\Delta F_{\mathrm{flare}}}{F_{\mathrm{\star}}} (t) \frac{\pi R_{\star} \int R_{\lambda} B_{\lambda} (T_{\mathrm{eff}}) d\lambda}{\pi \int R_{\lambda}B_{\lambda} (T_{\mathrm{flare}}) d\lambda} 
\end{equation}

By treating the integrated area under the flare as the effective area of emission, we can use (\ref{equation: RTL relation}) to write:
\begin{equation}
 L_{\mathrm{flare}} = 4\pi A_{\mathrm{flare}} T_{\mathrm{flare}}^{4}
\end{equation}
Substituting in for (\ref{equation: Flare area}), the bolometric flare energy is defined by
\begin{equation}
    L_{\mathrm{flare}} = 4\pi R_{\star} T_{\mathrm{flare}}^{4} \frac{\Delta F_{\mathrm{flare}}}{F_{\mathrm{\star}}} (t) \frac{\int B_{\lambda} (T_{\mathrm{eff}}) d\lambda}{\int B_{\lambda} (T_{\mathrm{flare}}) d\lambda}.
\end{equation}

\subsection{Pipeline Completeness}
\label{subsection: Completeness}
We perform injection-recovery and precision-recall tests to assess the performance of \textsc{ardor} across the flare parameter space as well as its ability to vet false positives.
\subsubsection{Precision-Recall}
\label{subsubsection: PR}
We construct precision-recall (PR) curves of both \ref{subsection: Tier 1} and \ref{subsection: Tier 2} by using flare injection (Tier 1) and known flares in TESS light curves (Tier 2). Constructing PR curves depends on the following classifications and definitions:
\begin{itemize}[noitemsep,topsep=0pt]
    \item \emph{True positives}: recovered injected flares.
    \item \emph{False negatives}: unrecovered injected flares.
    \item \emph{False positives}: recovered non-flare signals.
    \item \emph{Precision}: of recovered events, what fraction are flares?
    \item \emph{Recall}: of all flares, what fraction are recovered?
\end{itemize}
\vspace{+0.1cm}
In most machine-learning applications, PR curves are constructed by changing the classifier parameter ($\sigma_{T}$ for \ref{subsection: Tier 1}, $\chi^{2}_{\nu} $ for \ref{subsection: Tier 2}) and see how precision and recall respond. The PR Area Under the Curve (AUC) is a metric that quantifies \textsc{ardor}'s performance at each classification step. A PR AUC of $1$ is a "Perfect Classifier," and a PR AUC of $0.5$ is considered non-predictive. The closer the PR AUC is to 1, the more efficient the pipeline is at recovering true flares while minimizing false positives. We use the following set of classifiers for the PR curves for Tier 1 and Tier 2:
\begin{align}
\label{Equation: PR Metrics}
    \sigma_{T} &= \{ 2, 2.5, 3, 3.5, 4, 5 \}\, \mathrm{(Tier\,1)} \\        
        \chi^{2}_{\nu} &= \{ 0.1, 0.25, 0.5, 1, 10, 20 \} \, (\mathrm{Tier\,2})
\end{align}

For \ref{subsection: Tier 1}, the PR curves were constructed by randomly injecting fifteen flares into 200 randomly selected TESS light curves of G and M-type stars. The light curves used were visually vetted to ensure that no considerable flaring was present, which could erroneously label a real, non-injected flare as a false positive. Flares were automatically marked as a \emph{true positive}, \emph{false negative}, or \emph{false positive}, based on whether the flare was injected and successfully recovered.

\begin{figure}[!ht]
    \centering
    \includegraphics[width=0.47\textwidth]{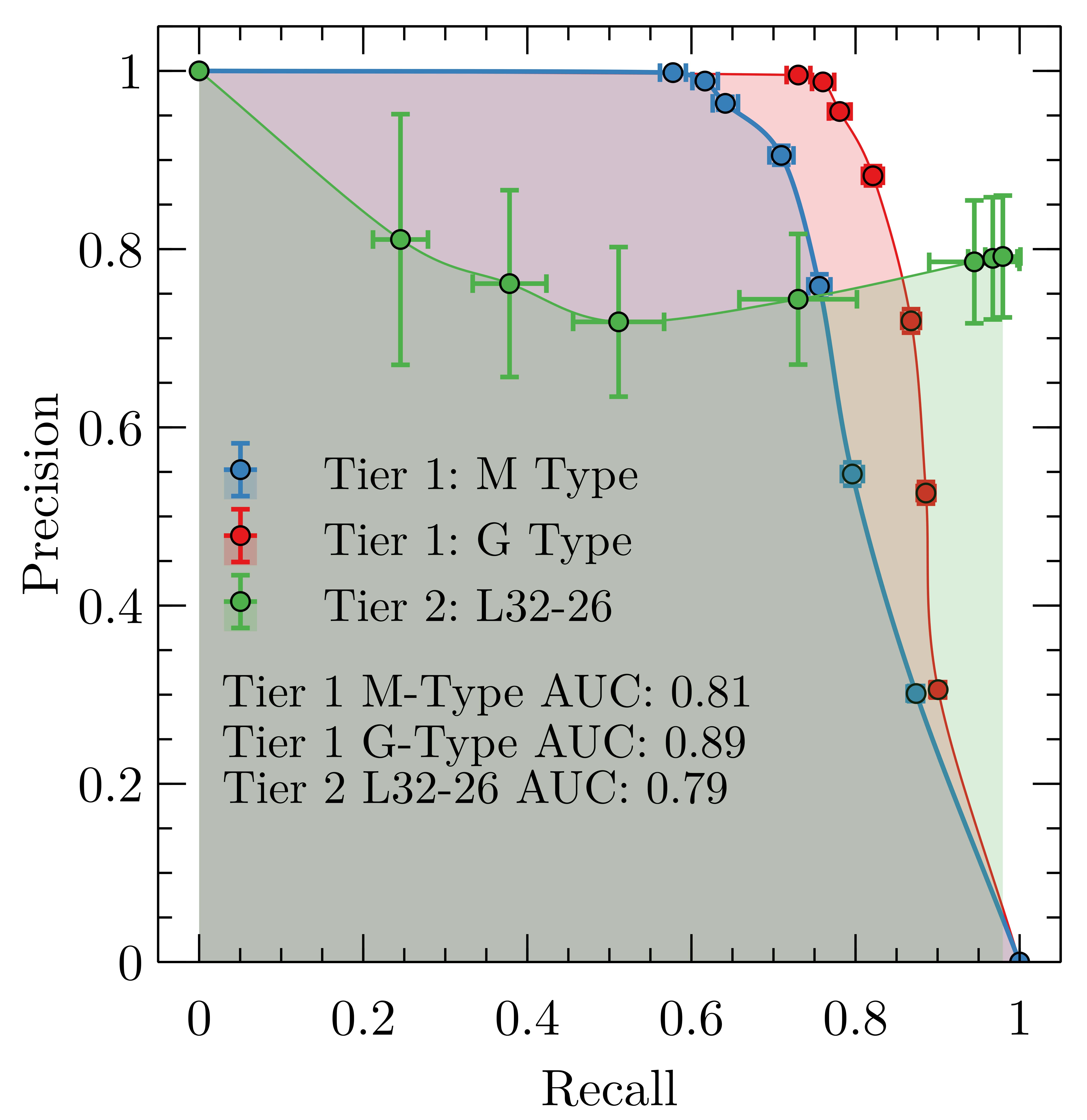}
    \caption{Precision-Recall curves of \ref{subsection: Tier 1} and \ref{subsection: Tier 2}. For Tier 1, as $\sigma_{T}$ increases, recall decreases while precision increases. For Tier 2, the precision decreases as recall increases until recall reaches 0.5 ($\chi^{2}_{\nu}=1$), where the curve inflects, and both precision and recall increase simultaneously. Errors are computed by assuming Poisson variance for each set of counts ($\propto\sqrt{N}$) and propagating, truncating where appropriate to ensure neither precision nor recall exceeds unity.}
    \label{fig: PR Curves}
\end{figure}

The above injection strategy was attempted for \ref{subsection: Tier 2}; however, the efficacy of \ref{subsection: Tier 2} depends on how strongly the decay profile is $\propto e^{-at}$. Since the FRED profile is by construction $\propto e^{-at}$ during the decay phase if a true flare passed Tier 1, any $\chi^{2}_{\nu} \sim 1$ threshold would pass. This does not inform how real flares will be retrieved that have decay phases deviating from an exponential profile, which $\chi^{2}_{\nu}$ is designed to measure. Thus, to construct a PR curve for \ref{subsection: Tier 2}, real TESS light curves were needed to proxy as a flare injection method. We used eight TESS light curves of the M3 dwarf L34-26, as it has a high density of flares per TESS light curve. We used a low Tier 1 threshold of $\sigma_{T}=2.5$ and visually vetted flares as either being true flares or false positives. Then, the $\chi^{2}_{\nu}$ thresholds described in \ref{Equation: PR Metrics} are used within Tier 2 to retrieve the vetted flares. The resulting curves and AUCs are shown in Figure \ref{fig: PR Curves}, with Tier 1 achieving AUCs of 0.81 and 0.89 for M-type and G-type light curves, respectively, and Tier 2 achieving an AUC of 0.79 for the visually vetted flares of L32-26. 

\subsubsection{Injection-Recovery}
\label{subsubsection: IR Param}
\begin{figure}[!ht]
    \centering
    \includegraphics[width=0.47\textwidth]{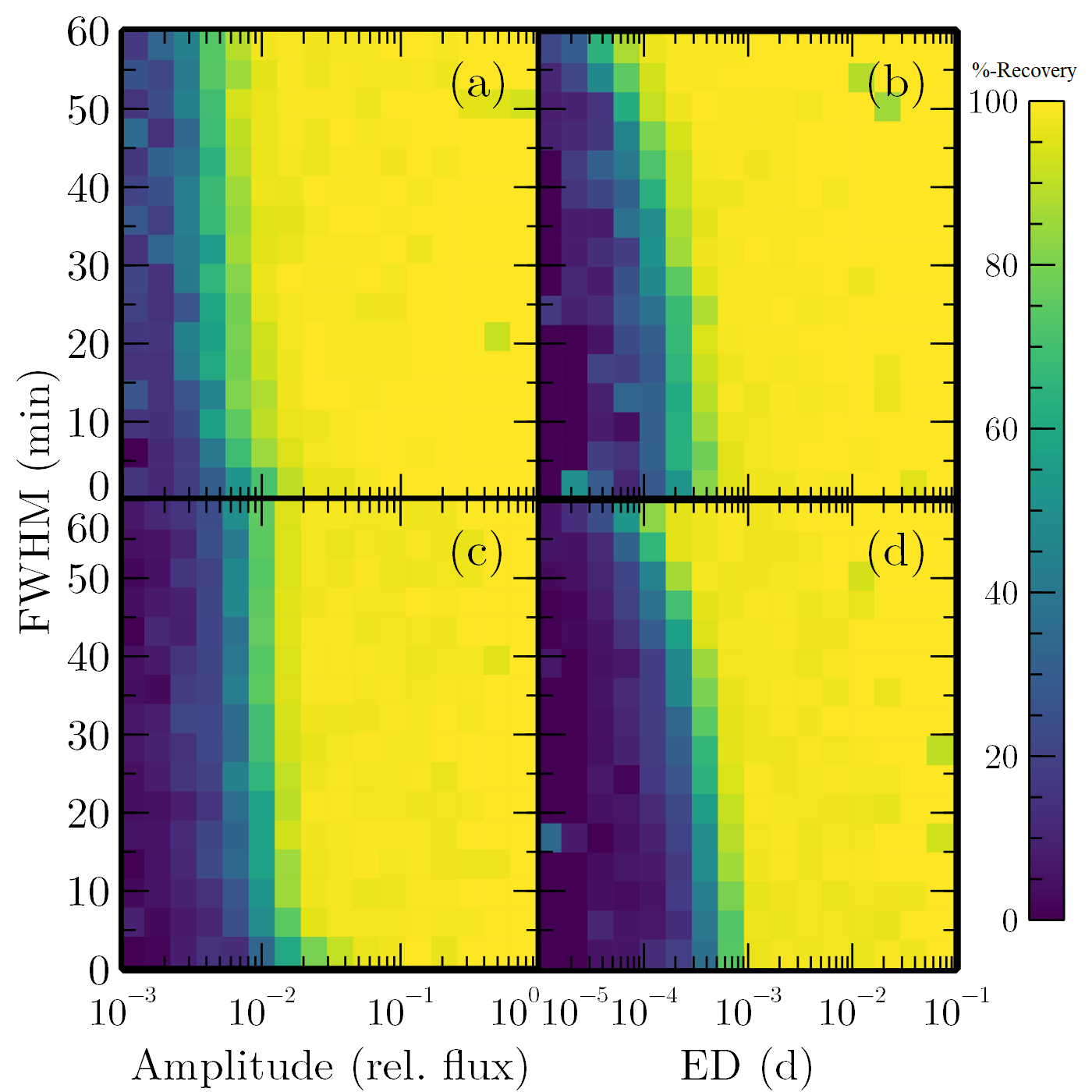}
    \caption{Injection-recovery metrics for Tier 1 of \textsc{ardor}. (a) Heat maps show the recovery rate of Tier 1 as a function of flare amplitude and FWHM for G-type stars. (b) Heat-map showing recovery rate as a function of integrated flare area and FWHM for \textsc{ardor}. (c-d) Identical to (a-b), but with M-type stars.}
    \label{fig: IR Param Figure}
\end{figure}

To assess how \textsc{ardor} performs across the flare parameter space, we compute the recovery rate of injected flares for Tier 1. We then construct heat maps that quantify the recovery rate across the FWHM and the amplitude. We sample amplitudes using a log-normal distribution with $\mu=1.25$ and $\sigma=-3$ and sample the FWHM using a uniform distribution ranging from 0.5 minutes to one hour. To bin the heatmaps, we take samples from a logarithmic interval of $[-3, 0]$ in amplitude space and a linear interval of $[0, 60]$ minutes in FWHM space. We then inject 15 flares into 200 randomly selected TESS light curves for both G and M-type stars. These are summarized in Figure~\ref{fig: IR Param Figure}.

For G-type stars, \textsc{ardor} maintains a high recovery rate up to amplitudes of $\sim 1\times10^{-2}$. M-type stars show an increase to $\sim 2\times10^{-2}$ due to higher noise. The recovery rate is independent of flare FWHM until the shortest duration flares with FWHM $\leq 5\, \mathrm{min}$.

\subsubsection{Performance at the Photometric Limit}
\label{subsubsection: Photometric Limit}
\begin{figure*}
    \centering
    \includegraphics[width=1\linewidth]{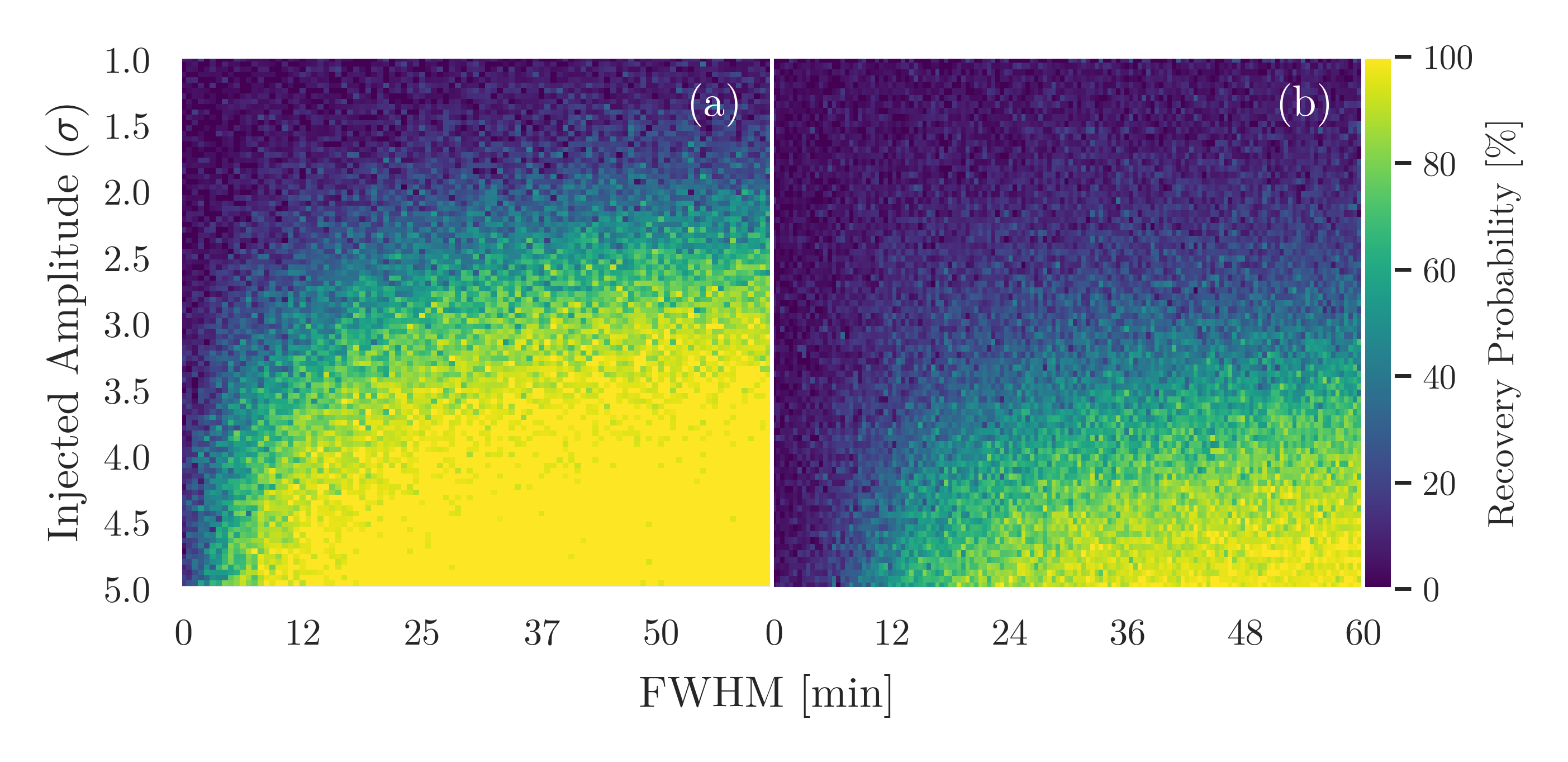}
    \caption{Injection-recovery comparison of low-amplitude flares for (a) tier 1 of \texttt{ardor} and (b) \texttt{AltaiPony}. Flare amplitudes are drawn from a uniform distribution from $[\sigma,5\sigma]$, with $\sigma$ representing the standard deviation of the photometric scatter within $\pm1000$ data points on either side of the flare. Flare FWHM are pulled from a uniform distribution of $(0,60]\,\mathrm{min}$. $\sigma$ is computed before the signal injection. The total number of flares sampled is (a) $N_{\mathrm{flare}}=232851$ and (b) $N_{\mathrm{flare}}=468000$.}
    \label{fig: Pipeline Comparison}
\end{figure*}

\texttt{ardor's} iterative refinement of flare candidates allows sacrificing recall over precision in tier 1, as tier 3 simultaneously vets noise structures while assigning an evidence metric for marginal cases. Photometric flare surveys necessarily undersample low-amplitude, short-duration flares. For example, numerous studies, e.g., \citet{Davenport2014, Gunther2020, ilin2021}, employ a high-precision heuristic that labels a signal as a flare candidate based on three consecutive $3\sigma$ data points. To confirm candidates, the signals are vetted visually. This leads to an inevitable bottleneck: loosening the acceptance function to low-amplitude flares decreases precision, resulting in more visual vetting of false positives. Although visual inspection is critical for eliminating false positives in pipeline development, it becomes increasingly impractical when scaling to $\sim10^{5}$ signals. Moreover, the reliability of this approach declines near the photometric limit, where sample fidelity hinges on the observer’s ability to distinguish signal from noise. To visualize the undersampling of low-amplitude flares in conventional photometric flare pipelines, we run a comparison between \texttt{ardor} and \texttt{AltaiPony} in Figure~\ref{fig: Pipeline Comparison} to compare recall in flare amplitude-duration space. We use the sample light curve from the G-type star, TIC 219089288, from TESS Sector 19, and perform injection-recovery tests using \texttt{ardor} and \texttt{AltaiPony}. For amplitudes, we sample from a uniform distribution between $[\sigma,5\sigma]$, where $\sigma$ is the photometric scatter in the detrended light curve within $\pm1000$ data points of the injected flare. Flare FWHM is again sampled from a uniform distribution between $(0,60]$ minutes. Flare recovery is pushed both in duration and amplitude space, allowing for a more complete sample in the low-amplitude limit of the flare frequency distribution.

\begin{figure*}
    \centering
    \includegraphics[width=0.975\textwidth]{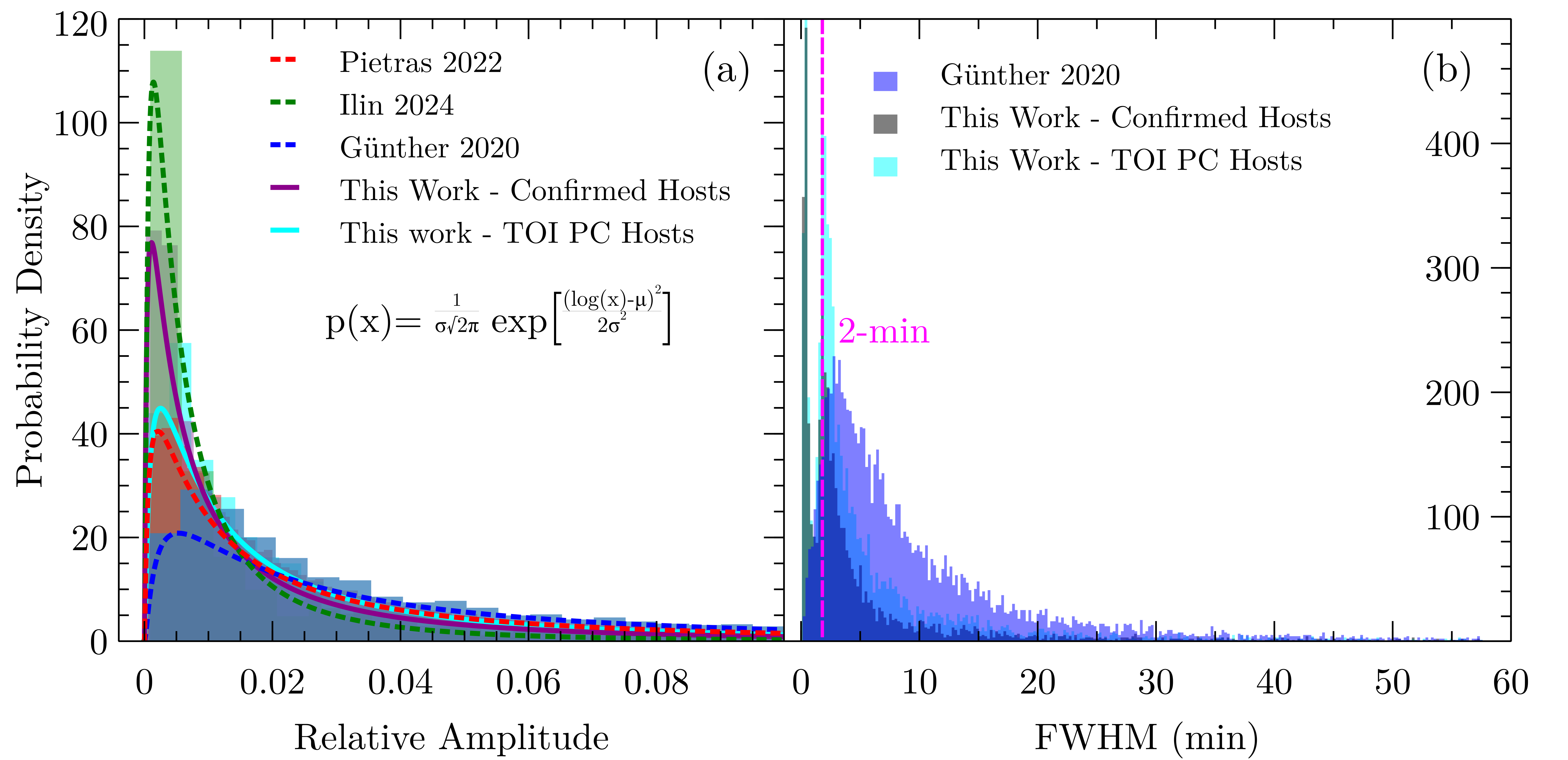}
    \caption{Empirical distribution of flare amplitudes from \citet{Gunther2020}, \citet{Pietras2022}, \citet{Ilin2024}, and this work. (a) Amplitude distributions of each catalog. The log-normal distributions corresponding to each sample are shown as either solid (this work) or dotted lines. (b) The FWHM distributions (as parameterized in \citet{Davenport2014}) of the flare samples in this work and in \citet{Gunther2020}.}
\end{figure*}
\subsubsection{Time Complexity}
\label{subsubsection: Time Complexity}
\begin{figure}
    \label{fig: Time_Complex}
    \includegraphics{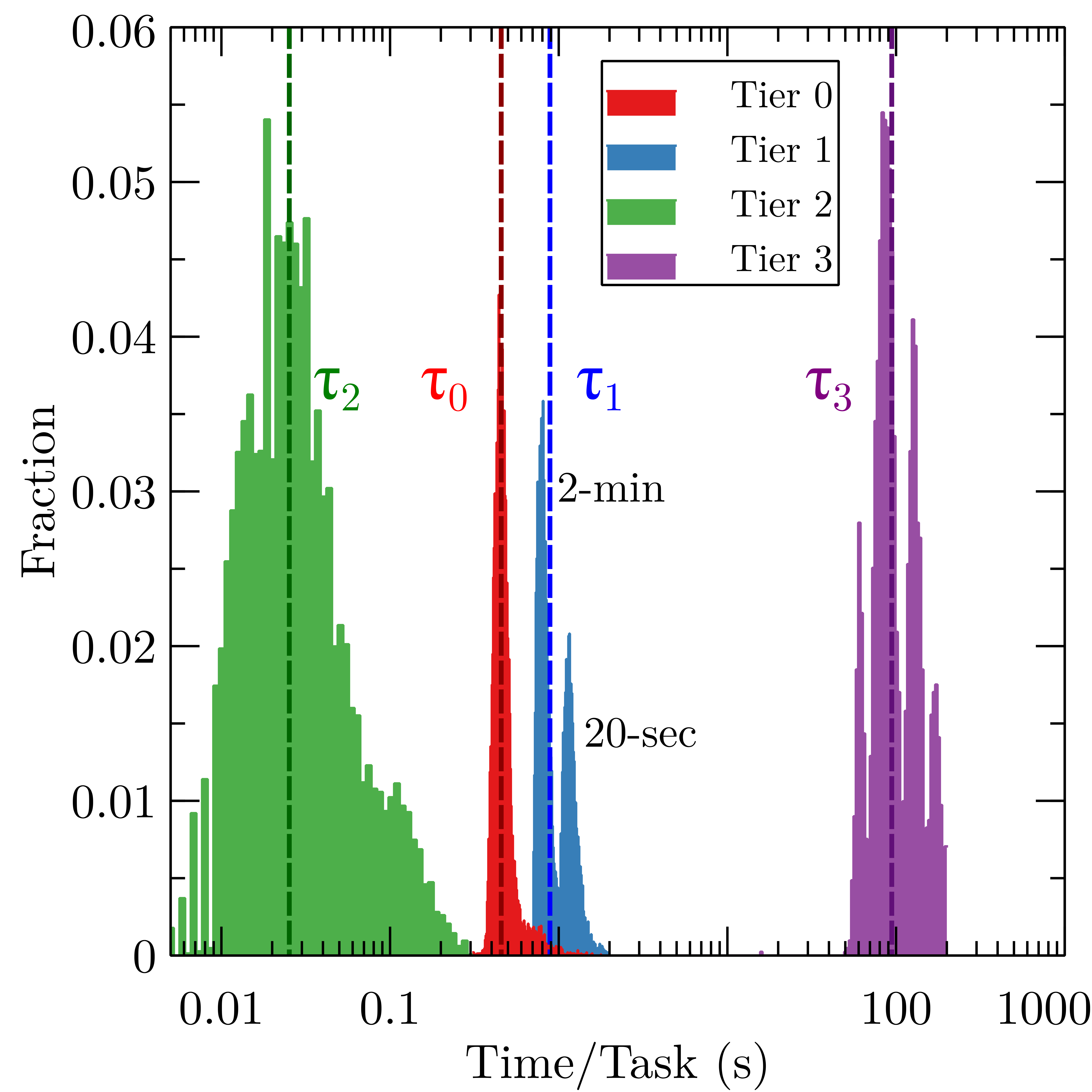}
    \caption{The time complexity of \textsc{ardor}. Reported times are for individual operations for each tier. For Tiers 0 and 1, this is the time taken to analyze the entire light curve. For Tiers 2 and 3, the times reported are for individual candidate flares. Median timescales for each tier $\tau_{n}$ are given as dashed lines in the respective color.}
\end{figure}

The computational performance of \textsc{ardor} is summarized in Figure~\ref{fig: Time_Complex}. The most time-intensive steps are detrending and performing MCMC/model comparison. Statistics derived from Tiers 0-2 are run in Python using a Ryzen 5 3600x. Tier 3 assumes a cluster of 15 CPU cores in parallel, which significantly decreases computation time. The bimodality observed in Tier 0 (detrending) is due to whether the ZG filter was applied to 20-second or 2-minute cadence data. The multi-peak characteristics in Tier 3 are due to the pipeline being run on different machines. As the pipeline was run on shared computation resources, the typical completion time varied as a function of time, depending on demand.

\subsection{Statistical Tests}
\label{subsection: Stat Tests}
The output of \textsc{ardor} is analyzed using goodness-of-fit tests and unbinned likelihood analysis to assess the clustering of phase-folded flares.

\subsubsection{Goodness-of-Fit Tests}
\label{subsubsection: GoF}
The primary data product of \textsc{ardor} is a list of identified flares with the corresponding orbital phase at the flare epoch. We construct empirical cumulative distribution functions (eCDF) for each target across the planetary phase. The eCDF is defined by
\begin{equation}
    F(\phi) = \frac{1}{N} \sum^{m}_{i=1} I(x_{i} \leq \phi), \quad 0\leq \phi \leq 1.
\end{equation}

To determine the deviation of each of the eCDFs from the expected uniform distribution, we employ three statistical tests: the one-sample Kolmogorov-Smirnov, Anderson-Darling, and Kuiper goodness-of-fit tests \citep{an1933, Anderson1954, Kuiper1960}. 

The one-sample Kolmogorov-Smirnov test (KS test) is a nonparametric test that determines if an eCDF is drawn from a given continuous CDF. The KS test statistic is defined as
\begin{equation}
\label{eq: KS Test}
    D_{n} = \sup_{\phi}| F(\phi) - U(\phi) |.
\end{equation}

\begin{figure}
    \label{fig: CDF}
    \includegraphics[width=0.47\textwidth]{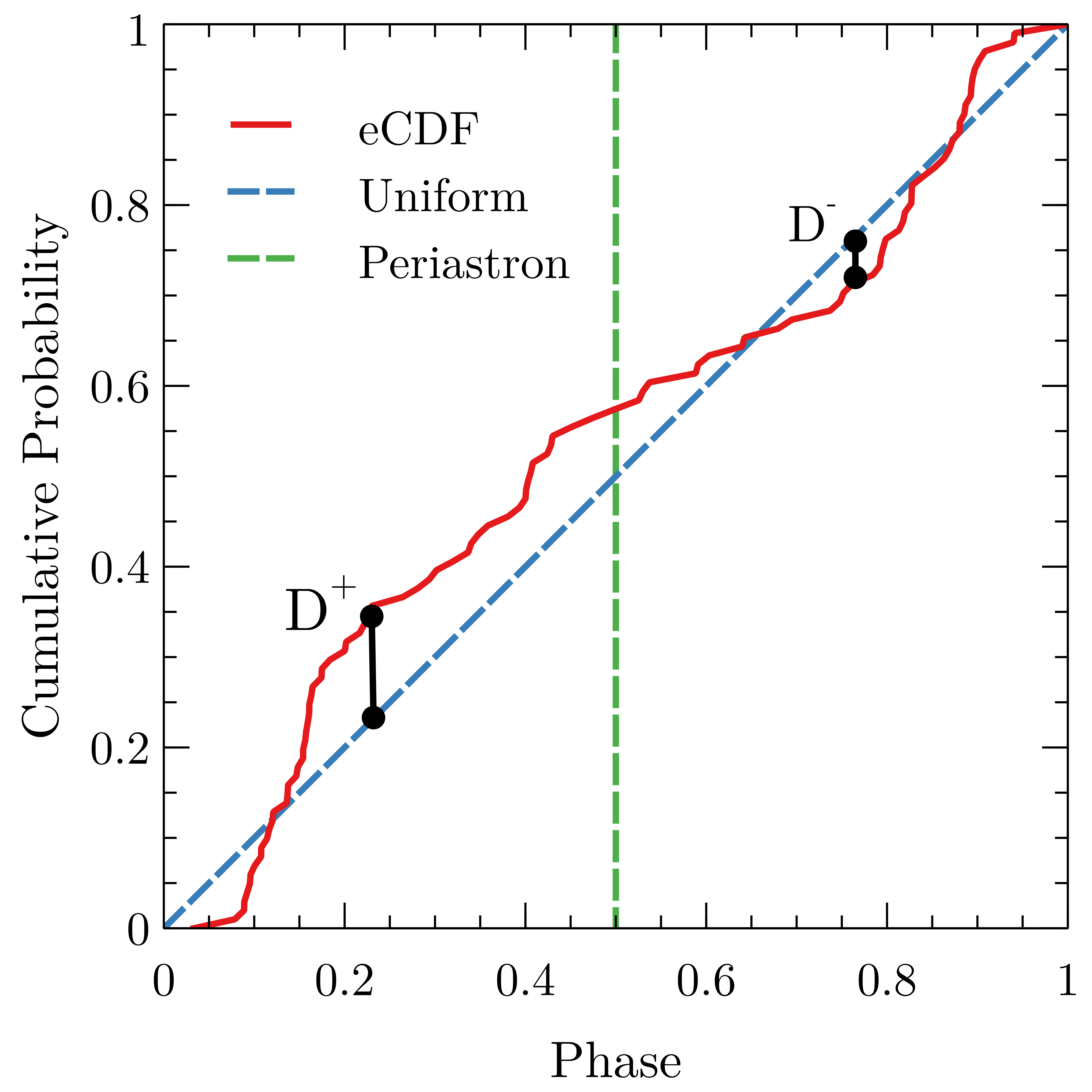}
    \caption{The empirical cumulative distribution function (eCDF) of the flares of the planet candidate TOI-2637.01. The test statistic for the KS test ($D^{+}$) and the Kuiper test ($V=D^{+}+D^{-}$) are labeled. The statistic for the AD test, $A$, does not have a convenient visual analog on the eCDF graph.}
\end{figure}

This is shown in Figure~\ref{fig: CDF} and is the maximum distance between the eCDF and the tested distribution. The KS test is most sensitive to the median of the distribution due to the boundary conditions requiring that $F(0) = 0$ and $F(1) = 1$. 

The Anderson-Darling (AD) test, another nonparametric test, has the following test statistic:
\begin{equation}
\label{eq: AD Test}
    A^{2} = N \int^{\infty}_{\infty} \frac{(F_{N}(\phi)-U(\phi))^{2}}{U(\phi)(1-U(\phi))} dU(\phi).
\end{equation}

The AD test exhibits greater sensitivity at the tails, but loses sensitivity at the median.

Lastly, we employ the Kuiper test. While the KS test statistic is the maximum absolute difference between the eCDF and CDF, the Kuiper test statistic is the sum of the maximum positive and negative differences between the two distributions.

\begin{equation}
\label{eq: Kuiper Test}
    V_{n} = D^{+}_{n} + D^{-}_{n}.
\end{equation}

The summation of discrepancies in either direction makes the Kuiper test invariant to cyclic transformation and maintains equal sensitivity across the whole range of the distributions. This is particularly useful when identifying induced flaring signals in systems where the epoch of periastron is not known.

Each test computes a statistic associated with an underlying distribution. Based on the value of the statistic within the distribution, a corresponding p-value is generated, returning the probability that the eCDF is drawn from the underlying CDF. The p-values from each test are a primary metric for identifying candidates for induced flaring. Several Python packages are used to compute the relevant test statistics and the resulting p-values. The KS test is implemented in the \textsc{scipy} statistics package. The AD test is implemented in the package \textsc{scikit-gof}, which provides implementations of goodness-of-fit tests not found in \textsc{scipy}'s library. Lastly, the Kuiper test is implemented in \textsc{astropy}'s statistics package. Each implementation is open-source and can be readily found on platforms such as \textsc{PyPi} or GitHub.

\subsubsection{Unbinned Likelihood Analysis}
\label{subsubsection: Unbinned Likelihood}

The models described in Section~\ref{section: Modeling} predict the stellar flare rate as a function of planetary phase, $\phi$, given a set of model parameters, $\theta$. To deduce the best-fit parameters for each model, unbinned likelihood analysis is performed, which computes the likelihood of an induced flare model $\mathscr{L}_{1}$. This is done by minimizing the likelihood function:

\begin{equation}
    -\log{\mathscr{L}_{1}} = -\sum_{i}\log{P(x_{i},\theta)}
\end{equation}

where $P(x_{i})$ is the probability of observing event $x_{i}$ with parameters $\theta$. The test statistic is:
\begin{equation}
\label{equation: TS}
    TS = -2(\log{\mathscr{L}_{0}} - \log{\mathscr{L}_{1}}).
\end{equation}
where ${\mathscr{L}_{0}}$ is the null model. Detection significance for unbinned likelihood analysis can be determined by taking the square root of the test statistic, $\sqrt{TS}_{VM}$, which we use as an additional detection metric. The likelihood function is minimized using \textsc{scipy}'s \emph{optimize.minimize}, which estimates the best-fit parameters of the resulting test statistics. Detailed descriptions of how this is implemented for each model are described in Section~\ref{subsubsection: Sim ULA}.

\section{Forward Modeling \& Simulations}
\label{section: Modeling}
We make three assumptions when modeling induced flares. First, induced flares are morphologically indistinguishable from quiescent flaring. This places no priors on the amplitude, energy, or equivalent duration. Secondly, SPIs only occur in sub-Alfvénic regimes of the planet's orbit; only planets entirely within their host's Alfvén surface or that pass under the surface due to, e.g., orbital eccentricity can magnetically interact with their host. Lastly, the interaction SNR is strongest at periastron. The SNR interaction will decrease with increasing distance from the host, with interactions terminating once the planet becomes super-Alfvénic. It is worth mentioning that there are models in which sufficiently conductive, non-magnetic planets may exhibit magnetic star-planet interactions \citep{Laine2013, Strugarek2017}. These are not considered for this simulation.

The following section describes a simulation that explores the feasibility of detecting phase-dependent, planet-induced stellar flares consistent with the above theoretical assumptions.

\subsection{Flare Probability Density Functions}
\label{subsection: Flare Probability Dist.}
The observable effect of magnetic SPIs is the alteration of the flare probability density function (PDF), $P(\phi)$, as a function of the planetary orbital phase, $\phi$, where the distributions peak when the planet is near periastron and sub-Alfvénic. In contrast, the null expectation is that stars flare independently of the orbital phase. Due to the ambiguity in how SPIs alter a typical star's flare probability density function, we offer two toy models. The first, described in Section~\ref{subsubsection: kappa prescription}, uses two parameters, $\kappa_{SPI}$, the SPI strength, and $\mu$, the SPI location. The next model, described in Section~\ref{subsubsection: inverse cubic prescription}, is motivated by an increase in the normalized PDF with the instantaneous separation of the star/planet system, using an inverse cubic law in the energy scaling of two magnetic dipoles. 

\subsubsection{The von Mises Prescription}
\label{subsubsection: kappa prescription}
Phase-correlated flaring can be quantified by phase-folding measured flare epochs with a period determined by the underlying physical process. This "clustering" can be modeled through the von Mises PDF, $V(\phi,\kappa_{SPI},\mu)$, parameterized by the reciprocal of variance, $\kappa_{SPI}$, and the circular mean, $\mu$:

\begin{equation}
\label{eq: von mises}
    V(\phi,\kappa_{SPI},\mu)=\frac{\exp[\kappa_{SPI}\cos(\phi-\mu)]}{2\pi I_{0}(\kappa_{SPI})}; -\pi\leq\phi\leq\pi.
\end{equation}
$I_{0}(\kappa_{SPI})$ is the modified Bessel function of order zero. To create bounds consistent with the normalized phase interval $[0,1)$, we linearly map the interval $[-\pi,\pi]\rightarrow[0,1)$ and renormalize the resulting PDF such that $\int^{\infty}_{-\infty}V(\phi,\kappa_{SPI},\mu)=1$. This is done using the \texttt{scipy} implementation, \texttt{scipy.vonmises.pdf}.

The periodic boundaries of the von Mises PDF characterize correlated flare signals that may extend across the boundaries of the periodic signal, where other unimodal distributions, such as a Gaussian, would be inappropriate. Throughout this work, the von Mises PDF will be used to describe our sample of phase-folded flare epochs with respect to different astrophysical periods. The simplicity of this model is advantageous in quantifying meaningful, unimodal deviations from expected uniform flaring, particularly in sparsely sampled targets. It makes no assumptions about the specific induced flaring mechanism except for its phase dependence. The von Mises PDFs used in this simulation are shown in Figure~\ref{figure: All_Simulations}.

Unbinned likelihood analysis for the von Mises distribution is computed by:
\begin{align}
\label{eq: VM Unbinned}
        -\log{\mathscr{L}_{1}}&=-\sum^{N_{\mathrm{flare}}}_{i}\log\left(V(\phi,\kappa_{SPI},\mu)\right)\\ -\log{\mathscr{L}_{0}}&=-\sum^{N_{\mathrm{flare}}}_{i}\log\left(U(\phi)\right)
\end{align}
where $U(\phi)$ is the uniform distribution evaluated at phase $\phi$. Equation~\ref{eq: VM Unbinned} is the function to be minimized. The test statistic is computed following Equation~\ref{equation: TS}.

\subsubsection{Inverse Cubic Prescription}
\label{subsubsection: inverse cubic prescription}
The second prescription assumes that the phase-folded flare PDF depends on two parameters: the instantaneous separation of the planet from its host, $r_{d}$, and the ratio $B_{r}=B_{planet}/B_{\star}$. The ratio of the magnetic fields determines the total integrated weight of the interaction term relative to the case of uniform fields. The function ${r_{d}}^{-3}$ over the orbital phase drives the shape of the curve. The instantaneous separation is computed by time-evolving the mean anomaly, $ M = nt + \omega_{p}$, and solving for the eccentric anomaly, $E$, using the Newton-Raphson (NR) method. We use the resulting $E$ to solve for the mean anomaly, $\nu$:
\vspace{-0.15cm}
\begin{align}
\label{equation: radial distance}
0=&E-\sin{E}-M\\\xRightarrow[NR]{}\nu(t)=&E(t)+\arctan{\left(\frac{\beta\sin{E(t)}}{1-\beta\cos{E}}\right)}
\end{align}
where $ \beta=e \left(1+\sqrt{1-e}\right)^{-1}.$ This can then be used to determine the orbital separation as a function of time and phase:
\begin{align}
\label{equation: orbital distance}
r_{d}(t)&=a\frac{(1-e^{2})}{\sqrt{1+e\cos{(\nu(t)+\omega)}}}\\\rightarrow r_{d}(\phi)&=\frac{r_{d}(t)}{T_{\mathrm{pl}}}\mod{T_{\mathrm{pl}}}.
\end{align}
We construct a normalized function, $C(\phi)$:
\begin{equation}
\label{equation: inverse-cubic}
C(\phi,a,e, \omega)= \frac{r_{d}(\phi)^{-3}}{\int^{1}_{0}{r_{d}(\phi)^{-3}} d\phi}
\end{equation}
 
Only relative deviations will impact the shape of $C$; thus, the $a$ dependency is dropped. To incorporate the correct behavior expected with $B_{r}$, we perform the following:
\begin{equation}
\label{equation: SPI_Inverse_Cubic}
    M(\phi,e,B_{r},\omega) = \frac{C(\phi)B_{r}+U(\phi)}{\int ^{1}_{0}\left(C(\phi)B_{r}+U(\phi)\right)\, d\phi}.
\end{equation}

The $B_{r}$ term applies a weight towards the relative strength of the planet's magnetic field over the star. The convolution with $U(\phi)$ also gives the correct behavior when $B_{r}\ rightarrow 0$. While this prescription does not predict induced flare signals for orbits with $e=0$, this does not eliminate the possibility of induced flares, as interaction with, e.g., a co-rotating hot spot on the stellar surface may induce flares at the synodic period between the planetary orbit and stellar rotation, as studied in \citet{Lanza2012, Ilin2022, Ilin2024}. This is discussed further in Section~\ref{subsection: Limitations Future Work}.

Unbinned likelihood analysis for the inverse cubic prescription is computed by doing the following:
\begin{align}
\label{eq: Cubic Unbinned}
        -\log{\mathscr{L}_{1}}&=-\sum^{N_{\mathrm{flare}}}_{i}\log\left(M(\phi,e,B_{r})\right)\\ -\log{\mathscr{L}_{0}}&=-\sum^{N_{\mathrm{flare}}}_{i}\log\left(U(\phi)\right)
\end{align}
where $U(\phi)$ is the uniform distribution evaluated at phase $\phi$. Equation~\ref{eq: VM Unbinned} is the function to be minimized. The test statistic is computed following Equation~\ref{equation: TS}.

\subsection{Detectability of Magnetic SPIs via Flare Injection}
\label{subsection: Sim Params}
We simulated empirical magnetic SPI flare signals in TESS photometry by injecting flares into model light curves following the geometric PDF $r_{\mathrm{norm}}$ for varying $e$ and $\omega_{p}$ for transiting planet geometries. Two simulations were performed on two-minute cadence TESS data: one on a model G-type light curve (TIC ID: 219101992, sector 16) and one on a model M-type light curve (TIC ID: 219822564, sector 14) with stellar masses and radii of $0.896\, M_{\odot}$, $0.92165\, R_{\odot}$ and $0.59\, M_{\odot}$, $0.603\, R_{\odot}$, respectively. 

Each light curve is sampled 10 times to mimic the typical data volume associated with TESS targets. This process is repeated 50 times to simulate 50 targets. The phase of the planet at the start of each simulated light curve is randomized with an arbitrary epoch of periastron. The orbital period of the planet is set at five days, resulting in a semi-major axis of $\sim 0.055\, \mathrm{AU}$ for the G-type star, and $\sim 0.048 \, \mathrm{AU}$ for the M-type star. The base stellar flare rates are derived from typical values for each type of star, assuming that the star is active. For example, even though only $\sim1\%$ of G-type stars flare in TESS flare studies, the flare rate is derived from G-type stars that are active \citep{Gunther2020, Medina2020, Crowley2022}. These flare rates are then modulated by the normalized flare PDF, $r_{\mathrm{norm}}$, depending on the phase of the planet.

\begin{deluxetable}{c|cccccc}
\label{table: Spectral stats}
\tablehead{\vspace{-0.2cm} &Flare Rate   & $T_{\mathrm{eff}}$  &$R_{\star}$ & $B_{\star}$ & $t_{obs}$ &  a\\ \vspace{-0.2cm}Type & & &  & &\\  &( $\mathrm{day}^{-1}$) & (K) & $(R_{\odot})$ & ($\mathrm{G}$) & ($\mathrm{days}$)  & (AU)}
\startdata
G  & $0.015$ & 5245 & 0.922 & 1 & 700  &0.055\\ 
M  & $0.5$ & 3837 & 0.603 & 100 & 700  &0.048 
\enddata
\caption{A table outlining the stellar parameters of the sample light curves used for the SPI simulation. Radius and $T_{eff}$ are derived from the TESS Input Catalog (TIC), retrieved using \textsc{astroquery}. Flare rate and field strengths are set to 'typical' values. The semi-major axis is determined by the stellar mass and the planetary period set at five days.}
\vspace{-1cm}
\end{deluxetable}
The magnetic SPI strength is parameterized by the $\kappa_{SPI}$ parameter for the von Mises prescription with the location set at $\mu=0.5$. We sample $\kappa=0,0.25,0.5,1,2,4,8$. For the inverse-cubic prescription, the SPI duration and strength are determined by the orbital geometry of the system as well as $B_{r}$. We sample four different eccentricities, $e=0, 0.05,0.2,0.5$, and three planetary field strengths for each eccentricity,  $B_{P} = 1\,,10,\,100\mathrm{\,G}$, the last of which is the strongest theoretical dynamo-driven planetary field strengths of young, massive exoplanets \citep{Reiners2010}. $B_{r}$ is computed using these values with the $B_{\star}$ values in Table~\ref{table: Spectral stats}. The injected flare amplitudes are sampled from the amplitude distribution by \citet{Pietras2022}, shown in Section~\ref{subsection: Flare Probability Dist.}. The FWHM is sampled from a uniform distribution spanning $(0,60]$ minutes.

\begin{figure*}
    \label{figure: All_Simulations}
    \centering
    \includegraphics[width=\textwidth]{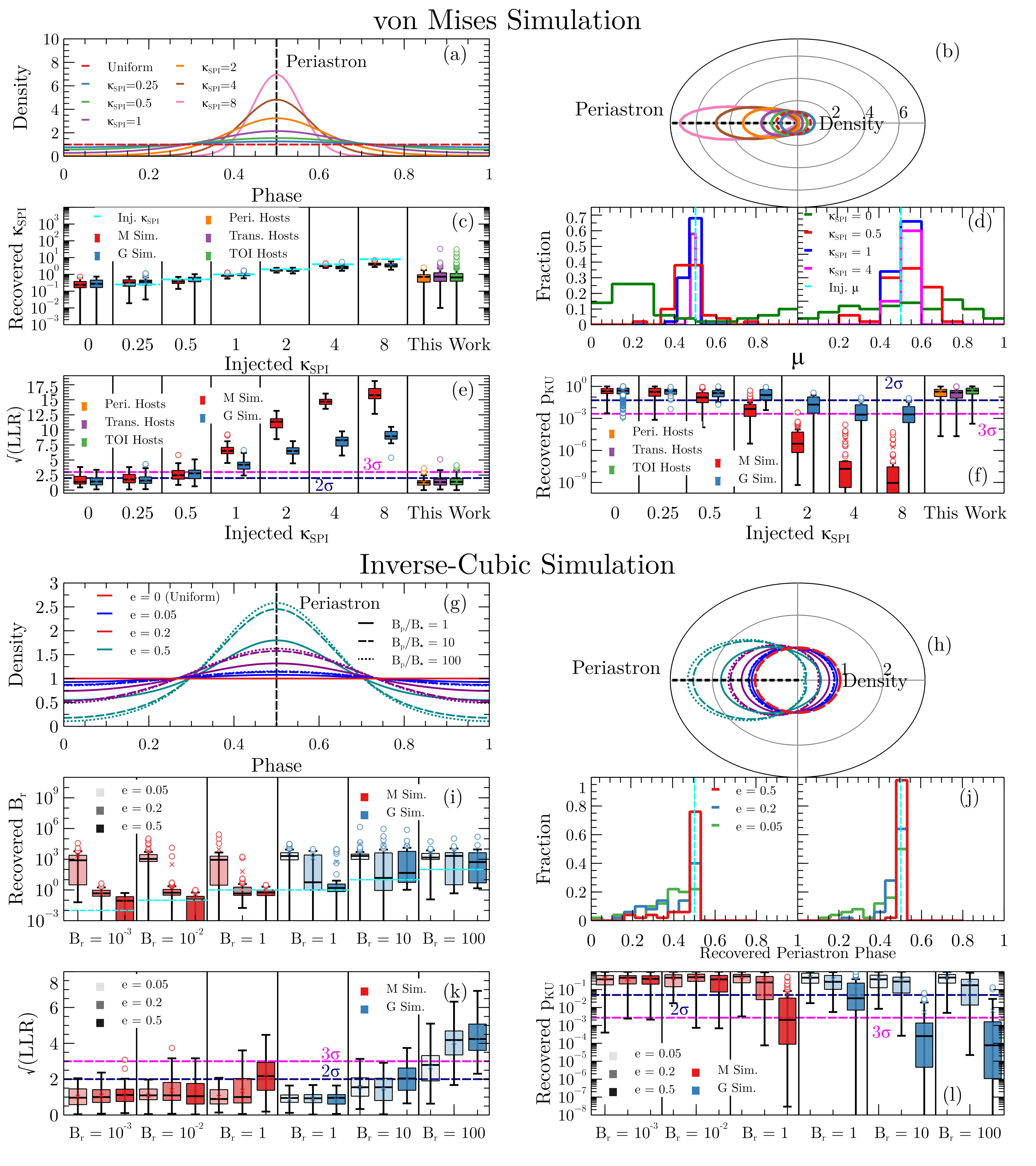}
    \caption{Summarized metrics of both simulations. (a) Underlying PDFs for the von Mises simulations. (b) The orbital view of the von Mises PDFs; the theta direction correlates to the true anomaly of the planet. (c) Recovered $\kappa_{SPI}$ for each sample of stars. The underlying $\kappa_{SPI}$ is denoted on the x-axis and the cyan bars. (d) The recovered $\mu$ parameter for each sample of stars. The left and right windows correspond to M and G-type samples, respectively. (e) Unbinned likelihood analysis SNR for each sample, with the underlying $\kappa_{SPI}$ shown on the x-axis. (f) Recovered $p_{KU}$ distributions for each star sample, with the underlying $\kappa_{SPI}$ shown on the x-axis. (g-l) Identical to (a-f),  but for the inverse cubic prescription. The eccentricity is assumed to be known for the unbinned likelihood analysis. Where relevant, $2\sigma$ and $3\sigma$ detection thresholds are denoted by the dark blue and magenta dashed lines, respectively. Box plots show the median and $\pm1.5$ interquartile range.}
\end{figure*}
\subsection{Simulation Results and Analysis}
\label{subsection: Simulation Analysis}
To quantify the effect that changing the magnetic SPI parameters has on the detectability of induced flares, the goodness-of-fit tests described in Section~\ref{subsubsection: GoF}
as well as unbinned likelihood analysis described in Section~\ref{subsubsection: Unbinned Likelihood} are performed on each target. This simulation serves to contextualize the detections found in real photometric data.

\subsubsection{Goodness-of-Fit Tests}
\label{subsubsection: Sim GoF}

Each GOF test described in \ref{subsection: Stat Tests} is employed on each simulated star, with the resulting $p_{KU}$ distributions shown in Figures (f) and (l) in Figure~\ref{figure: All_Simulations}. These distributions allow us to constrain the probability of various detection thresholds for a given SPI strength. Since the GOF tests are nonparametric, they only indicate non-uniform flaring at some significance and do not offer evidence for a particular model. However, these tests are helpful for the preliminary identification of systems of interest. The samples we derive in this work closely resemble the injected uniform distributions across every metric, except for a few outliers that exceed the $3\sigma$ detection threshold (see Section~\ref{section: Discussion}). This is consistent with assumed uniform flaring.

For the von Mises simulation, the Kuiper test showed a higher power in SPI detection as injected $\kappa_{SPI}$ increased, as shown in the sub-figure (f) in Figure \ref{figure: All_Simulations}. There existed at least one $3\sigma$ detection in all injected $\kappa_{SPI}$ in the M-type samples, with a significant shift at $\kappa_{SPI}=1$, where half of the stars sampled exhibit a $3\sigma$ detection. The lower flare rate of the G-type sample resulted in an overall decrease in power compared to the M-type sample, and even at the highest injected SPI signal, half of the sample remained below the $3\sigma$ detection threshold. 

The simulated inverse cubic PDFs have a lower density than the von Mises simulations. While M-type stars flared more frequently than the most active G-type stars, the instantaneous separation model proposed in \ref{section: SPMI Models} has an interaction strength dependent on the ratio of the planetary field to the stellar field. Even in the best-case scenario, a weak M-type field of $\sim100\,\mathrm{G}$ with a maximally driven planetary dynamo gives $B_{r}=1$. In contrast, the weaker stellar fields of G-type stars in this model are more easily perturbed, but assuming that SPIs do not alter a star's quiescent flare rate, an equivalent baseline will yield fewer flares. For the M-type sample, the $p_{KU}$ distributions in figure (l) of Figure~\ref{figure: All_Simulations} show consistent $3\sigma$ detections with $e=0.5,\,B_{r}=1$. In contrast, the G-type sample shows significant detections at $B_{r}=10,\,e=0.5$, with a planetary field similar to Jupiter of $B_{P}=10\,\mathrm{G}$. In our model, G-type stars are more likely to yield an observable SPI signal.

\subsubsection{Unbinned Likelihood Analysis}
\label{subsubsection: Sim ULA}

We use unbinned likelihood analysis to determine the best fit $\mu$ and $\kappa_{SPI}$ parameters. Injecting uniform flares did not result in any significant detections in the $\kappa_{SPI}$ space, with $\kappa_{SPI}\ll10^{-2}$ favoring the uniform model over a particular $\kappa_{SPI}$ value, as seen in subplot (c) in Figure \ref{figure: All_Simulations}. The $\kappa_{SPI}$ values from the flare survey in this work are consistent with the $\kappa_{SPI}$ distribution. For the location of the von Mises SPI in sub-plot (d) in Figure~\ref{figure: All_Simulations}, clustering becomes increasingly pronounced at the injection phase with increasing $\kappa_{SPI}$. Although they are not shown, the empirical flare distributions show no preference with respect to the planetary phase. The detection signal, derived from the square root of the log-likelihood ratio between the uniform model and the von Mises model, shows a noticeable increase in detection significance at $\kappa_{SPI}=0.25$, suggesting that the unbinned likelihood analysis has a greater sensitivity in determining a significant detection in contrast to the GOF test, in addition to the best-fit parameters.

$B_{r}$ and $\omega$ were retrieved for the inverse cubic model. To contextualize $\omega$ into the normalized phase bounds of $[0,1]$, we linearly map the derived $\omega$ and place the periastron at 0.5 phase. In this case, we assume precise knowledge of $e$ and use it to constrain $B_{r}$. In particular, in the case of the M-type sample, the convergence of $B_{r}$ is poor in the low eccentricity limit, which is expected as the PDF converges to the uniform case for both $e\rightarrow0$ and $B_{r}\rightarrow0$. In contrast to the GOF tests, the M-type sample converges on the injected $B_{r}$ value more efficiently than the M-type sample due to the additional likelihood of higher flare counts. The per $B_r$ on phase for both samples converges quickly to 0.5, as seen in sub-plot (j) in Figure~\ref{figure: All_Simulations}. The minimization of the likelihood function finds the solution by evolving from 0 to periastron, so solutions $<0.5$ are returned since they are degenerate with those $>0.5$. The $\omega$ returned by $e=0.05$ shows only a slight preference $\sim0.5$. The detection significance for the inverse cubic prescription mirrors the GOF tests, with a subtle power increase in the $B_{r}=100$ case in the G-type sample. As before, the M-type sample becomes significantly discernible from the uniform case at the injected $ B_r =1$ case.

\subsubsection{Occurrence Rate of Significant Flare Clustering from Uniform Flare Samples}
\label{subsubsection: Stochastic Detections}

\begin{figure*}
    \centering
    \includegraphics[width=1\linewidth]{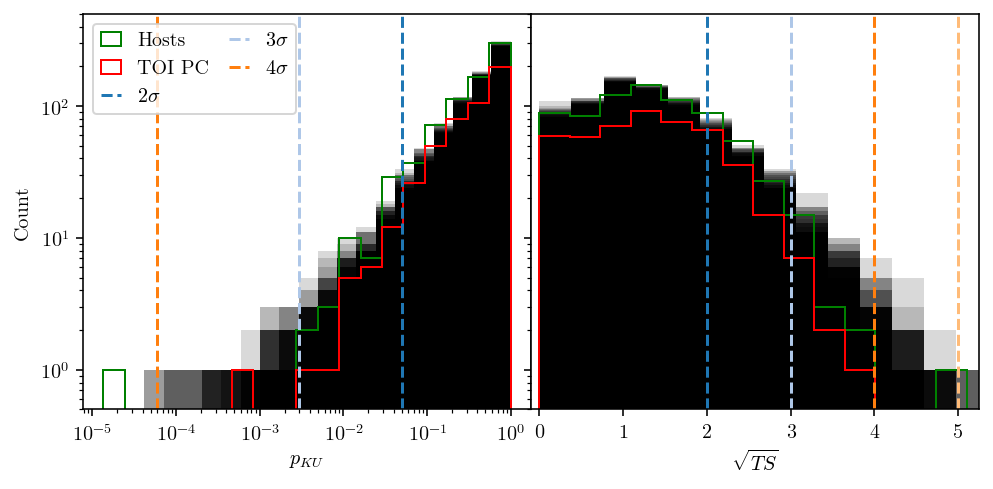}
    \caption{Histograms of 100 resampled $p_{\mathrm{KU}}$ and $\sqrt{TS}_{VM}$ values from uniformly sampled targets. The observed distributions of exoplanet hosts and TOI planet candidate hosts are shown in green and red, respectively. The dashed lines show significance markers.}
    \label{fig: TS_KU_compare}
\end{figure*}

Significant flare clustering may emerge stochastically from uniformly drawn events with a large number of samples. We investigate the occurrence rate of significant flare clustering by resampling the flares observed in Section~\ref{section: Results} using a uniform distribution. Each target is resampled with the same number of flares observed, with the flare phase drawn from a uniform distribution. We then perform the KU and ULA tests and determine the frequency of significant results. We repeat this 100 times to assess the occurrence rates of $2,3,4,\&\,5\sigma$ detections. The results of this simulation are summarized in Figure~\ref{fig: TS_KU_compare} and Table~\ref{table: Resampled Flares}.
\\
\begin{deluxetable}{c|c|cccc}
\label{table: Resampled Flares}
\tabletypesize{\footnotesize}
\tablecaption{Results of resampling flares from uniform distributions by significance of the Kuiper test and unbinned likelihood analysis. The total number of synthetic hosts tested is $N = 33600$. The values represent the proportion of targets found in each significance threshold. For example, the '$2\sigma$' column indicates the proportion of samples within the interval $2\sigma\leq x<3\sigma$.}
\tablehead{\colhead{Test Result} \vline & Sample & \colhead{$2\sigma$} &\colhead{$3\sigma$} &\colhead{$4\sigma$} & \colhead{$5\sigma$}}
\startdata 
\multirow{2}*{$p_{\mathrm{KU}}$} & Synthetic&0.047&2.6e-3&4.4e-5&0\\
&Observed&0.090&0&7.2e-3&0\\
\hline
\multirow{2}*{$\sqrt{TS}_{VM}$} & Synthetic&0.15&0.027&2.0e-3&2.9e-5\\
&Observed&0.16&0.018&0&3.6e-3\\
\enddata
\end{deluxetable}
The populations of significant detections for both the observed and synthetic populations are generally consistent. The KU test shows an excess of $2\sigma$ detections and no $3\sigma$ detections, in contrast with the simulated populations. Additionally, the $4\sigma$ and $5\sigma$ for the $p_{KU}$ and ULA analysis, respectively, are orders of magnitude higher in the observed data than in the synthetic data. This is due to the outlier, TOI-1062\,b, which is discussed in Section~\ref{subsubsection: TOI-1062 b}. The synthetic uniform-flare samples are consistent with the observed flare distributions, except for the $4$–$5,\sigma$ outliers present in the data.

\subsection{The SPI Metric \texorpdfstring{$\beta_{\mathrm{SPI}}$}{bSPI}}
\label{subsection: SPI Metric}
\begin{figure}
    \centering
    \includegraphics[width=0.47\textwidth]{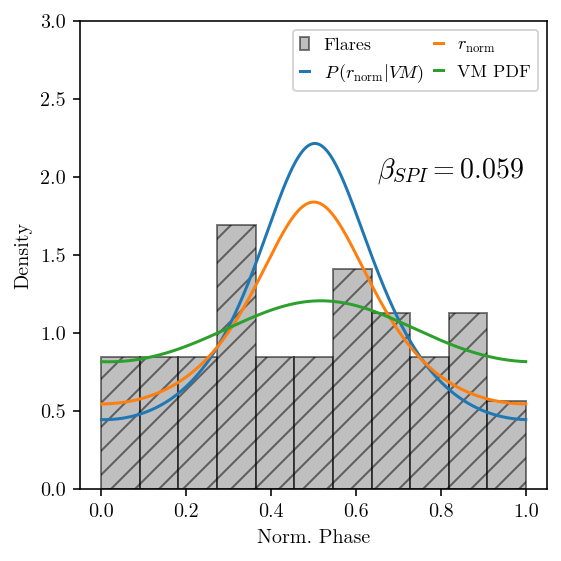}
    \caption{An example visualizing the procedure from Equations~\ref{eq: beta SPI} using Gl 674\,b. The flare distribution is shown in gray. The resulting von Mises PDF derived using ULA is shown in green. The geometric SPI PDF, $r_{\mathrm{norm}}$, is shown in orange. Phase is normalized to the interval from $[0,1]$, and periastron is set at phase $\phi=0.5$. The resulting $\beta_{\mathrm{SPI}}=0.059$, indicating marginal flare clustering at periastron.}
    \label{fig: beta_SPI}
\end{figure}

Given a sample of flare epochs phase-folded to the planetary period, the location of the clustering with respect to periastron is substantial evidence in identifying induced-flare candidates. Real stars may exhibit flare clustering inconsistent with induced flare models, but will emerge as significant detections in statistical tests by chance or a signal from another source. To aggregate the data and induce flare models, we construct the metric $\beta_{\text{SPI}}$ using the following algorithm:
\begin{align}
\label{eq: beta SPI}
    V(\kappa,\mu,\phi)&=\frac{\exp{(\kappa\cos{\theta})}}{2\pi I_{0}(\kappa)}\\
    r_{d}(a,e,\omega_{p},\phi)&= \frac{a(1-e^{2})}{(1+e\cos{[\nu(E,e,\omega_{p},\phi)]})}\\
    r_{\mathrm{norm}}&=\frac{r^{-3}_{d}}{\int_{0}^{2\pi}r^{-3}_{d}\,d\phi}\\
    \beta_{\mathrm{SPI}}&=\log{\left[\int^{2\pi}_{0}(V-U)r_{SPI}\,d\phi\right]}.
\end{align}
$\beta_{\mathrm{SPI}}$ has multiple desirable properties. The joint probability of $V$ and $r_{SPI}$ produces a function enhanced when the clustering is close to periastron, and diminished when at apastron. Integrating provides a metric that correlates with the expected SPI strength. Additionally, at $e=0$, $r_{SPI}\rightarrow0$ for all phase. Thus, flare clustering candidates with $e=0$ are not consistent with periastron-dependent induced flares. To supplement our analysis, we use $\beta_{\mathrm{SPI}}$ to contextualize flare clustering with respect to the orbital configuration and line-of-sight. We show an example of the underlying functions driving $\beta_{\text{SPI}}$ in Figure~\ref{fig: beta_SPI} using Gl 674\,b ($e=0.20\pm0.02$, $a=0.039$) as an example \citep{Bonfils2007}.

Transiting planets require an additional consideration. Assuming the interaction that produces induced flares is local, the induced flare SNR will be diminished in some interval around the secondary eclipse at $\nu=270^{\circ}$. The orbital phases we are blind to depend on the interaction mechanism in tandem with the local magnetic environment. Without further model development, such as MHD simulations, it is challenging to define particular bounds. In general, a simple model to account for line-of-sight geometries around the secondary eclipse is:
\begin{equation}
V(\kappa,\mu,\phi)=
\begin{cases}
0 & \text{if }-\phi_{\mathrm{0}}<\phi<\phi_{0}\\V(\kappa,\mu,\phi) & \text{otherwise}
\end{cases}
\end{equation}
where $\phi_{\mathrm{0}}$, an angle symmetric around superior conjunction at $\nu=270^{\circ}$, is determined by the interaction mechanism and environment. The flare study by \citet{Ilin2024} places a constraint of $\phi_{0}=90^{\circ}$. Indeed, the white-light continuum emission of solar flares is close to the stellar surface, so white-light flares "behind the limb" will be blocked by the photosphere \citep{Oliveros2012}. This is reasonable assuming SPMIs are constrained to a one-to-one correlation between orbital phase and stellar rotational phase. However, the induced flare mechanisms proposed by \citet{Lanza2012, Saur2013, Lanza2018} depend on the reconnection of the planetary field with extended structures or energy perturbations via Alfvén wings, leading to interactions that are non-local between the orbital and rotational phases. For example, the Alfvén waves produced by a sub-Alfvénic magnetized planet are predicted to span many solar radii. They can connect with parts of the stellar surface, exhibiting significant phase displacement between the orbital and surface phases \citep{Saur2013}. Instead of estimating $\phi_0$, we contextualize the line-of-sight geometry for transiting magnetic SPI candidates for a set of line-of-sight cones in the orbital flare plots in Figures~\ref{fig: Peri_Polar_Plots}. As particular values of $\phi_{0}$ will change both between systems and in time, this question should be investigated rigorously using MHD simulations, which can reveal the extent of non-locality of SPMIs.

\subsection{Model Limitations and Future Work}
\label{subsection: Limitations Future Work}

The inverse-cubic model is physically well-motivated, but has properties that may be counterintuitive or poorly behaved. First, it predicts that sub-Alfvénic planets with circular orbits will not exhibit induced flaring. Suppose that the existence of the planet enhances the stellar activity compared to an identical star without a planet. Flare clustering should not be preferential at a particular point in the orbit. One may argue that the "bend-and-snap" model investigated by \citet{Lanza2012, Ilin2024} could lead to induced flares for a circular geometry. In that model, the interaction depends on the synodic period between the planet's orbit and the stellar rotation, and it can only manifest while the structure exists. Additionally, the interaction would be smeared over a long temporal baseline, since continuous interaction with different structures would appear at arbitrary phases during the planetary orbit, randomly sampling different probability maxima across the phase space. In contrast, sufficiently eccentric orbits in the "bend-and-snap" framework would still predict flares at periastron, as it enhances the interaction probability with a coronal structure, while having a significantly reduced interaction probability when the planet is at apastron. 

One limitation of the inverse cubic model is the reduction in PDF response as $B_{r}$, which scales as $1/x$. This can be justified since the function behaves reasonably when $B_r$ is within physically plausible values, e.g., $B_r < 100$; however, convergence using unbinned likelihood analysis performs poorly as $B_r$ values become large. The strict dependence of $e>0$ also limits its application to many systems. Due to these limitations, the analysis in Section~\ref{section: Results} omits fitting for the inverse cubic model, which requires a more rigorous analytical development or magnetohydrodynamics simulations.

We detect SPI signals depending on the expectation that the SPI manifests as a unimodal clustering of flares at a close-orbiting planet's periastron. This does not place priors on the parametric distributions (e.g., amplitude, energy, FWHM) of planet-induced flares, in contrast to, e.g., \citet{Lanza2012, Lanza2018, Ilin2024}. The uncertainties surrounding the mechanism(s) that could drive SPIs, combined with the complex time-evolving nature of stellar and planetary magnetic environments, can unintentionally lead to 'p-hacking' when selecting a specific subsample within the flare population. If induced flares \emph{are} preferential at periastron, magnetic SPIs should manifest probabilistically with a long photometric baseline with sufficient SNR even when drawn from a particular parameter space.

\section{Data}
\label{section: Data}
Data were acquired by downloading a list of all known exoplanet hosts with known periastron $\leq 0.1 \,\mathrm{AU}$ (or $a \leq 0.1 \,\mathrm{AU}$ if eccentricity is unknown) through the NASA Exoplanet Archive composite planetary systems table \dataset[doi:10.26133/NEA13]{https://dx.doi.org/10.26133/NEA13}. The periastron constraint increases the likelihood that the planets are sub-Alfvénic, which is expected to induce magnetic interactions between the star and the planets. Using each host's TESS Input Catalog (TIC) number, all 2-minute and 20-second cadence data from the Science Processing Operations Pipeline (SPOC)  were downloaded via \textsc{astroquery} from the MAST archive. A total of 2,083 hosts had at least one data set associated with their corresponding TIC ID, and 10,558 files were analyzed. An identical procedure was followed for TESS objects of interest (TOI) hosts. The TIC IDs for all TOI planet candidates (PCs) were retrieved using the NASA Exoplanet Archive's TESS Project Candidate table. To estimate if TOI PCs are sub-Alfvénic, the period is used to calculate the semi-major axis of a circular orbit, with the mass of the star estimated using the mass-radius relationship of main-sequence stars and assuming that the planetary mass is negligible, such that:
\begin{equation}
\label{equation: TOI Condition}
    T \leq 2\pi \sqrt{\frac{(0.1\,\mathrm{AU})^3}  {(G (R_{\star}/R_{\odot})^{4/3}M_{\odot})}}.
\end{equation}

At the time of this paper, this included 3,519 targets with 17,423 files. All \texttt{.fits} files for both samples were analyzed using \textsc{ardor}.

\subsection{Determination of Periastron}
\label{subsection: Determination of Periastron}
The NASA Exoplanet Archive's composite planetary systems table reports the epoch of periastron, $t_{\mathrm{peri}}$, and the argument of periastron, $\omega_{\star}$. Although $\omega_{\star}$ is computed from $t_{\mathrm{peri}}$, they are not always reported simultaneously. We use the reported $t_{\mathrm{peri}}$ to phase-fold the flare epochs when available. When both $\omega_{\star}$  and $t_{\mathrm{transit}}$ are reported, we follow the NASA Exoplanet Archive's algorithm to determine $t_{\mathrm{peri}}$:
\begin{align}
    \omega_{\star}&=180+\omega_{p}; \,\nu=\pi/2-\omega_{p}\\
    E&=2\arctan{\left[\sqrt{\frac{1-e}{1+e}}\tan{\frac{\nu}{2}}\right]}\\
    t_{\mathrm{peri}}&=\frac{P}{2\pi}(E-e\sin{E})
\end{align}
We generate a $\pm1\sigma$ credible interval by propagating all uncertainties contained in this algorithm using a Monte Carlo method with 5000 samples, assuming that the uncertainties in the literature parameters are approximately Gaussian.

\section{Results}
\label{section: Results}

Flares from the search were compiled in catalogs for known hosts and TOI planet candidate hosts. An example of the first eight entries for the exoplanet host catalogs is shown in Table~\ref{table: Flare Parameters Exo}. These lists include derived flare parameters, bolometric flare energy, and the Bayes factor $\Delta\log{Z}$. These are identical in both the known exoplanet host and TOI catalogs.

Since the 20-second cadence data always has 2-minute data associated with it, we removed degenerate flares with flare epochs within $\sim 5 \,\mathrm{min}$ of each other. We keep the parameters with the greater Bayes factor for a given pair of degenerate flares. 
\begin{deluxetable}{c|ccccc}[!ht]
\label{table: Flare Parameters Exo}
    \tabletypesize{\scriptsize}
    \tablecaption{A table showing the flare parameters derived from the output of Section~\ref{subsection: Tier 3}.}
    \tablehead{\vspace{-0.2cm} & Epoch & Amp. & FWHM & Energy & 
        \\ \vspace{-0.2cm} Host ID & & & & & $\Delta\log{Z}$\\ & (BJD) & (rel. flux) & (days) & (erg)}
    \startdata
    51 Peg & 2829.42 & $\left(1.80^{0.9}_{0.8}\right) \mathrm{e} {\text{-}4}$ & $\left(1.80^{0.3}_{0.4}\right) \mathrm{e} {\text{-}3}$ & 2.55e33 & 11.19 \\
    51 Peg & 2852.78 & $\left(3.10^{1.7}_{1.2}\right) \mathrm{e} {\text{-}4}$ & $\left(8.80^{2.7}_{2.4}\right) \mathrm{e} {\text{-}4}$ & 2.16e33 & 3.2 \\
    61 Vir & 1579.94 & $\left(6.92^{18}_{1.95}\right) \mathrm{e} {\text{-}4}$ & $\left(1.36^{0.2}_{0.8}\right) \mathrm{e} {\text{-}4}$ & 4.74e33 & 14.4\\
    61 Vir & 1581.36 & $\left(1.23^{0.7}_{0.4}\right) \mathrm{e} {\text{-}3}$ & $\left(5.46^{1.8}_{1.4}\right) \mathrm{e} {\text{-}4}$ & 3.21e32 & 9.0 \\
    61 Vir & 3042.92 & $\left(2.71^{0.18}_{1.1}\right) \mathrm{e} {\text{-}4}$ & $\left(6.66^{2.0}_{1.9}\right) \mathrm{e} {\text{-}4}$ & 9.19e31 & 4.25\\
    61 Vir & 3058.50 & $\left(2.28^{1.1}_{0.9}\right) \mathrm{e} {\text{-}3}$ & $\left(2.90^{0.9}_{0.7}\right) \mathrm{e} {\text{-}4}$ & 2.78e32 & 7.7 \\
    AU Mic & 1326.37 & $\left(2.33^{1.1}_{0.8}\right) \mathrm{e} {\text{-}3}$ & $\left(1.64^{0.5}_{0.4}\right) \mathrm{e} {\text{-}3}$ & 7.61e32 & 13.4 \\
    AU Mic & 1326.70 & $\left(2.25^{0.04}_{0.7}\right) \mathrm{e} {\text{-}2}$ & $\left(6.00^{0.6}_{0.4}\right) \mathrm{e} {\text{-}3}$ & 1.11e34 & 167 \\
    $\vdots$ & $\vdots$ & $\vdots$ & $\vdots$ & $\vdots$ & $\vdots$
    \enddata 
    \tablecomments{Flare epochs are given (BJD-2457000). The uncertainty values of the flare epochs are of order $\sim1\mathrm{e}{-5}\;\mathrm{days}$, and are reported in the full table. The uncertainties in flare energy are not computed. However, they could reasonably be computed using the uncertainty provided in flare temperature, stellar host, and the derived parameters provided in this table.}
\end{deluxetable}
\subsection{Notable Flares and Frequently Flaring Systems} 
\label{subsection: Notable Flares}

This section identifies high-amplitude, high-energy `superflares` and frequently flaring targets. Both cases have implications for habitability, e.g., atmospheric loss via photoevaporation \citep{Owen2017, Gunther2020}.

\subsubsection{Notable Flares}
\label{subsubsection: Notable}
\begin{figure}
    \centering
    \label{figure: High_Amp_Flare}
    \includegraphics[width=0.47\textwidth]{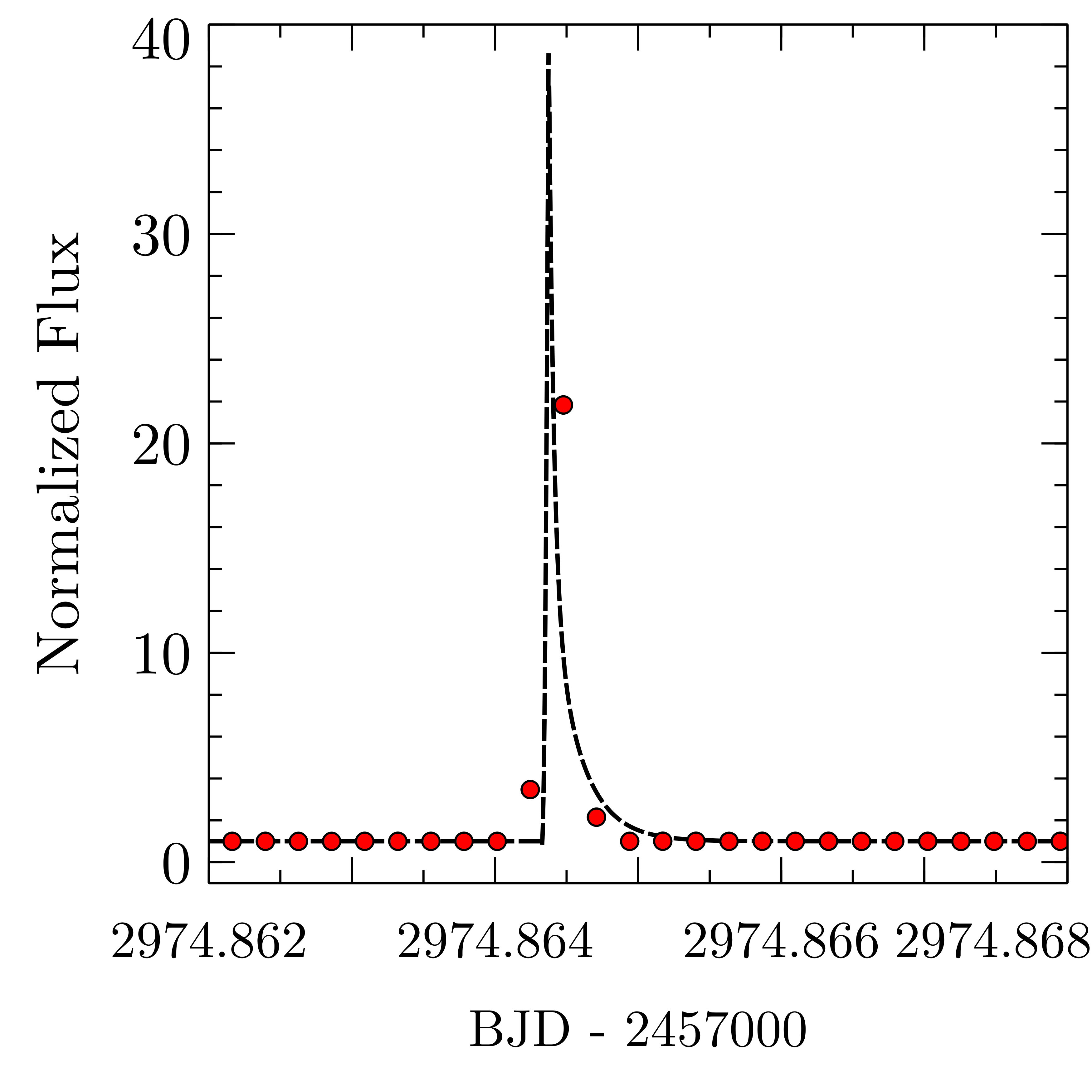}
    \caption{The highest amplitude flare across all systems analyzed with \textsc{ardor} on L 98-59. The best fit of the empirical flare model is shown in the black dashed line.}
\end{figure}

\begin{figure}
    \centering
    \label{figure: High_Amp_Flare_TPF}
    \includegraphics[width=0.47\textwidth]{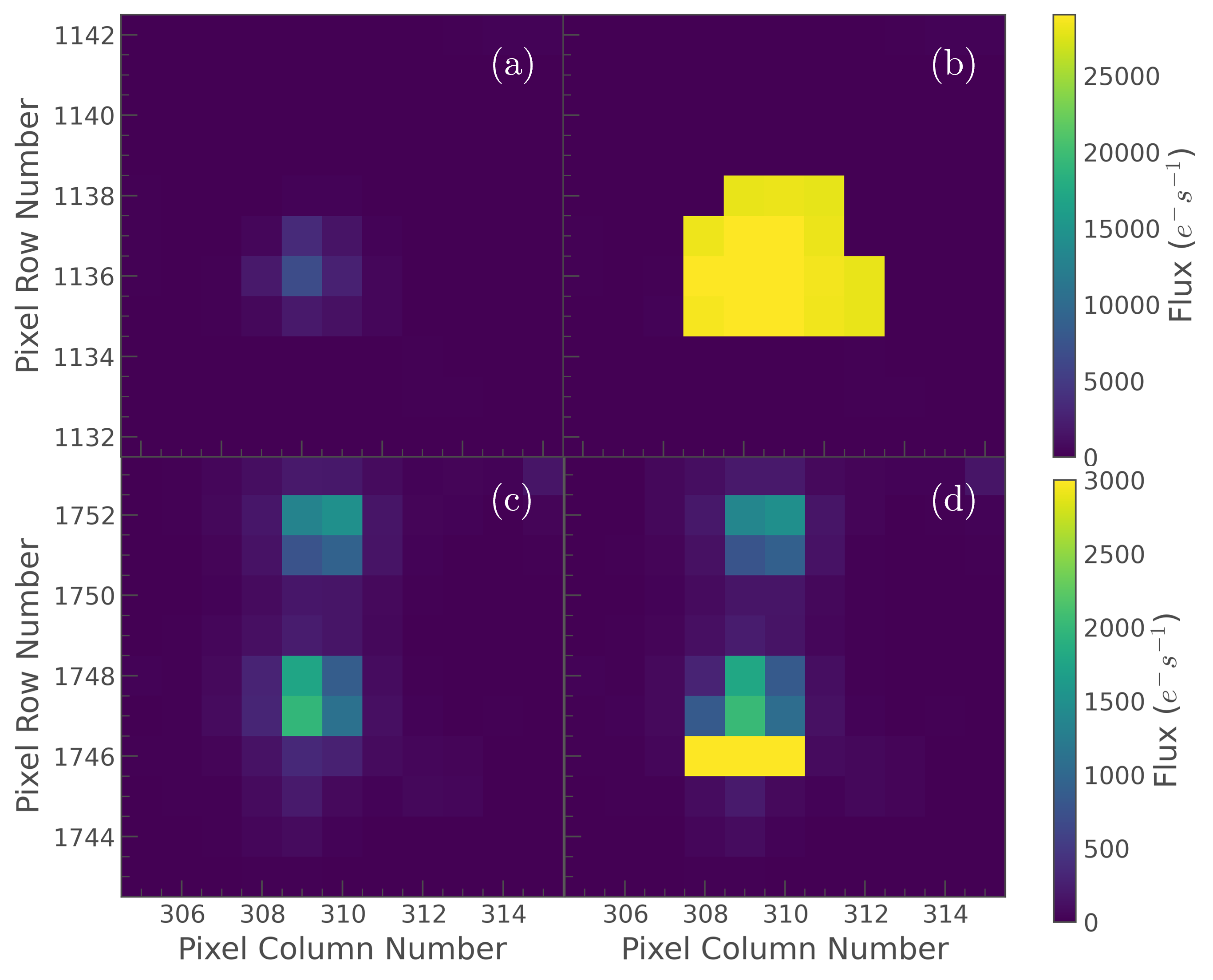}
    \caption{Target pixels of the highest amplitude and highest energy flares. (a) L 98-59 centroid before the flare. (b) The target pixels during the flare. (c-d) Same as (a-b), but for the highest energy flare on TOI-1347.}
\end{figure}

The highest amplitude flare found in our flare sample was from L 98-59, a bright M dwarf that hosts four close-in planets\citep{Demangeon2021}. The flare reached an amplitude of 38.7 times the quiescent flux, with an estimated bolometric energy of $6.44\times 10^{34}\,\mathrm{erg}$. L 98-59 flares semi-frequently, with 59 flares recorded in this catalog over 21 TESS sectors ($\sim489\,\mathrm{days}$). The photometry and model are shown in Figure \ref{figure: High_Amp_Flare}, while the target pixels for the event are shown in Figure \ref{figure: High_Amp_Flare_TPF}.

TOI-1347, an active main-sequence G-type star hosting two close-in super-Earths, produced the most energetic flare in our survey, reaching an energy of $6.36\times10^{36}\,\mathrm{erg}$ with an amplitude of 3.6 times the quiescent flux. The system produces high-amplitude flares frequently, with 46 flares detected across 34 TESS sectors, having an average flare amplitude of 0.15 relative flux units. Across all stellar parameters, the system is an analog to a young Sun, with $R_{\star}=0.84\, R_{\odot}$, $M_{\star}=0.928\, M_{\odot}$, $[Fe/H]=0.02$ and $t_{\mathrm{age}}=0.89\,\mathrm{Gyr}$ \citep{Paegert2021}. The transiting planets around this system have been newly confirmed \citep{Rubenzahl2024}. Follow-up observations could be significant in evaluating their capacity to maintain atmospheres. TOI-1347 is also essential for contextualizing solar activity within the broader population of main-sequence, Sun-like stars.

\subsection{Flare Frequency Distributions}
\label{subsection: FFDs}

Flare frequency distributions (FFDs) are used to quantify the flaring characteristics of different stars \citep{Lacy1976, Gunther2020,ilin2021}. FFDs are generated by counting the number of flares, $N_{\mathrm{flares}, E_{i}}$ below a particular bolometric energy threshold ($E_{i}\leq E_{0}$) divided by the observation time ($t_{\mathrm{obs}}$), and taking the logarithm, which yields an approximation for the frequency of flares, $\nu$, of at least a specific energy over time:
\begin{equation}
\label{equation: FFD}
    \log_{10}(\nu_{E_{i}\leq E_{0}})=\log_{10}\left(\frac{N_{\mathrm{flare}, E_{i}\leq E_0}}{t_{\mathrm{obs}}}\right).
\end{equation}

\begin{figure}[!ht]
    \centering
    \label{figure: FFD_TOIs}
    \includegraphics[width=0.47\textwidth]{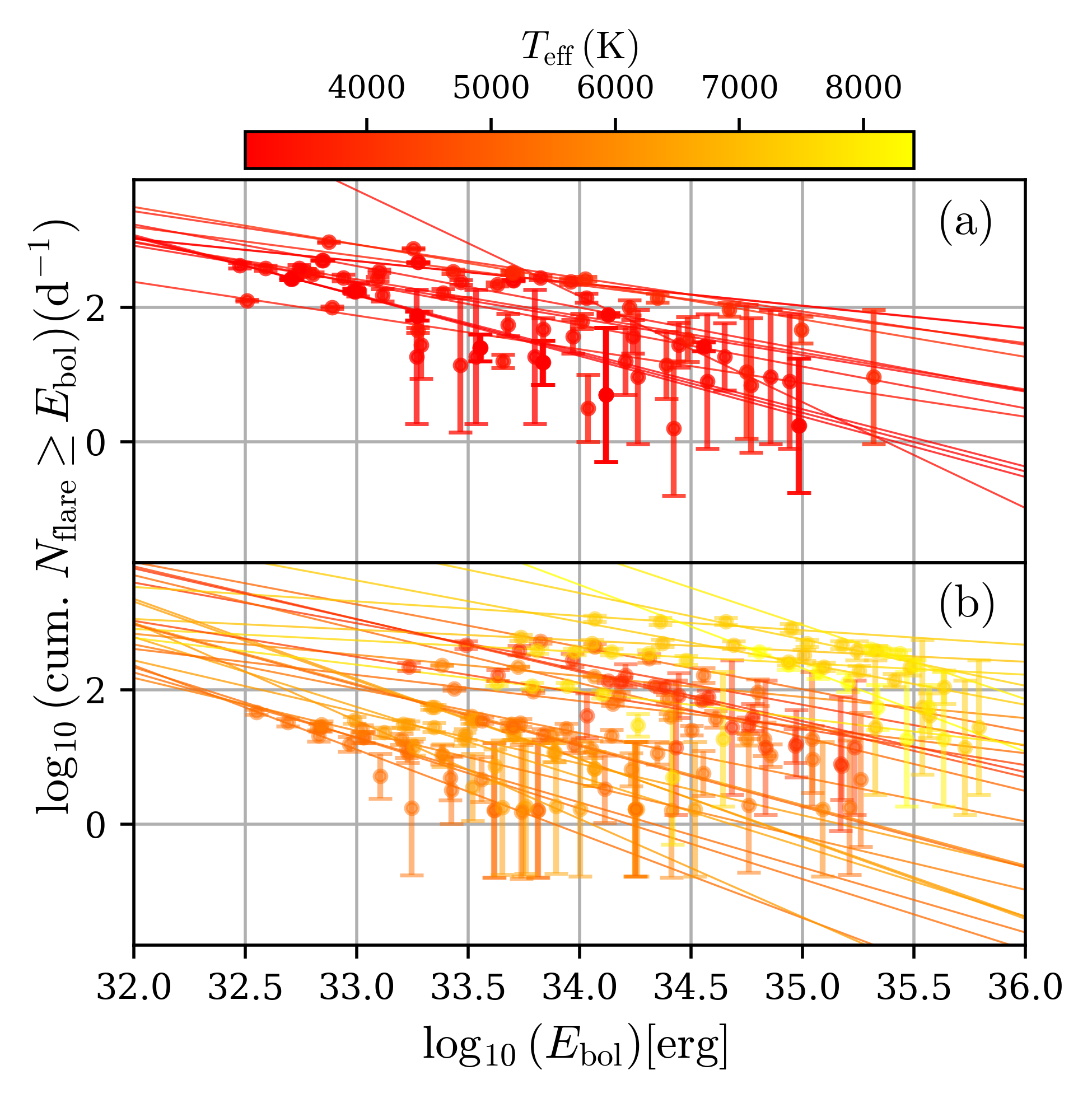}
    \caption{Flare frequency distributions (FFDs) of all TOI PC hosts with a minimum of $N_{flare}\geq15$ and $\overline{\Delta\log(Z)}\geq5$, binned by $T_{eff}$, $[Fe/H]$ and $t_{age}$. (a) FFD of all M-dwarf TOI PC hosts ($\leq 4000\,\mathrm{K}$). (b) FFD of all TOI PC hosts with $T_{eff}\geq4000\,\mathrm{K}$.}
\end{figure}

\begin{figure*}
    \centering
    \label{figure: FFD_Hosts}
    \includegraphics[width=0.975\textwidth]{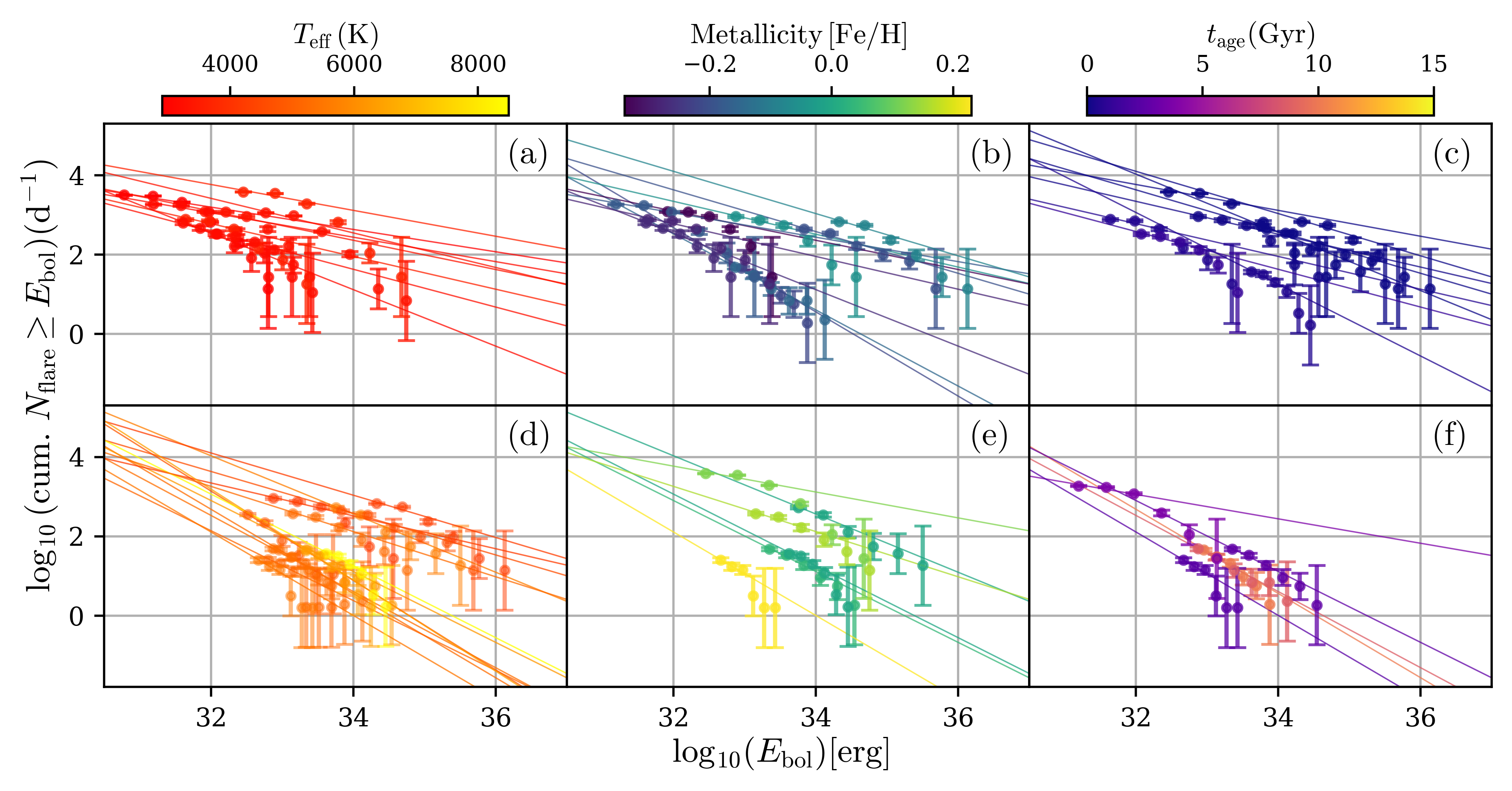}
    \caption{Flare frequency distributions (FFDs) of all exoplanet hosts with a minimum of $N_{flare}\geq15$ and $\overline{\Delta\log(Z)}\geq5$, binned by $T_{eff}$, $[Fe/H]$ and $t_{age}$. (a) FFD of all M-dwarf hosts ($\leq 4000\,\mathrm{K}$). (b) FFD of exoplanet hosts with $[Fe/H]<0$. (c) FFD of exoplanet hosts with stellar ages less than $2\,\mathrm{Gyr}$. (d) Identical to (a), but with $T_{eff}\geq4000\,(\mathrm{K})$. (e) Identical to (b), but with $[Fe/H]\geq0$. (f) Identical to (c), but with $t_{age}>2\,\mathrm{Gyr}$.}
\end{figure*}

FFDs can be fitted using a linear fit in the log-log space parameterized by:
\begin{equation}
\label{equation: FFD Linear}
     \log_{10}(\nu_{E_{i}\leq E_{0}})\approx\log_{10}(\beta)+\alpha\log_{10}(E_{0})
\end{equation}
where $\beta$ is related to the frequency of flaring and $\alpha$ relates to the typical energy released by the star. $\alpha$ is assumed to be negative, consistent with higher energy flares occurring less frequently than lower energy flares. To generate high-fidelity FFD plots, only flares with $\Delta\log Z\geq 5$ and stars with $N_{flare,\Delta\log Z\geq5}\geq 15$ are plotted in both Figure \ref{figure: FFD_Hosts} and Figure \ref{figure: FFD_TOIs}. 

\begin{deluxetable}{c|ccccc}
    \label{table: FFD Params Exo}
    \tablecaption{First seven entries of derived $\alpha,\beta$ for known exoplanet hosts in the sample. Flares with $N_{flare}\geq 10$ are reported in the table. Flares with $\Delta\log{Z}<5$ have been removed. $T_\mathrm{eff}$, derived from the TESS Input Catalog (TIC), is also reported.}
    \tablehead{Sys. ID & $N_{\mathrm{flare}}$ & $\alpha\pm \delta\alpha$ & $\beta \pm \delta\beta$ & $T_{\mathrm{eff}}\,(\mathrm{K})$& $\overline{\Delta\log Z}$}
    \startdata
HD 143105    & 12  & $-0.5\pm0.2$   & $17\pm8$  & 6380 & 7.1  \\
TOI-1347    & 14  & $-0.7\pm0.2$   & $23\pm7$  & 5464 & 9.5  \\
WASP-100    & 15  & $-0.7\pm0.2$   & $25\pm7$  & 6900 & 25 \\
Kepler-1850 & 14  & $-0.4\pm0.2$   & $17\pm8$  & 4239 & 7.5  \\
HD 153557    & 32  & $-1.1\pm0.2$    & $39\pm6$  & 4837 & 9.0  \\
TOI-220     & 24  & $-1.1\pm0.2$    & $37\pm7.$  & 5298 & 8.8  \\
HD 163607    & 12  & $-1.8\pm0.8$    & $60\pm30$ & 5522 & 6.9  \\
\vdots&\vdots&\vdots&\vdots&\vdots&\vdots
    \enddata
\end{deluxetable}
The error bars in the FFD plots are computed following the procedure described in \citet{Hawley2014}. This assumes flares to follow a Poisson distribution as a function of energy, which then follows:
\begin{equation}
    \Delta \nu=\frac{\sqrt{N_{\mathrm{flares}}}}{t_{\mathrm{obs}}}.
\end{equation}
Propagating by taking the derivative of the logarithm of the flare rate: 
\vspace{-.1cm}
\begin{align}
    \label{equation: FFD uncertainty}
    \Delta \log_{10}(\nu)=\Delta\nu \,\cdot\nu^{-1}=\\\left(\frac{\sqrt{N_{\mathrm{flares}}}}{t_\mathrm{obs}}\right)\left(\frac{N_{\mathrm{flares}}}{t_\mathrm{obs}}\right)^{-1}=\frac{1}{\sqrt{N_{\mathrm{flares}}}}.
\end{align}

\begin{deluxetable}{c|ccccc}
    \label{table: FFD Params TOI}
    \tablecaption{First eight entries of derived $\alpha,\beta$ for TOI PC hosts. Flares with $N_{flare}\geq 10$ are reported in the table. Flares with $\Delta\log{Z}<5$ have been removed. $T_\mathrm{eff}$, derived from the TESS Input Catalog (TIC), is also reported.}
    \tablehead{Sys. ID & $N_{\mathrm{flare}}$ & $\alpha\pm \delta\alpha$ & $\beta \pm \delta\beta$ & $T_{\mathrm{eff}}$& $\overline{\Delta\log Z}$}
    \startdata
119.01 & 17  & $-1.0\pm0.3$   & $33\pm9.0$  & 5330  & 14 \\
131.01 & 17  & $-0.6\pm0.2$   & $23\pm6.0$  & 4170  & 20 \\
171.01 & 18  & $-0.8\pm0.2$   & $29\pm8.0$  & 5240  & 14 \\
201.02 & 17  & $-1.0\pm0.3$    & $36\pm10.0$ & 6460  & 12 \\
211.01 & 14  & $-2.0\pm0.6$    & $69\pm20.0$ & 5870 & 7.7  \\
212.01 & 34  & $-0.55\pm0.09$  & $21\pm3.0$  & 3330  & 23 \\
214.02 & 27  & $-1.3\pm0.2$    & $43\pm7.0$  & 5350  & 8.0  \\
218.01 & 291 & $-0.331\pm0.007$ & $13.6\pm0.2$  & 3150  & 25 \\
\vdots&\vdots&\vdots&\vdots&\vdots&\vdots
    \enddata
\end{deluxetable}
\vspace*{-\baselineskip}
This assigns larger uncertainty to uncommon events (i.e., high-energy flares) due to undersampling. Uncertainties are incorporated when determining $\alpha$ and $\beta$ for each system and are determined using \texttt{scipy}'s \emph{curve\_fit} routine. For each host with $N_{\mathrm{flare}}\geq10$ and flares with $\Delta\log{Z}\geq5$, we provide $\alpha$ and $\beta$ values with stated uncertainties in Table~\ref{table: FFD Params Exo} and Table~\ref{table: FFD Params TOI}.

\subsection{Phase Correlated Flares}
\label{subsection: GoF Results}
Flares can be correlated with a broad range of periodic astrophysical phenomena, including SPMIs and stellar rotation. To identify phase correlation within our sample, we perform goodness-of-fit and unbinned likelihood analyses on each target, phase-folding it to three different periods: the stellar rotational period,$P_{\star,\text{rot}}$, planetary orbital period, $P_{\text{orb}}$, and the synodic period, $P_{\text{syn}}$ between the orbital period and rotational period. We used stellar rotational periods available on the NASA Exoplanet Archive, in tandem with a derived stellar rotation catalog using the \texttt{SpinSpotter} pipeline with the same TESS photometry used to construct the flare catalog \citep{Holcomb2022}. In searching for magnetic SPIs, we present a metric, $\beta_{\mathrm{SPI}}$, that accounts for both the empirical flare distribution and the orbital geometry of periastron-dependent induced flares. Additionally, in the context of periastron-dependent magnetic SPIs, we present both statistically significant flare clustering, agnostic of orbital geometry, as well as targets with $\beta_{\mathrm{SPI}}>0.2$.

\subsubsection{Flare Correlation with close-in Planetary Periastron}
\label{subsubsection: Correlation with Periastron}

We investigate flare correlation between flare epochs and periastron approaches of close-in, sub-Alfvénic exoplanets as evidence of magnetic SPIs. The p-value distributions for each GoF test described in Section~\ref{subsubsection: GoF} for each sub-sample of hosts are summarized in Figure~\ref{figure: p_value_distributions}. Most targets show no significant deviation from uniformity, with p-values consistent with the null hypothesis across all goodness-of-fit tests. The $\sqrt{TS}_{VM}$ distributions for each sample are shown in Figure~\ref{figure: VM_TS}. Similar to the p-value distributions, most hosts exhibit flaring with low detection significance. Notably, 13 systems exhibit a $\sqrt{TS}>3$, with one system achieving a $\sqrt{TS}>5\sigma$. A subset of notable systems is shown in Table~\ref{table: results} and discussed in detail in Section~\ref{section: Discussion}. For eccentric systems, we compute $\beta_{\text{SPI}}$ following the prescription in Section~\ref{subsection: SPI Metric}, which is summarized in Figure~\ref{figure: beta_SPI Dist}.

\begin{figure*}
    \centering
    \label{figure: CandidateLCs_Peri}
    \includegraphics[width=0.86\textwidth]{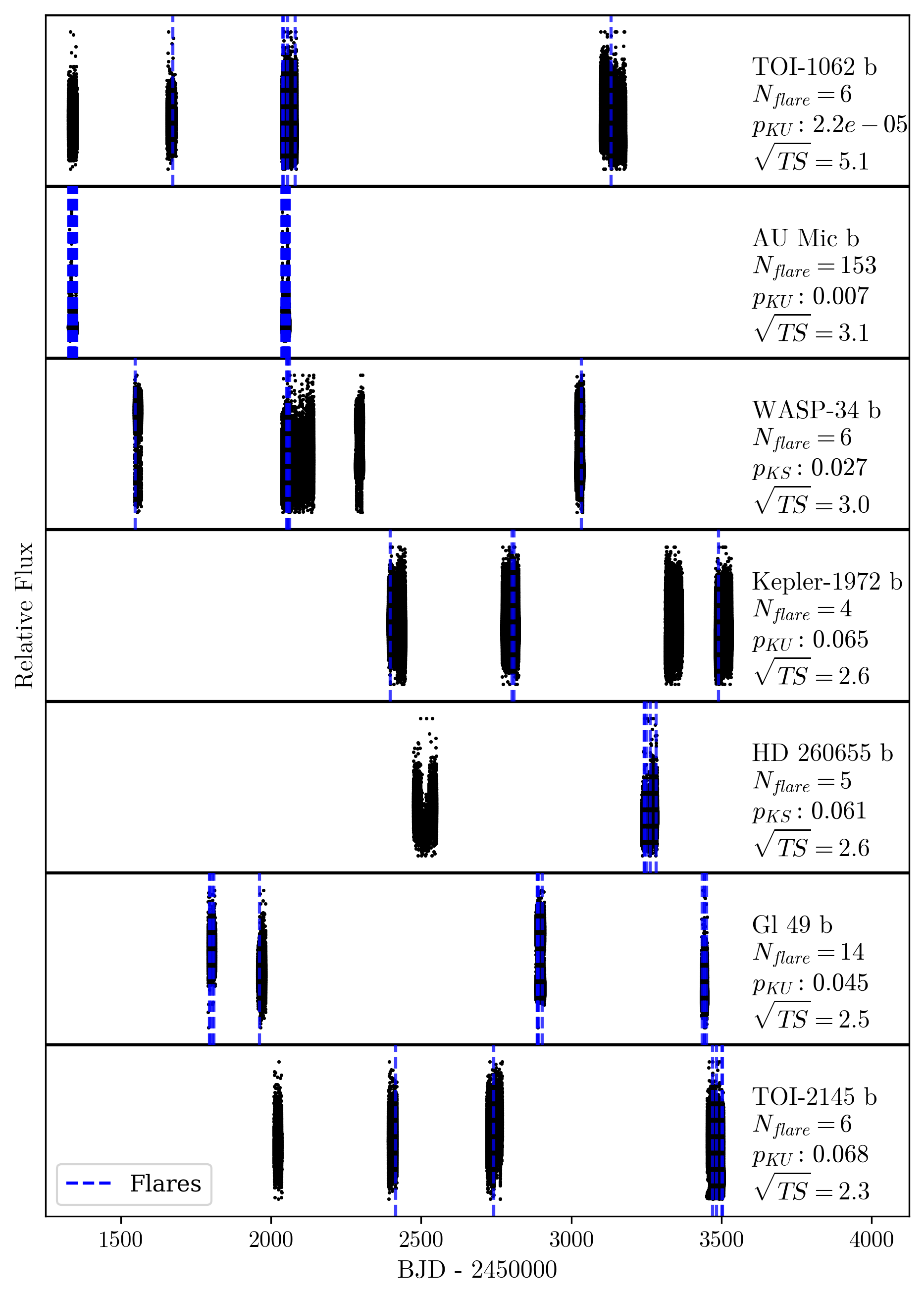}
    \caption{All TESS photometry available for targets with a constraint on the argument of periapsis with any $p_{K}<0.05$ and $\sqrt{TS}_{VM}>2$. The targets are ranked with the highest $\sqrt{TS}_{VM}$. We denote flare epochs by the dashed blue lines.}
\end{figure*}

\begin{figure*}
    \centering
    \label{figure: CandidateLCs_Trans}
    \includegraphics[width=0.875\textwidth]{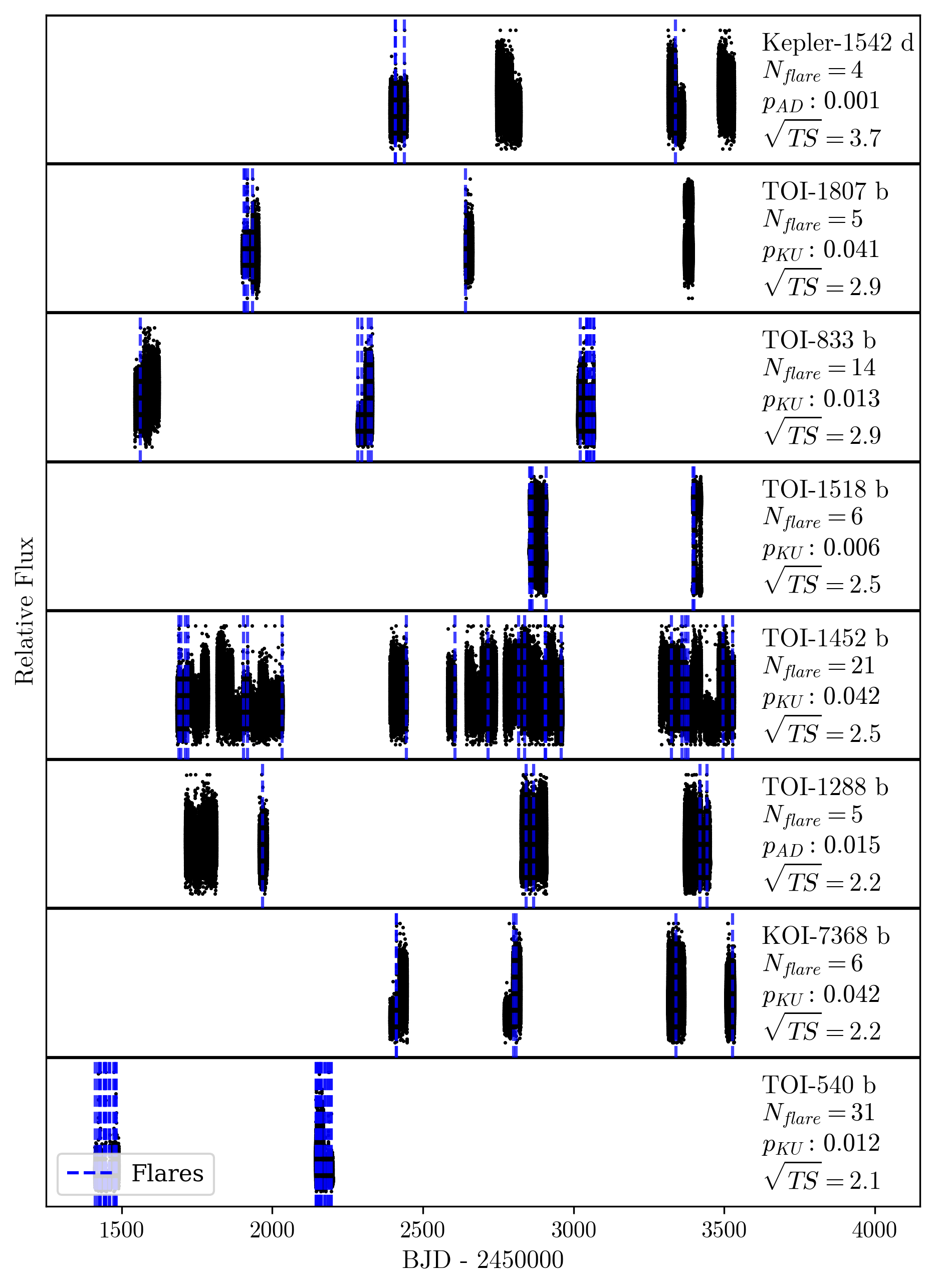}
    \caption{All TESS photometry available for targets without a constraint on the argument of periapsis with a p-value $<0.05$ across any of the GOF tests. We denote flare epochs by the dashed blue lines.}
\end{figure*}

\begin{figure*}
    \centering
    \label{figure: CandidateLCs_TOI}
    \includegraphics[width=0.9\textwidth]{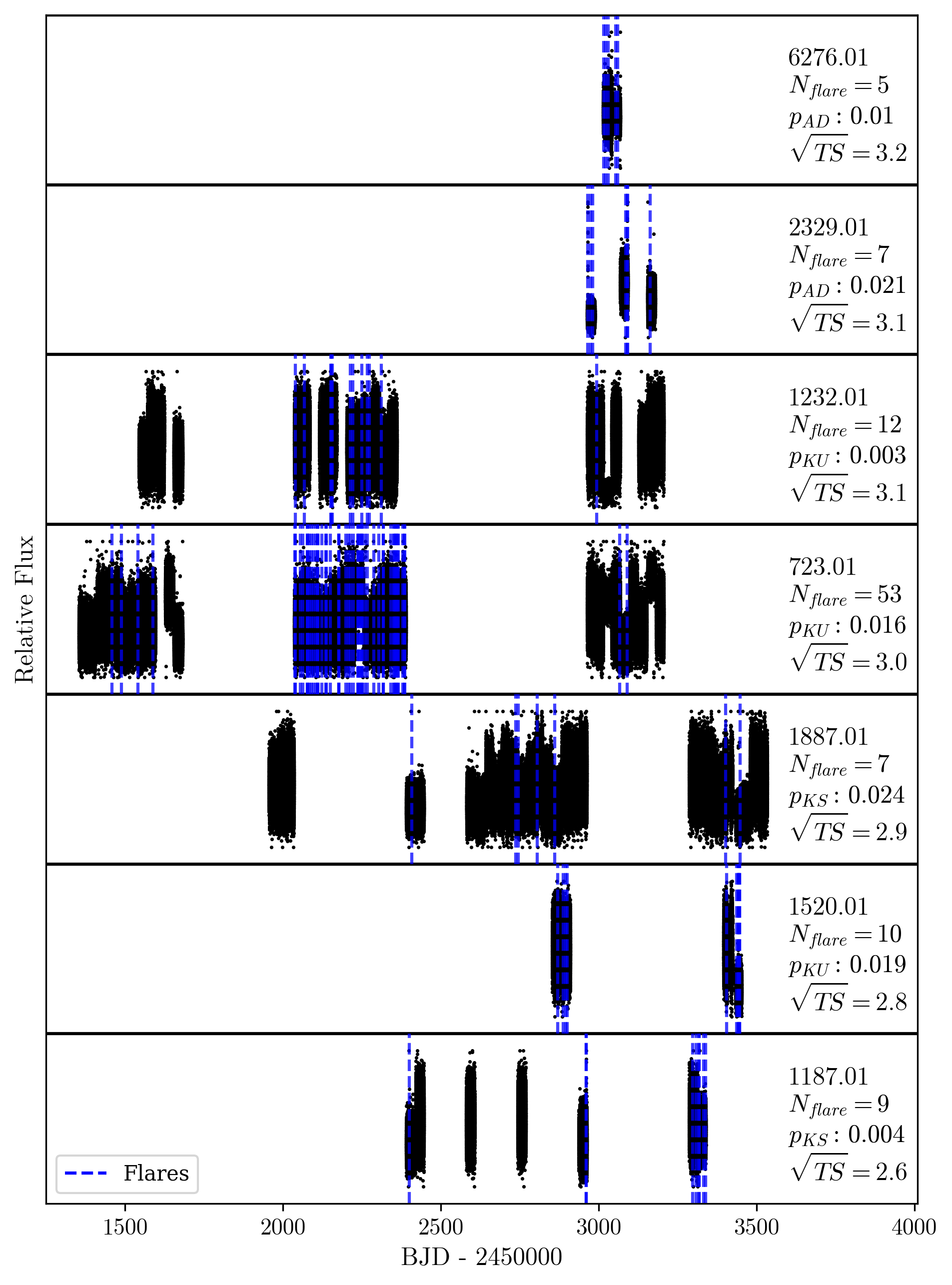}
    \caption{All TESS photometry available for TOI planet candidates with a p-value $<0.025$ across any GOF test. We denote flare epochs by the dashed blue lines.}
\end{figure*}
\movetabledown=4cm

\begin{rotatetable*}
\label{table: results}
\centering
\begin{deluxetable*}{c|ccccccc|ccccc}
    \tabletypesize{\footnotesize}
    \tablecaption{\small Targets exhibiting statistically significant flare clustering. Candidates have at least one metric with $>2\sigma$ significance, $N\geq4$ flares and $\overline{\Delta\log{Z}}>5$. Orbital parameters $a$ and $e$ give geometric context for each system. Closest approach distance, $r_{p}$, the estimated Alfvén radius of the host, and the probability that the planet's periastron is sub-Alfvénic, $P(r_{c}\leq R_{A})$*, are also reported.}
    \tablehead{
        \vspace{-0.2cm}&&&&&&&&a&&$r_{c}$&$R_{A}$&\\
        \vspace{-0.2cm}Sys. ID&$p_{\mathrm{KU}}$&$p_{\mathrm{KS}}$&$p_{\mathrm{AD}}$&$\overline{p_{\mathrm{KS}}}$&$\overline{p_{\mathrm{AD}}}$&$N$&$\sqrt{TS}_{\mathrm{VM}}$&&e&&&$P(r_{c}\!\leq\!R_{A})$\\
         &\colhead{}&\colhead{}&\colhead{}& & & &&(AU)& & (AU)  & (AU) &
        }
    \startdata
    & \multicolumn{12}{c}{\normalsize Hosts of Interest with known Argument of Periapsis} \\
    \hline
    TOI-1062\,b& \textbf{2.2e-5} & \textbf{1.9e-4} & \textbf{1.3e-3} & 2.4e-3 & 1.5e-2 & 6  & \textbf{5.1} &$0.052^{+0.024}_{-0.025}$\citett{1}&$0.177\pm0.042$\citett{1}&$0.043\pm{0.02}$&$0.065^{+0.2}_{-0.04}$&0.69  \\
    AU Mic\,b&6.9e-3&1.2e-2&1.7e-2&1.2e-2&1.7e-2&153&3.1&$0.065\pm0.0012$\citett{2}&$0.006\pm0.001$\citett{2}&$0.052^{+0.003}_{-0.0013}$*&$0.015^{+0.002}_{-0.006}$*&0.99*\\
    WASP-34\,b&3.1e-2&2.7e-2&9.6e-2&2.7e-2&9.6e-2&9&2.6&$0.0524\pm0.0004$\citett{3}&$0.04\pm0.01$\citett{4}&$0.0503^{+6e-4}_{-4e-4}$ &$0.058^{+0.17}_{-0.04}$&0.57\\
    Kepler-1972\,b&6.5e-2&5.2e-1&4.1e-1&5.2e-1& 2.0e-1&4& 2.64 &$0.08\pm0.001$\citett{5}&$0.067^{+0.07}_{-0.04}$\citett{6}&$0.075\pm0.002$\citett{5}&$0.11^{+0.3}_{-0.08}$&0.64\\
    HD 260655\,b&1.4-1&6.1e-1&1.1e-1&6.1e-2&1.1e-1&5&2.35&$0.02933\pm0.0002$\citett{7}&$0.039^{+0.04}_{-0.02}$\citett{7}&$0.0281^{0.001}_{-0.002}$&$0.03^{+0.07}_{-0.018}$&0.5\\
    Gl 49\,b&4.4e-2&3.0e-1&3.2e-1&3.0e-1&3.2e-1&14& 2.5&$0.090\pm0.001$\citett{8}&$0.36^{+0.099}_{-0.096}$\citett{8}&$0.058^{+0.008}_{-0.01}$&--&--\\
    TOI-2145\,b&4.4e-2&3.0e-1&3.2e-1&3.0e-1&3.2e-1&6& 2.5&$0.111^{+0.001}_{-0.002}$\citett{9}&$0.208^{+0.03}_{-0.05}$\citett{9}&$0.087^{+0.003}_{-0.002}$&$0.22^{+0.6}_{-0.15}$&0.73\\
    \hline
    & \multicolumn{12}{c}{\normalsize Hosts of Interest Without Known Argument of Periapsis} \\
    \hline
    TOI-1807\,b&4.1e-2&7.0e-2&2.0e-1&1.1e-1&1.1e-1&5&3.1&$0.0122\pm0.003$\citett{9}&$0.006\pm0.001$\citett{9}&$0.052^{+0.003}_{-0.0013}$&$0.02^{+0.04}_{-0.012}$&0.74\\
    TOI-1518\,b&6.4e-3&2.3e-1&2.6e-1&1.4e-1&2.3e-1&6& 2.5 &$0.0389\pm0.001$\citett{11}&$<0.01$\citett{11}&$0.0389\pm0.001$&--&--\\
    TOI-1452\,b&4.2-2&6.3e-2&1.0e-1&1.8e-2&2.3e-2&21&2.5&$0.061\pm0.003$\citett{12}&$0$\citett{12}&$0.061\pm0.003$&--&--\\
    TOI-1288\,b&1.2e-1&2.7e-2&1.5e-2&2.1e-2&5.5e-3&5& 2.2&$0.0374\pm0.0007$\citett{9}&$0$\citett{9}&$0.0374\pm0.0007$&$0.077^{+0.27}_{-0.05}$&0.70\\
    KOI-7368\,b&4.1e-2&2.6e-1&3.4e-1&2.1e-1&2.8e-1&6& 2.2&$0.0678$\citett{13}&$0$\citett{13}&$0.0678$&$0.025^{+0.05}_{-0.014}$&0.10\\
    TOI-540\,b&1.2e-2&2.3e-1&2.0e-1&4.4e-2&3.6e-2&31& 2.1&$0.0122\pm+0.0004$\citett{14}&$0$\citett{14}&$0.0122\pm+0.0004$\citett{14}&$0.04^{+0.09}_{-0.025}$&0.68\\
    \hline
    & \multicolumn{12}{c}{\normalsize TESS Objects of Interest Planet Candidates} \\
    \hline
    6276.01  & 1.9e-1 & 2.6e-1 & 9.8e-3 & 1.9e-2 & 2.7e-1 & 5&3.2&0.048**&--&--&--&-- \\
    2329.01  & 1.7e-1 & 1.7e-1 & 2.1e-2 & 1.8e-1 & 1.6e-1 & 7 & 3.1 & 0.0658**&--&--&--&-- \\
    1232.01 & 3.2e-3 & 1.0e-2 & 8.3e-3 & 3.6e-3 & 1.1e-2 & 12 &3.1&0.117**&--&--&--&-- \\
    723.01 & 1.6e-2 & 3.4e-2 & 1.9e-2 & 2.4e-2 & 5.9e-2 & 53 &3.0&0.0229**&--&--&--&--\\
    1887.01  & 6.8e-2 & 2.4e-2 & 4.8e-2 & 2.0e-1 & 1.9e-1& 7& 2.9& 0.0287**&--&--&--&-- \\
    1520.01 & 1.9e-2 & 7.3e-2 & 5.5e-2 & 2.3e-2 & 9.0e-2 & 10&2.8&0.0693**&--&--&--&-- \\
    5051.01 & 3.8e-2 & 7.9e-2 & 1.6e-2 & 2.6e-2 & 5.5e-2 & 9 &2.7&0.0418**&--&--&--&-- \\
    1187.01 & 1.2e-2 & 3.9e-3 & 7.5e-2 & 5.2e-2 & 3.4e-2 & 9 &2.6&0.041**&--&--&--&-- \\
    \enddata
    \tablerefs{\citet{Otegi2021}\citett{1},\citet{Wittrock2023}\citett{2},\citet{Smalley2011}\citett{3},\citet{Stassun2017}\citett{4},\citet{Fulton2018}\citett{5},\citet{Leleu2022}\citett{6}.\citet{Luque2022}\citett{7}, \citet{Perger2019}\citett{8}, \citet{Polanski2024}\citett{9}, \citet{Giacalone2022}\citett{10}, \citet{Cabot2021}\citett{11}, \citet{Cadieux2022}\citett{12}, Kepler Objects of Interest Table \citett{13}, \citet{Ment2020}\citett{14}, TESS Objects of Interest Table \citett{15}}
    \tablecomments{*The Alfvén surface estimates for AU Mic are computed using typical equatorial surfaces derived in \citet{Alvarado2022} using MHD simulations\\ **Semi-major axes of TOI PCs are estimated from the public SPOC reports from the TOI catalog.}
\end{deluxetable*}
\end{rotatetable*}

\begin{figure}
    \centering
    \label{figure: beta_SPI Dist}
    \includegraphics[width=0.47\textwidth]{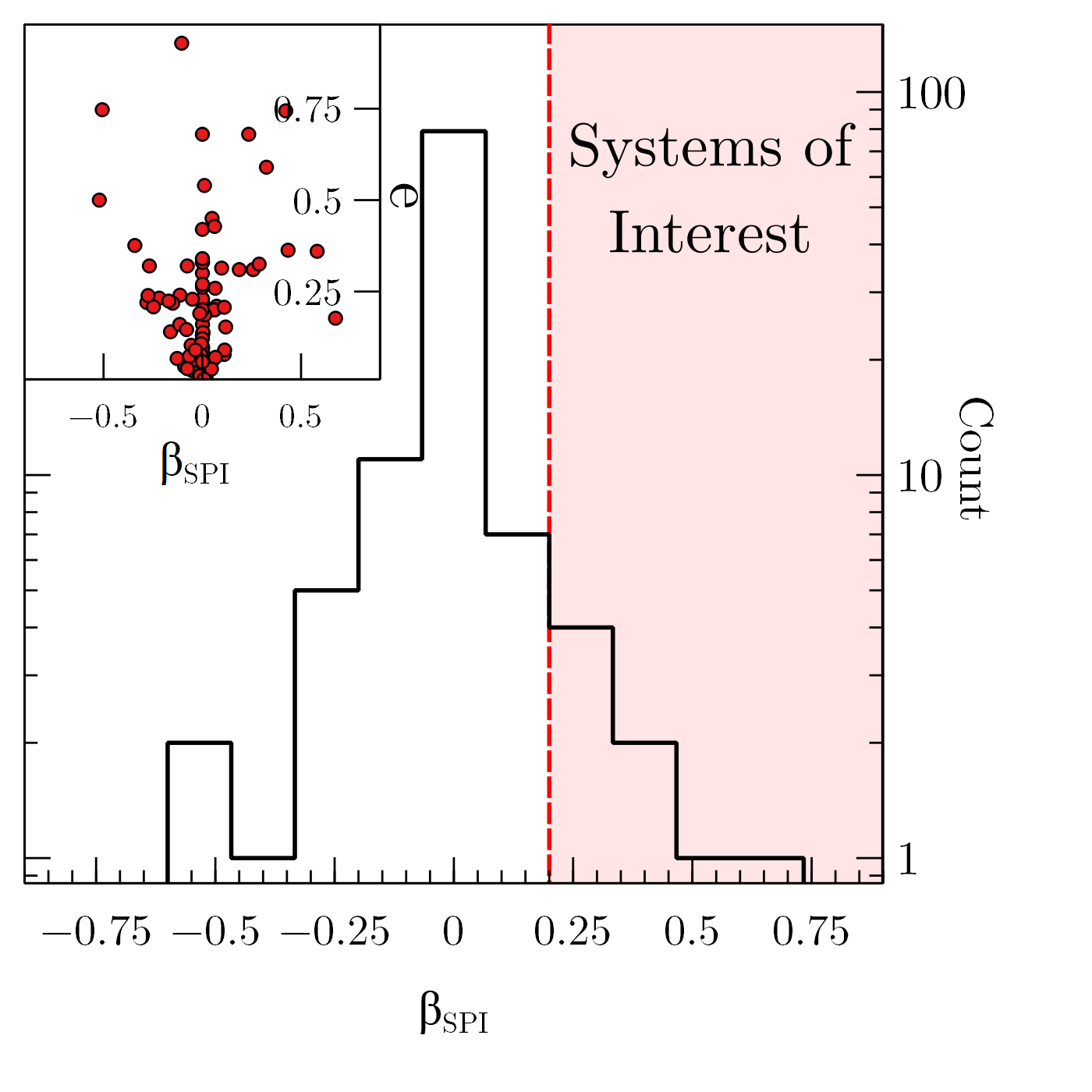}
    \caption{Histogram and scatter plot of the distribution of $\beta_{\text{SPI}}$ for eccentric systems with at least four flares. Systems with $\beta_{\text{SPI}}>>0$ show flare clustering consistent with periastron-dependent SPMIs. $\beta_{\text{SPI}}\approx0$ are systems which are consistent with uniform flaring. Systems with $\beta_{\text{SPI}}<<0$ have flare clustering around apastron. The scatterplot shows most systems are consistent with uniform flaring ($\beta_{\text{SPI}}=0$) as a function of orbital distance. Systems with $\beta_{\text{SPI}}>0.2$ are designated as systems of interest.}
\end{figure}

Small perturbations in the flare PDF through SPMIs require a large observational baseline to emerge as a statistically significant signal, as seen in Section~\ref{subsection: Simulation Analysis}. We identify flare clustering in systems consistent with periastron-dependent SPMIs using $\beta_{\text{SPI}}$, even if the sample eCDF does not deviate significantly from uniform flaring. We present systems with $\beta_{\mathrm{SPI}}>0.2$, chosen based on the tail of the distribution in Figure~\ref{figure: beta_SPI Dist}, as additional induced flare candidates. We quantify the additional observation time required to yield a $3\sigma$ detection, $t_{3\sigma}$, assuming the flare sample is representative of the underlying PDF. Using the computed empirical flare rate for all TESS sectors analyzed per target, we inject flares drawn from the best-fit PDF and calculate the observational baseline required to achieve a $3\sigma$ detection in either the goodness-of-fit tests or the unbinned likelihood analysis. We repeat this 100 times for each target and take the median value. The results are highlighted in Table~\ref{table: Beta SPI Candidates}.

\begin{figure*}[p]
\centering
\label{fig: Beta_Polar_Plots}
\gridline{\fig{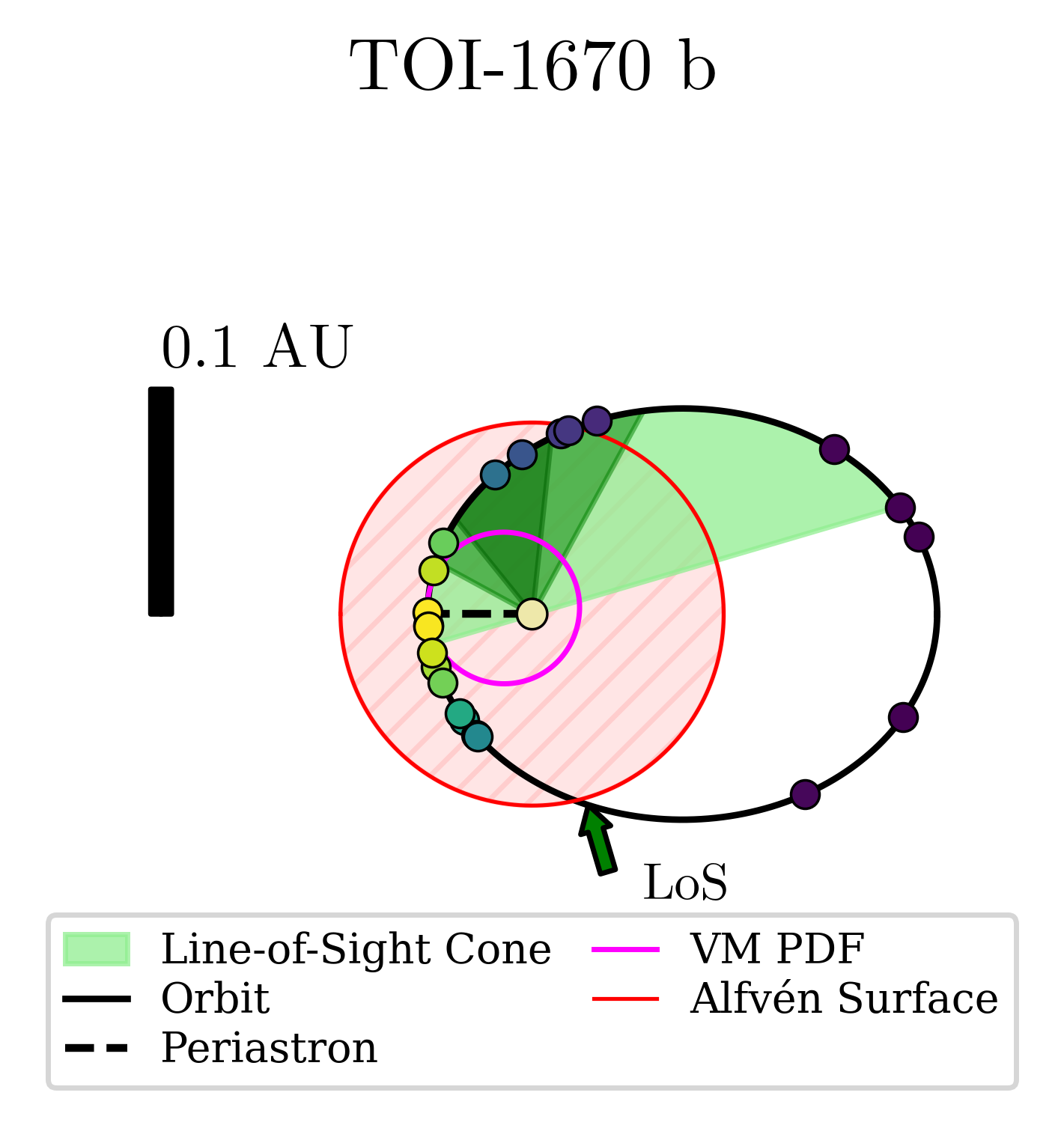}{0.3\textwidth}{(a)}
          \fig{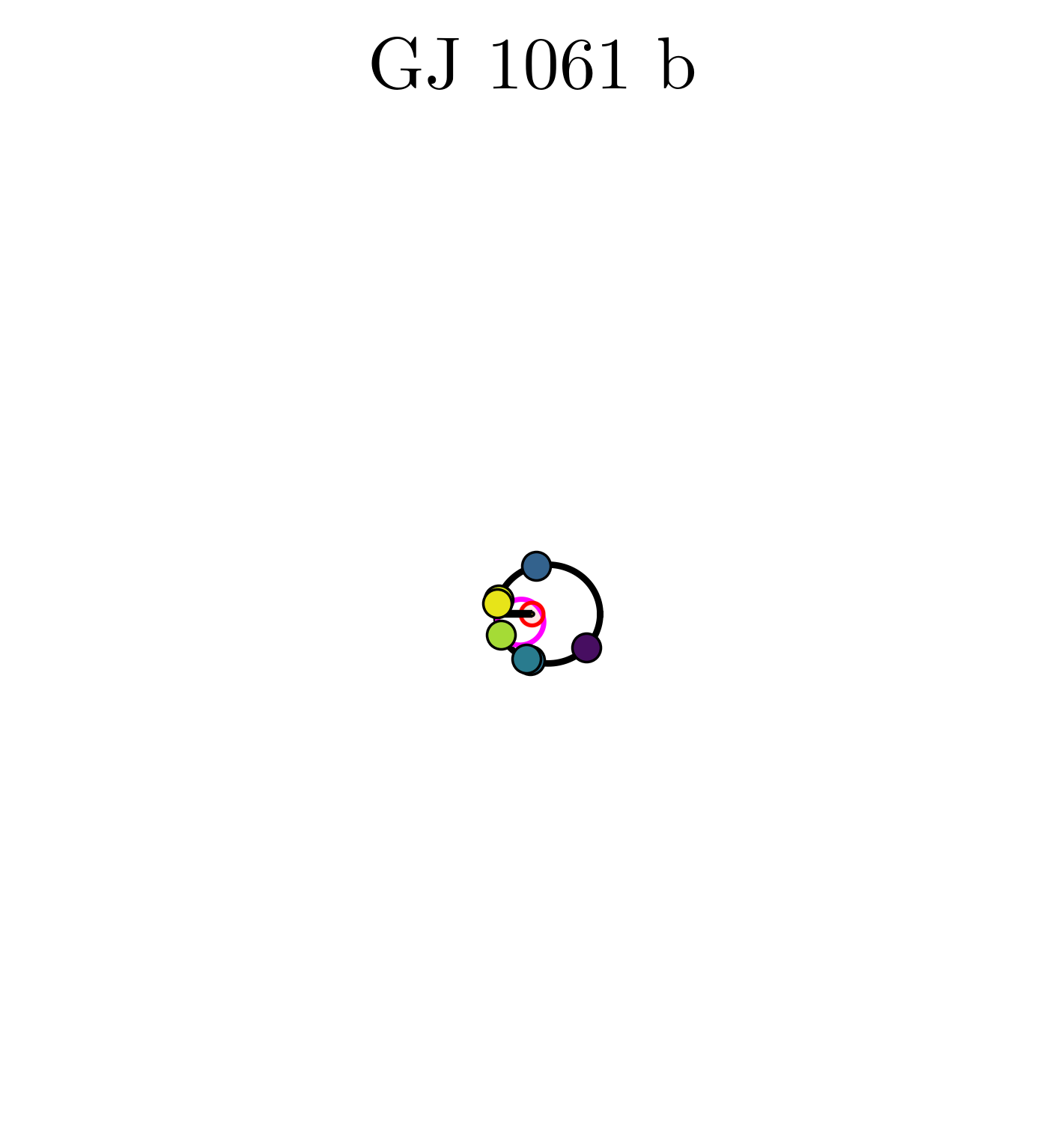}{0.3\textwidth}{(b)}}
\gridline{\fig{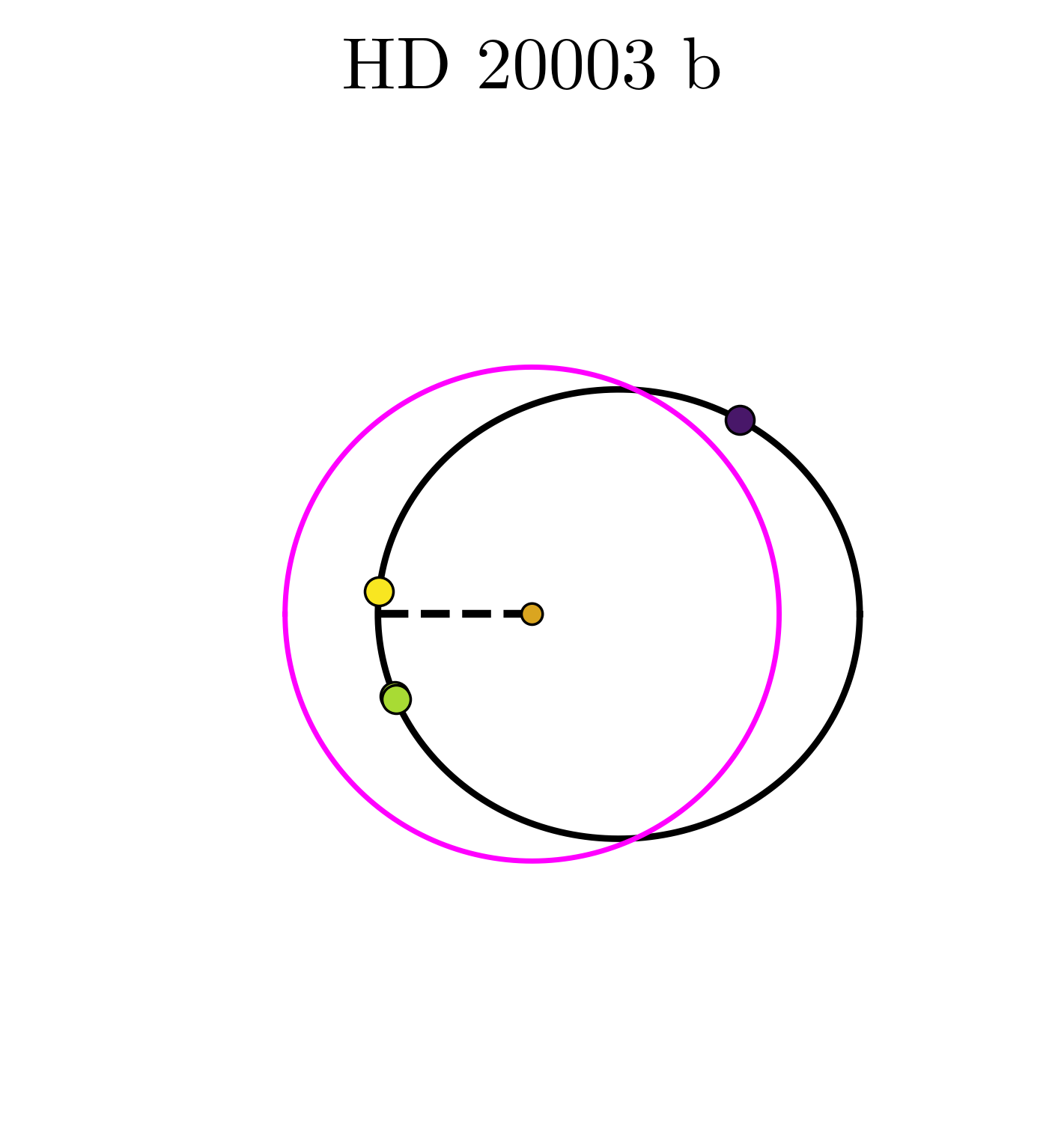}{0.3\textwidth}{(c)}
          \fig{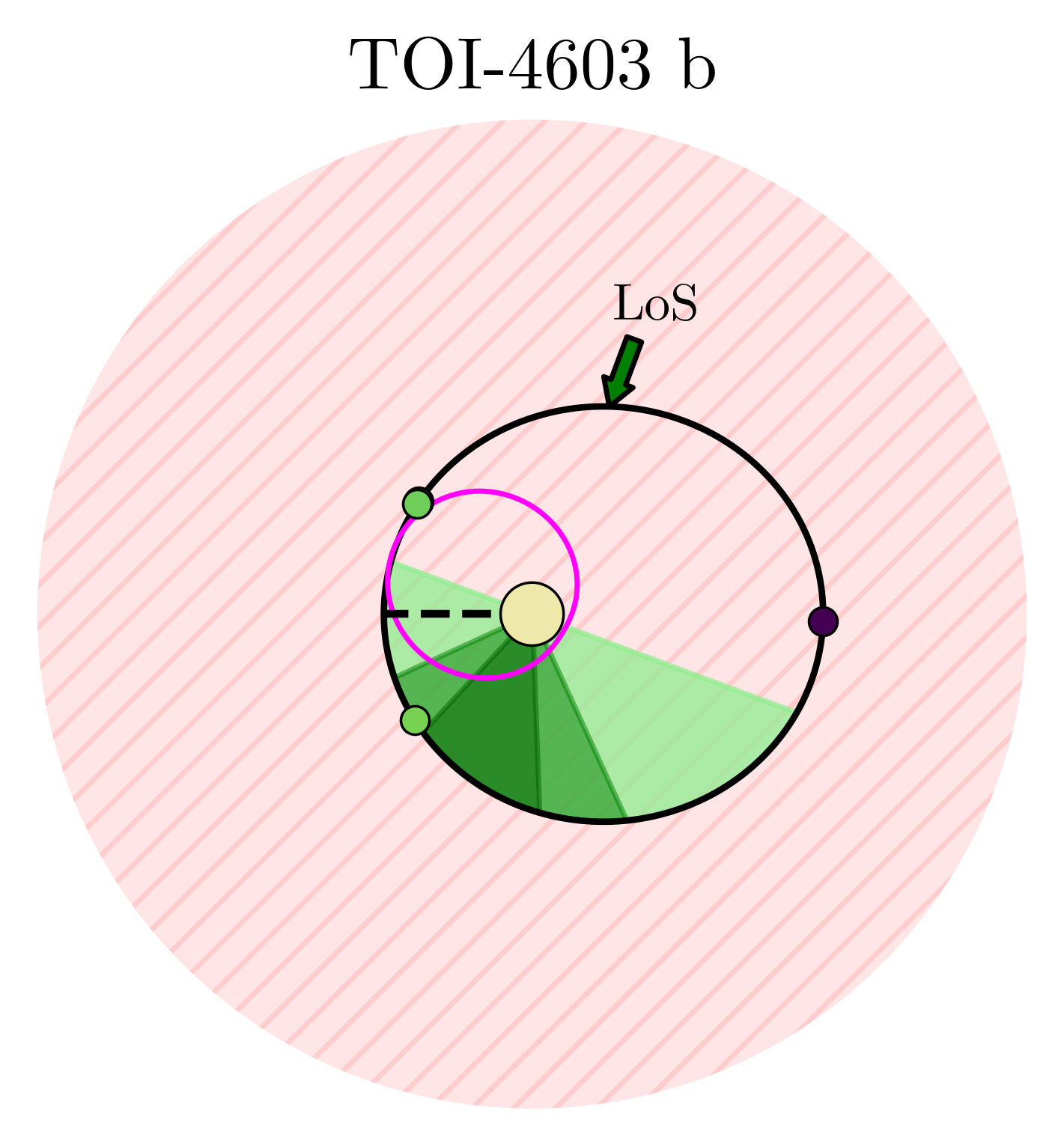}{0.3\textwidth}{(d)}}
\gridline{\fig{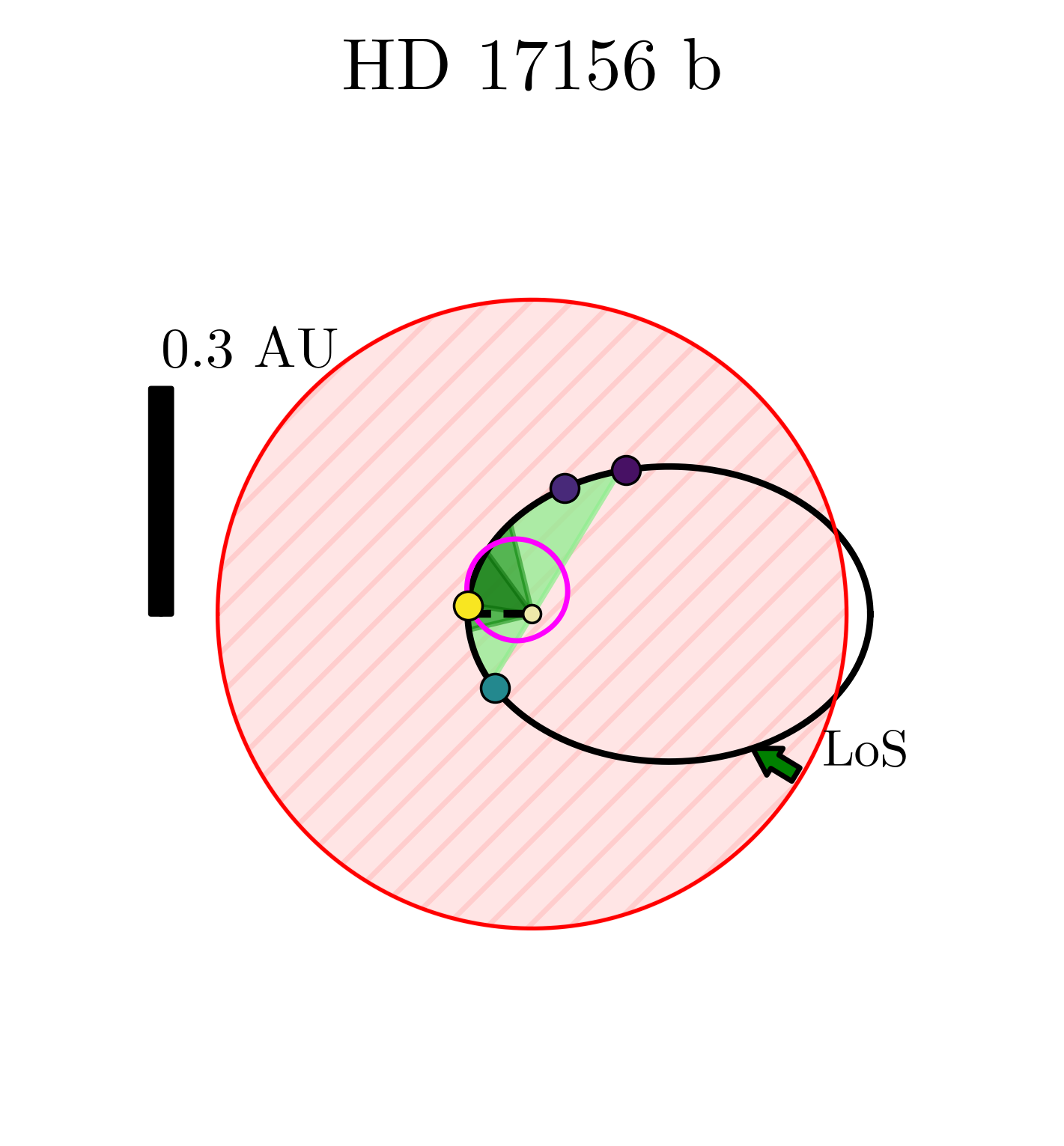}{0.3\textwidth}{(f)}
          \fig{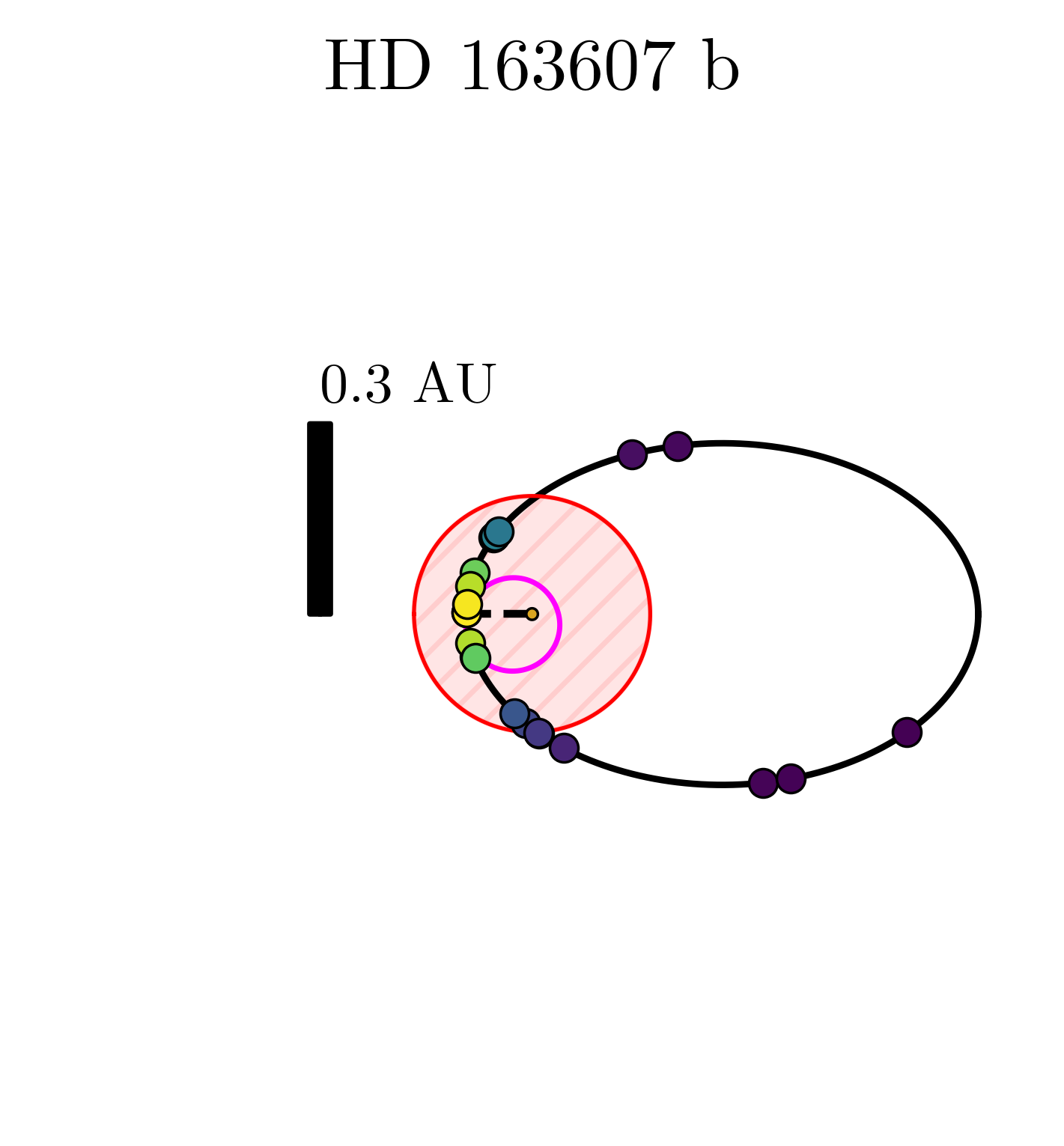}{0.3\textwidth}{(g)}}
\caption{Polar projections of all systems of interest with $\beta_{\mathrm{SPI}} > 0.2$. Phase-folded flare epochs are denoted by colored markers imposed on the orbital path (black line). Lighter markers correspond to flares closer to periastron, while darker markers correspond to flares closer to apastron. The cyan line shows the best-fit von Mises PDF projected onto the orbit. The red shaded region denotes the best-guess Alfvén surface from Section~\ref{subsubsection: Alfvén Surface}. The shaded green regions represent different possible line-of-sight constraints when the planet is behind the stellar limb, illustrating $\pm90^{\circ}$, $45^{\circ}$, $22.5^{\circ}$ cones. The green arrow denotes the observational line-of-sight. The orbital diagrams follow the same scale in (a), except for (f) and (g), where new scaling is used due to the high eccentricity of the systems.}
\end{figure*}

\begin{deluxetable}{c|cccccc}
    \label{table: Beta SPI Candidates}
    \tablecaption{Systems with $\beta_{\mathrm{SPI}}>0.2$. We report $t_{3\sigma}$ if applicable. Additionally, we report $\Delta\phi$, the angle between the best-fit $\mu$ parameter and $\omega_{p}$.}
    \tablehead{Sys. ID & $N_{\mathrm{flare}}$ & $\beta_{\text{SPI}}$ & $p_{\text{KU}}$ & $\sqrt{TS}_{VM}$& $t_{3\sigma}$ &$\Delta\phi$\\&&&&&[d]&[deg]}
    \startdata
TOI-1062\,b    & 6 & 0.63 & 1.9e-4 & 5.1 & -- & 6\\
Gl 49\,b     & 14 & 0.43 & 0.05 & 2.48 & 654 & 31\\
HD 163607\,b & 21 & 0.42 & 0.28 & 1.62  & 1466& 28\\
TOI-4603\,b  & 4  & 0.29 & 0.60 & 0.98 & 2844& 31\\
GJ 1061\,b   & 7  & 0.26 & 0.57 & 1.29 & 513& 45\\
HD 20003\,b  & 4  & 0.58 & 0.24 & 1.45 & 4582& 6\\
HD 17156\,b  & 4  & 0.23 & 0.70 & 0.88 & 4891& 59\\
TOI-1670\,b  & 22 & 0.32 & 0.24 & 1.33 & 2006& 13
\enddata
\end{deluxetable}

\begin{figure*}[p]
\centering
\label{fig: Peri_Polar_Plots}
\gridline{\fig{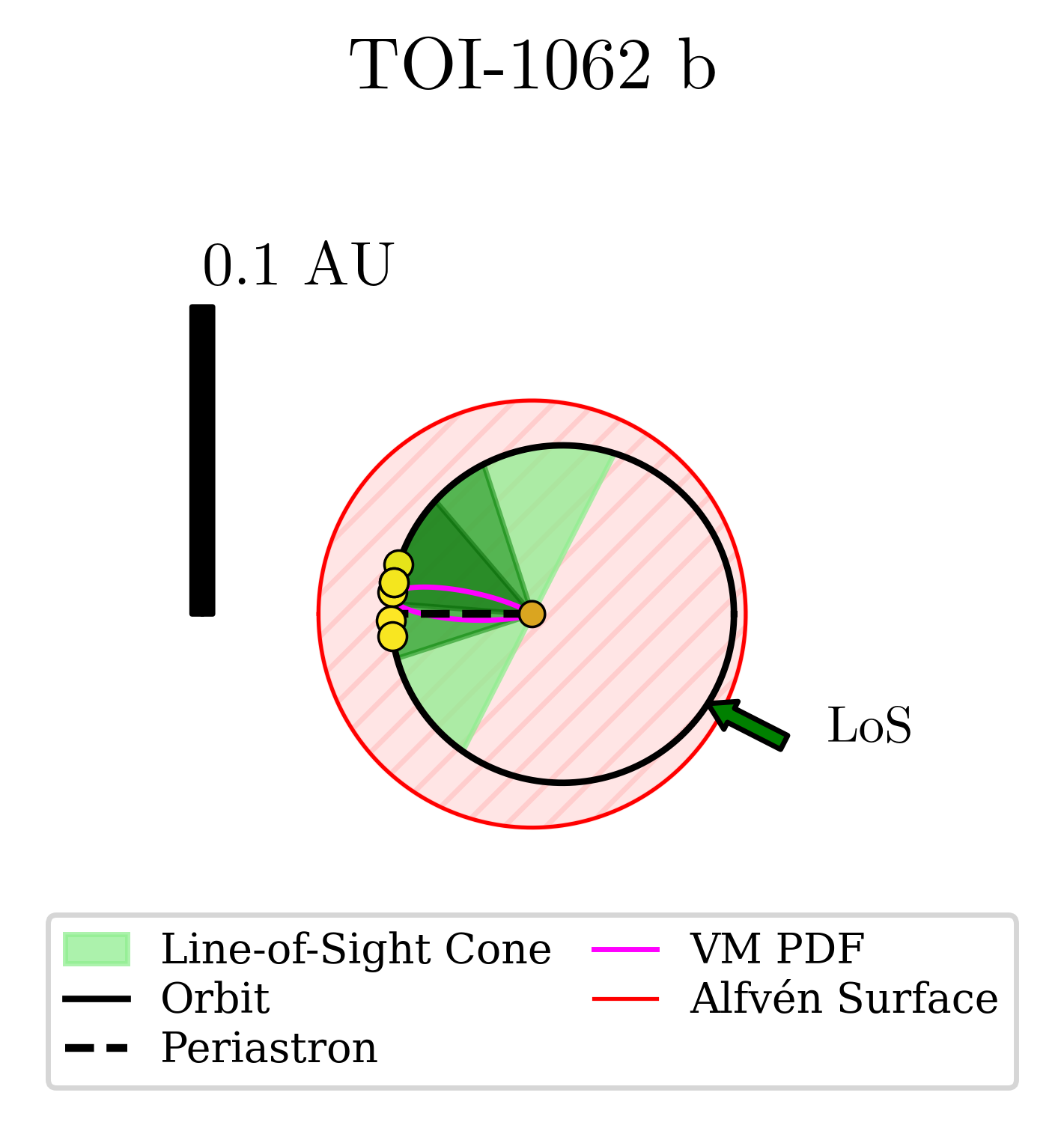}{0.3\textwidth}{(a)}
          \fig{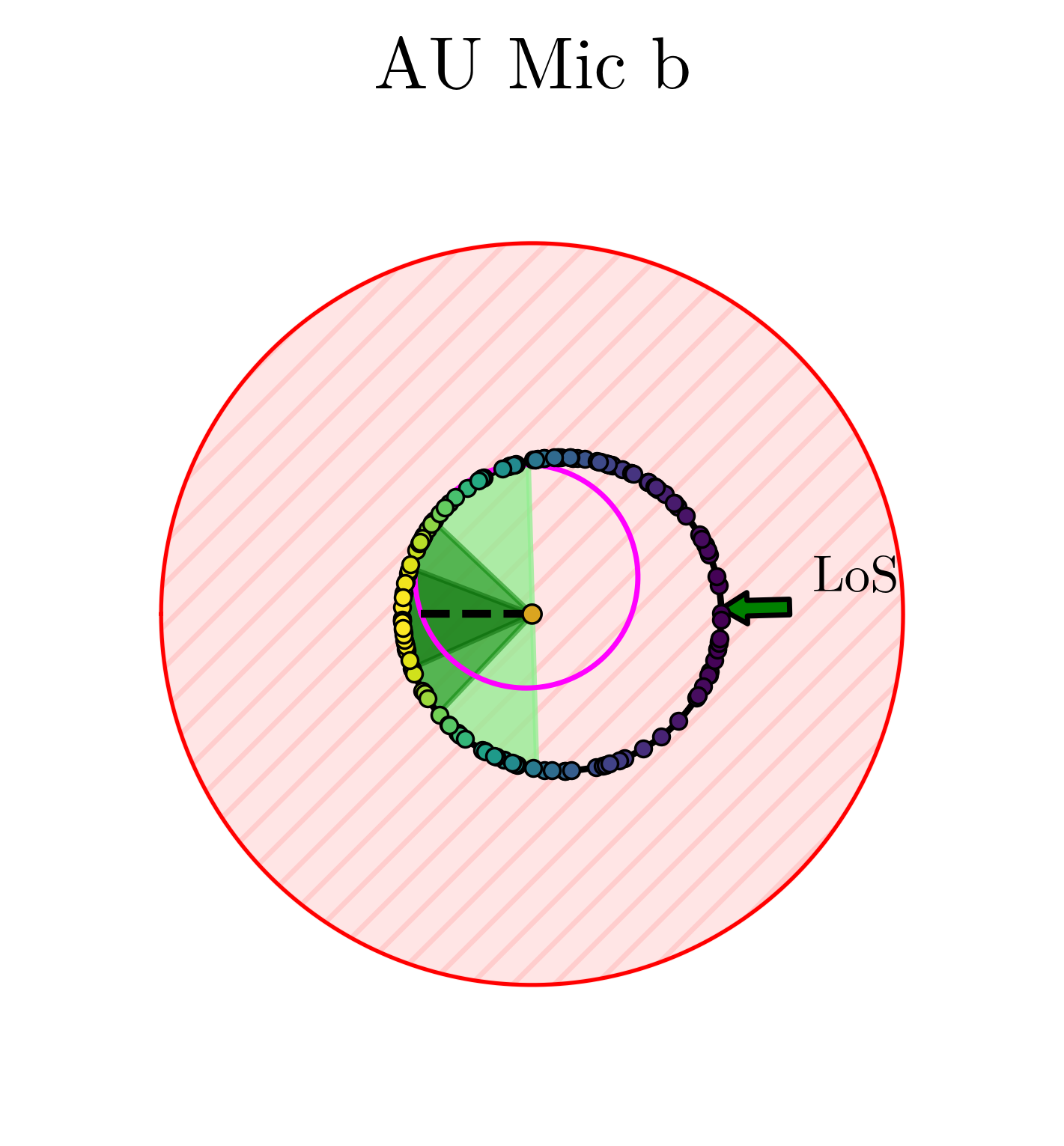}{0.3\textwidth}{(b)}}
\gridline{\fig{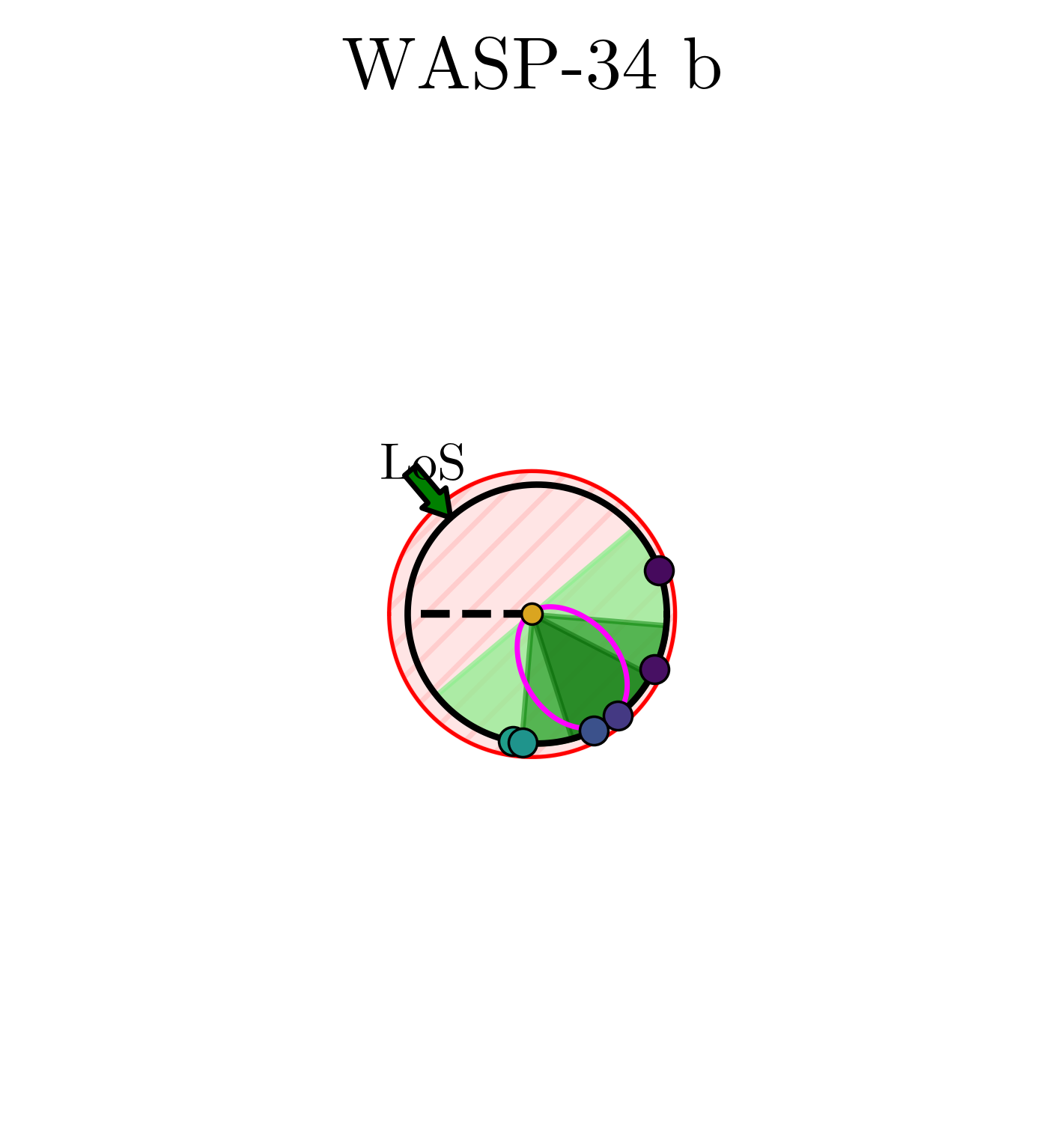}{0.3\textwidth}{(c)}
          \fig{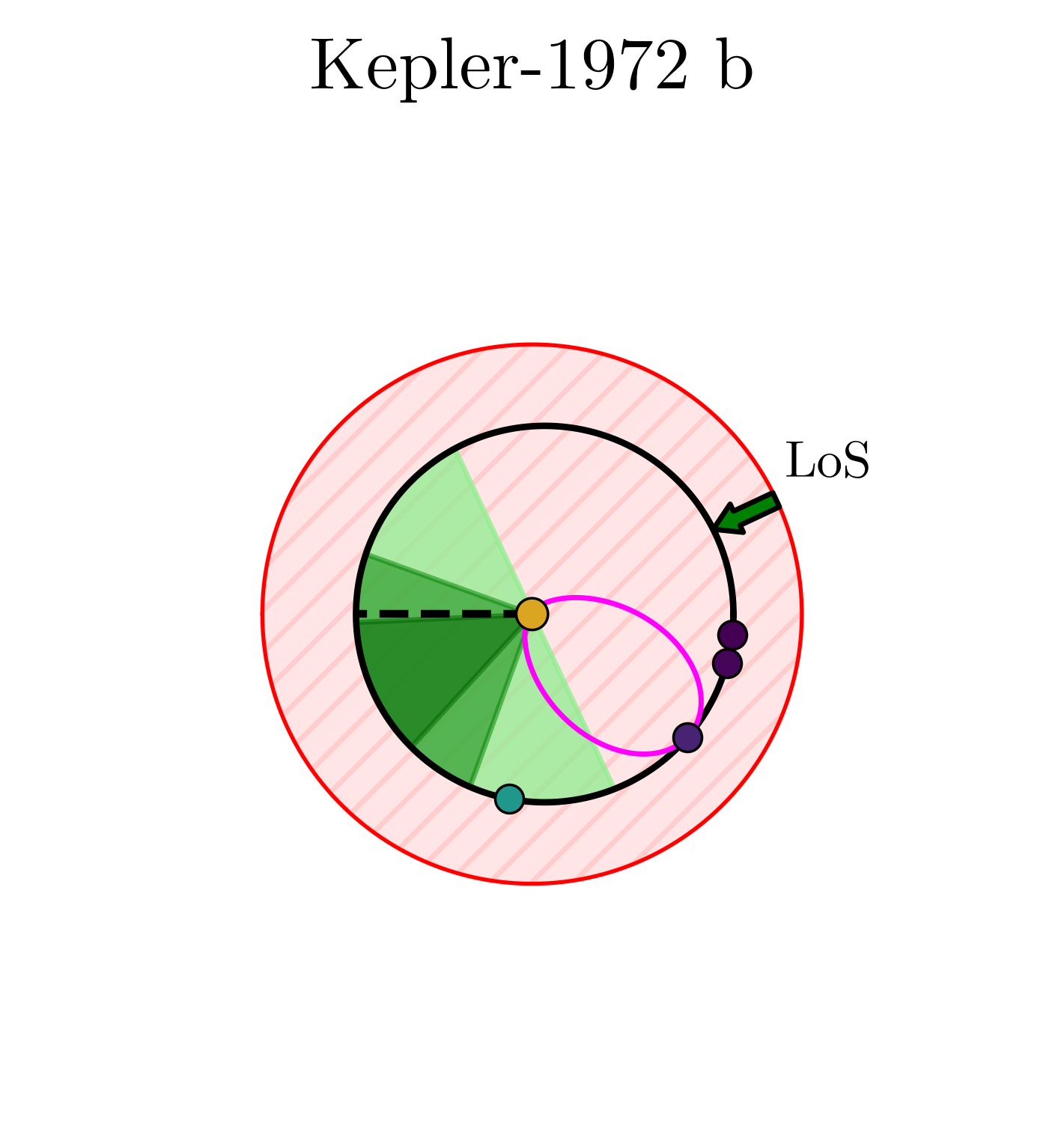}{0.3\textwidth}{(d)}
          \fig{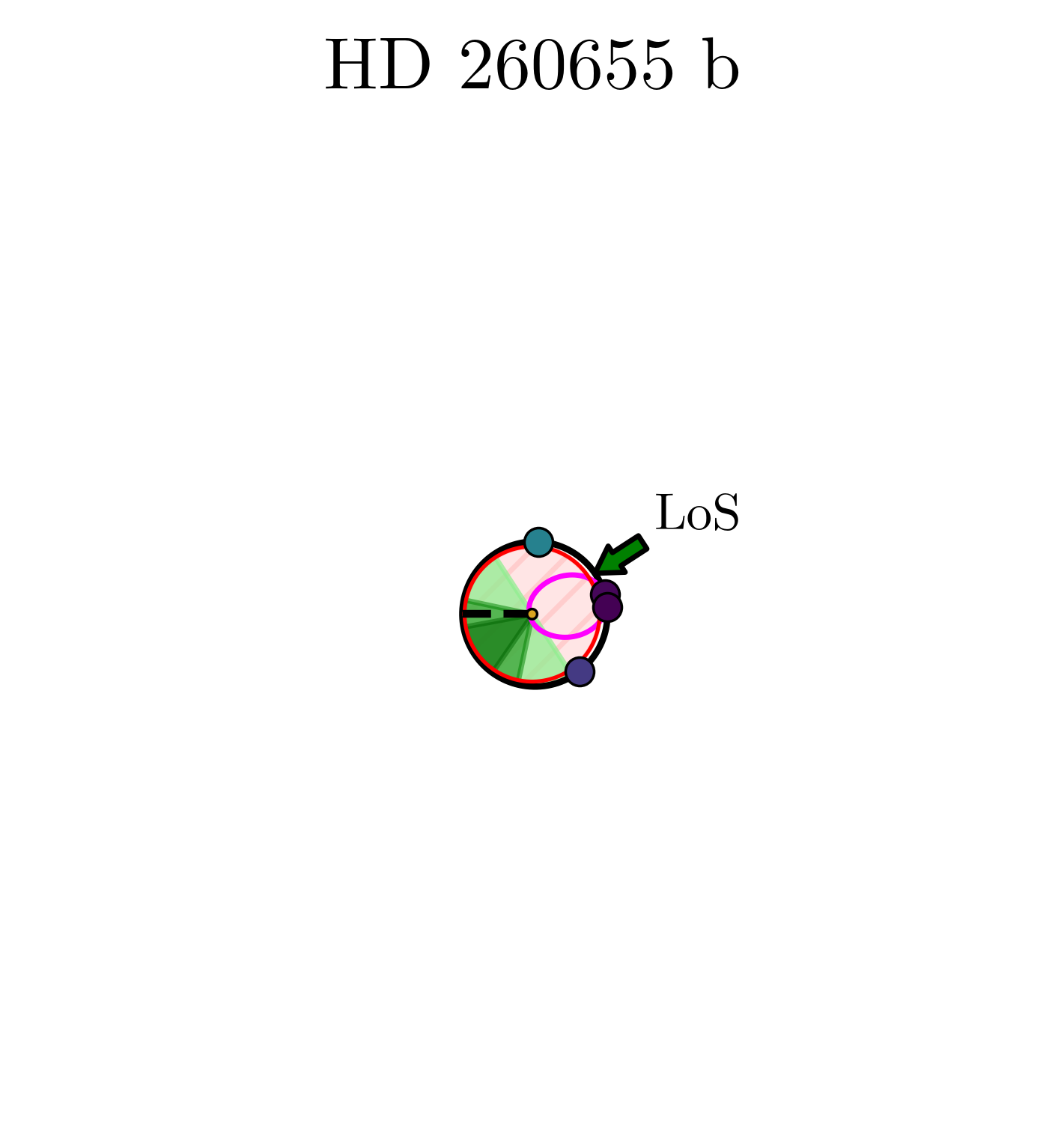}{0.3\textwidth}{(e)}}
\gridline{\fig{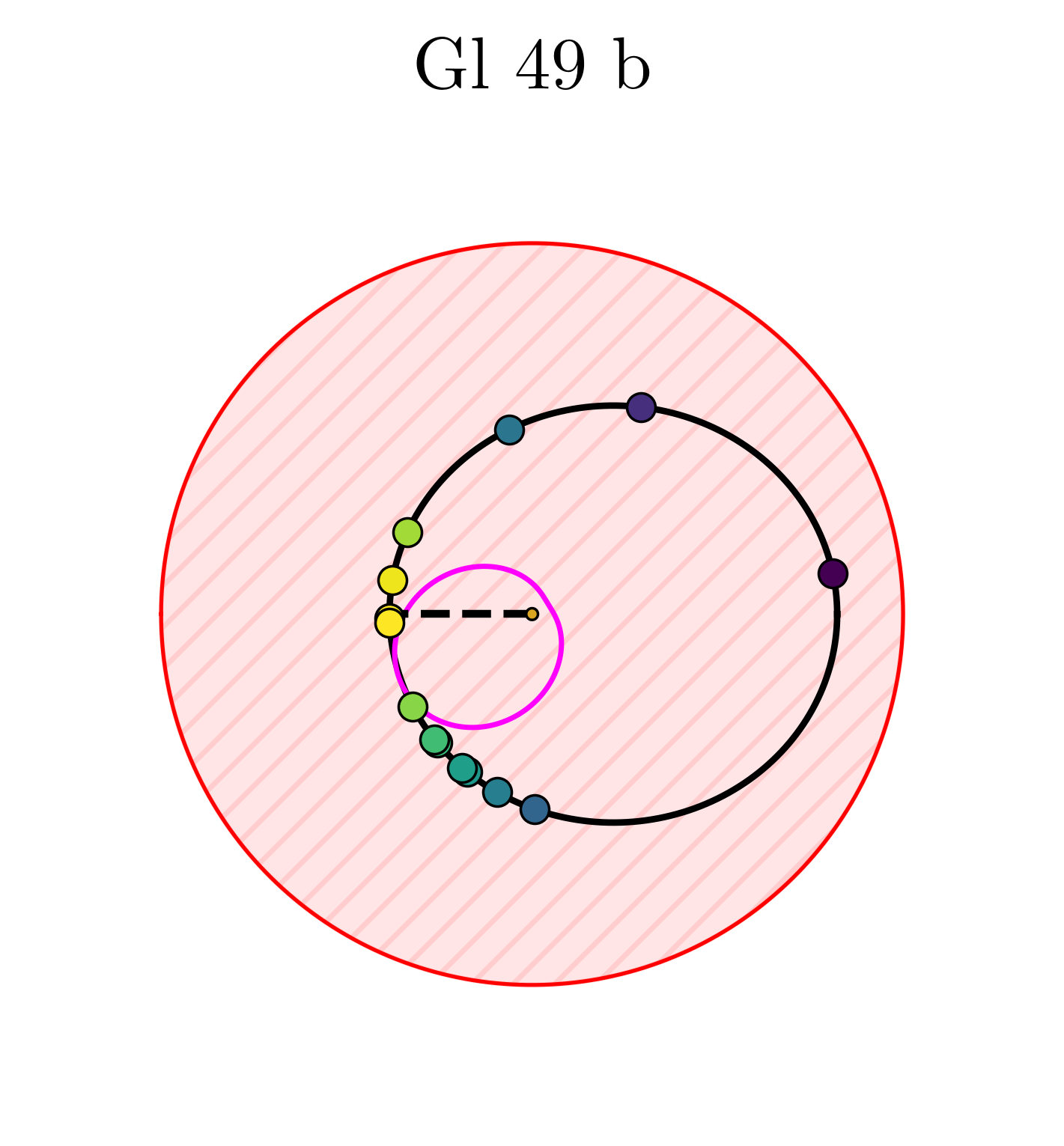}{0.3\textwidth}{(f)}
          \fig{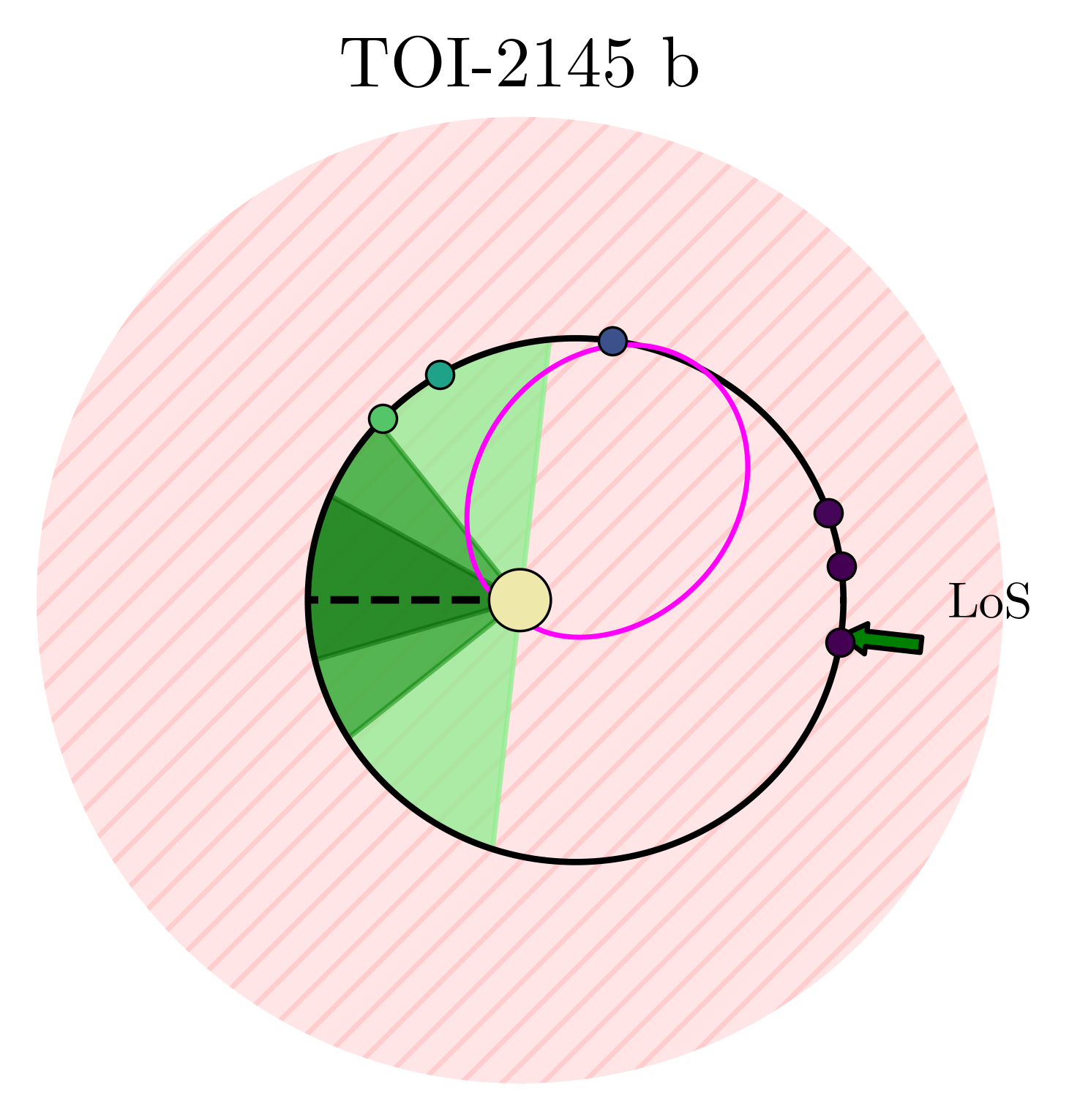}{0.3\textwidth}{(g)}}
\caption{Polar projections of all systems of interest with a constraint on $\omega$. The blue dotted-dashed line represents the planetary orbit. All figures are to scale with the marker in (a) denoting the length of $0.1\,\mathrm{AU}$. The cyan line shows the best-fit von Mises PDF projected onto the orbit. The dashed purple lines are the individual flare epochs derived in this work. The estimates of the Alfvén surface are shown in red. The solid red line indicates the best-guess value, while the $\pm1\, \sigma$ confidence intervals are denoted by the light and dark-red regions, respectively. $\omega$ is denoted by the dashed black-line set $180^{\circ}$ from the origin, with the black shaded regions indicating $\pm1\sigma$ confidence intervals.}
\end{figure*}

\begin{figure}
    \centering
    \label{figure: p_value_distributions}
    \includegraphics[width=0.47\textwidth]{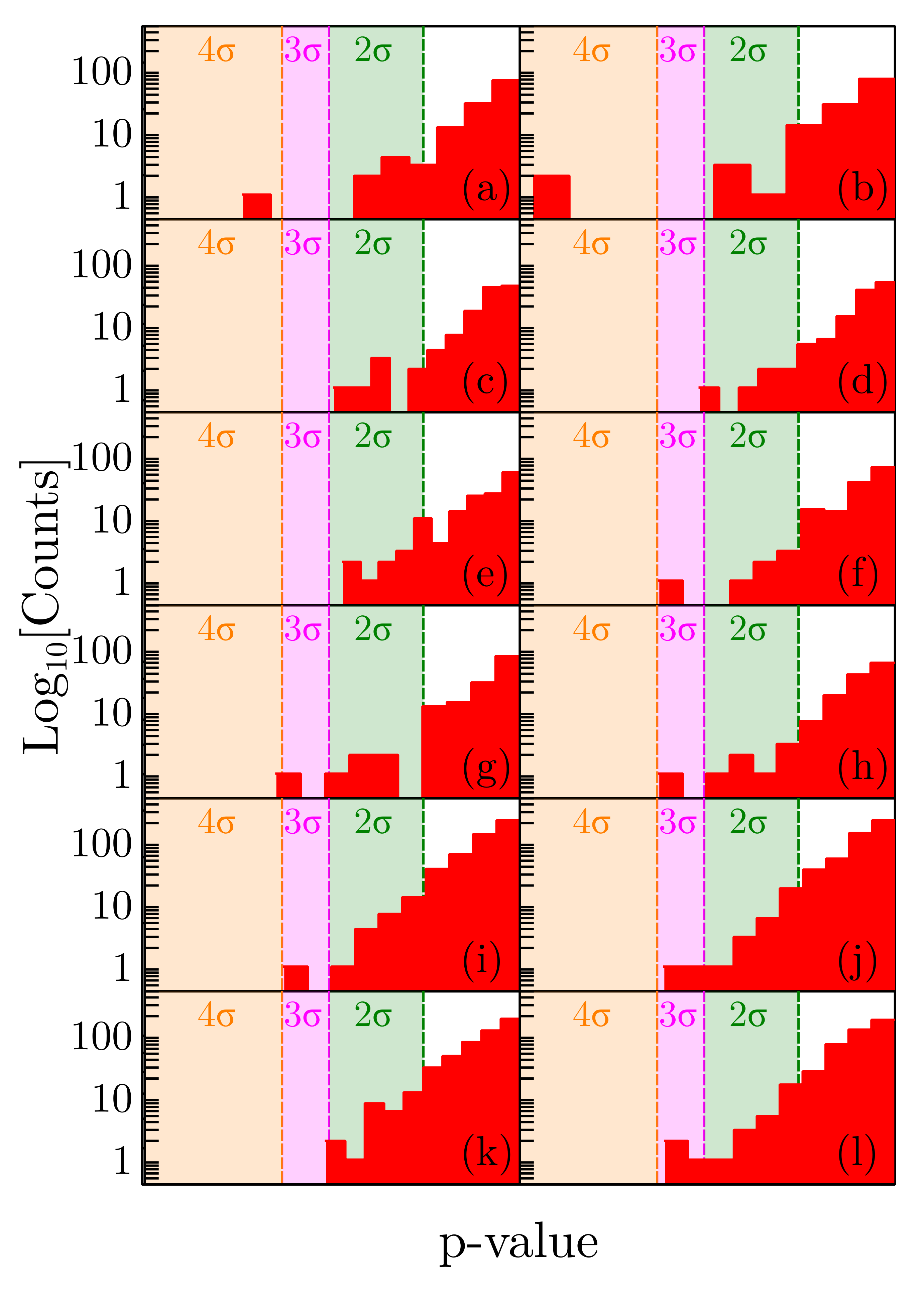}
    \caption{Histograms of the p-value for each GoF test. (a) $p_{KU}$ distribution for exoplanet hosts with a constrained $\omega$ value. (b) Similar to (a), but the $p_{KS}$ distribution. (c) Similar to (b), but for $p_{AD}$. (d) Median $p_{AD}$ value distributions sampling different periastron phases. (e-h) Same as (a-d), but for exoplanet hosts with no constraint on $\omega$. (i-l) Same as (a-d), but for TOI hosts.}
\end{figure}

\begin{figure}
    \centering
    \label{figure: VM_TS}
    \includegraphics{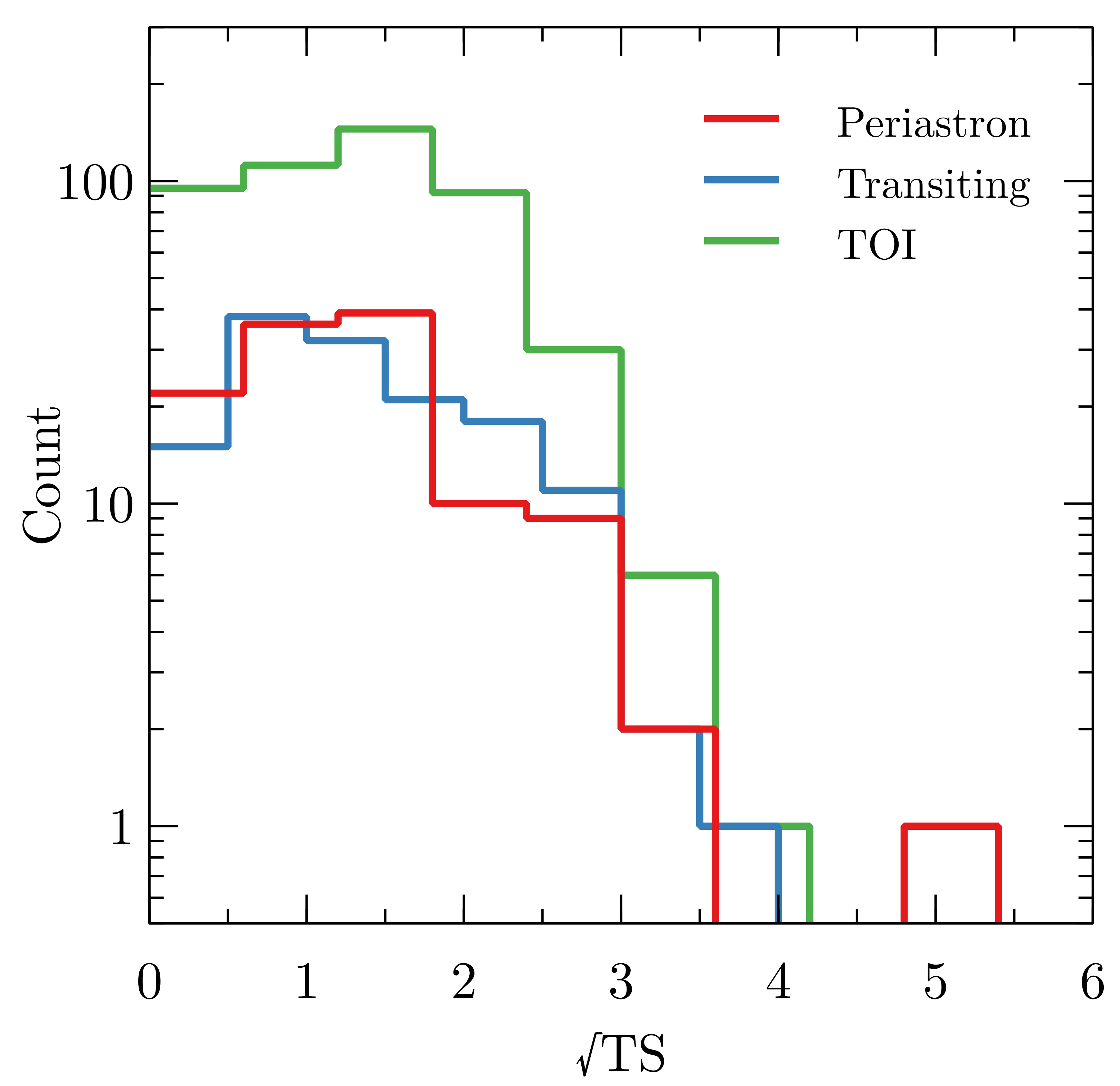}
    \caption{Distribution of the $\sqrt{TS}_{VM}$ (\ref{equation: TS}) values for each of the three sub-samples in this work.}
\end{figure}

\subsubsection{Flare Correlation with Stellar Rotational Period}
\label{subsubsection: Stellar Rotation}
To supplement the available stellar rotational periods available on the NASA Exoplanet Archive, we use the \texttt{SpinSpotter} algorithm described in \citet{Holcomb2022}. The stellar rotation periods are identified by generating the autocorrelation function (ACF) for each set of TESS photometry for each target across all sectors. Initial period guesses are generated by using a Fast Fourier Transform on the ACF and fitting a series of $k$ parabolas on each peak in the ACF. "Rotators" and "non-rotators" are delineated through an acceptance function requiring $R^{2}_{\mathrm{avg}}\geq0.9$, $0.4\leq B_{\mathrm{avg}}\leq0.5$, and $A_{\mathrm{avg}}/B_{\mathrm{avg}}>0.15$, where $R^{2}_{\mathrm{avg}}$ is the average correlation coefficient of each fit, $B_{\mathrm{avg}}$ is the average width of each parabolic fit, and $A_{\mathrm{avg}}$ is the average height of each parabola. Due to the TESS cadence, only $P_{\star}\leq14$ are considered. This limitation is described in depth in \citet{Holcomb2022}. A transit mask is applied for each system with transiting planets to avoid confusion in the ACF period. We construct stellar rotational period catalogs for the known exoplanet host sample as well as all TOI planet candidate hosts.

\begin{figure*}
    \centering
    \label{figure: St Rot Compare}
    \includegraphics[width=\textwidth]{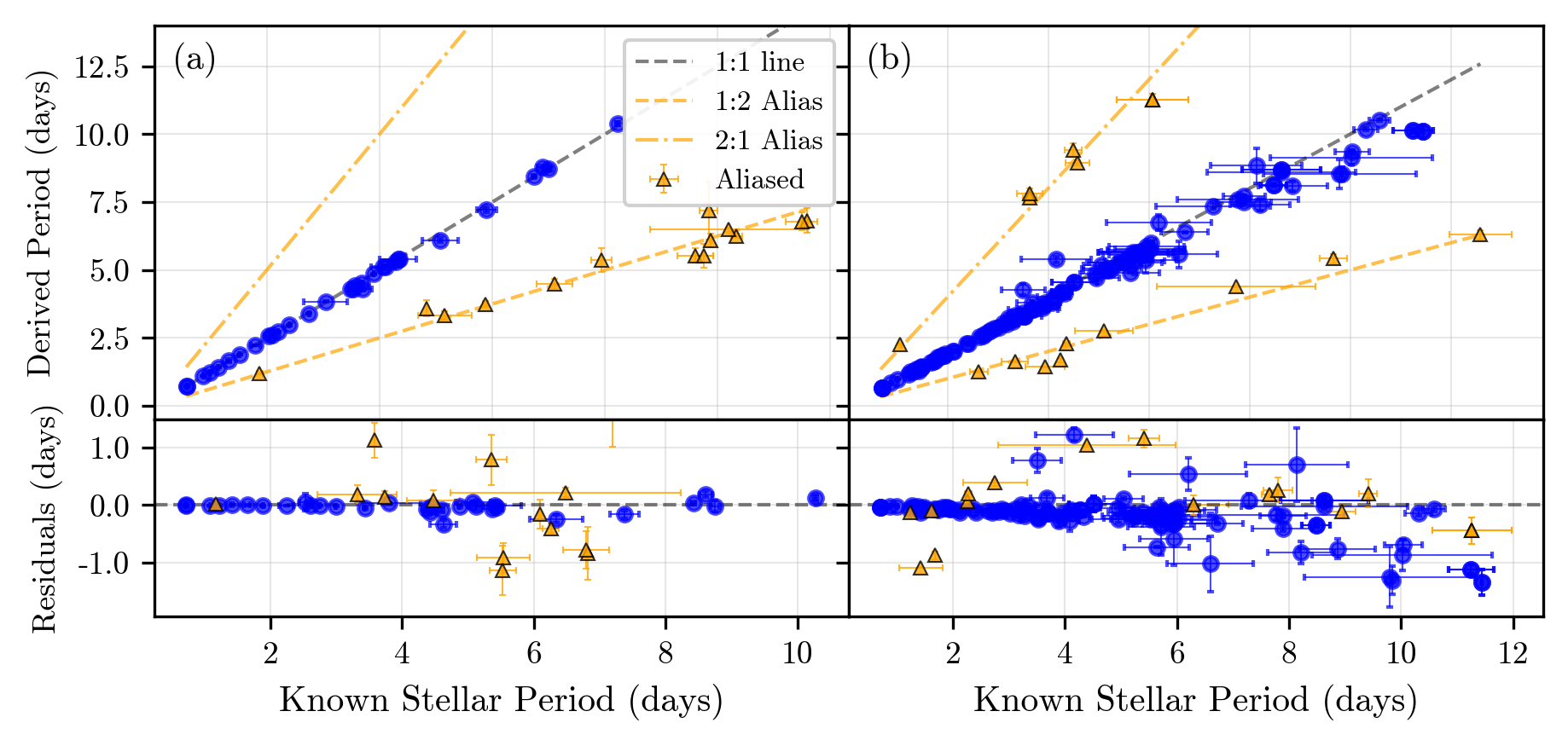}
    \caption{Comparison of stellar rotation periods of exoplanet hosts and TOI PC hosts derived using \texttt{SpinSpotter} and those found on the NASA Exoplanet Archive and \citet{martins2020}, respectively. (a) Derived stellar periods compared to known stellar periods of exoplanet hosts. Orange lines/markers denote aliased targets. Aliasing is corrected in the residual plot. (b) Identical to (a), but for TOI PC hosts.}
\end{figure*}

We identified 174 stellar rotational periods of exoplanet hosts consistent with the above acceptance function, 121 of which do not have a stellar rotation period entry in the NASA Exoplanet Archive. We repeated the procedure for TOI planet candidate hosts of the stars analyzed in our flare sample, retrieving 667 periods. We compared the results with those of the previous analysis by \citet{martins2020}, who maintained a TOI rotation catalog through April 2022 and identified 202 targets with "unambiguous rotation." We can recover 134 of these detections, noting that TOI planet candidate hosts that were designated as false positives have not been analyzed. The comparison between our findings and the values in the literature is summarized in Figure~\ref{figure: St Rot Compare}.

Aliasing can occur in the ACF when two spots are present on opposite hemispheres of the star. Across both samples, we identify 35 aliased targets, shown in Figure~\ref{figure: St Rot Compare} in orange. Aliased targets exhibit greater scatter and error estimation, likely due to differential rotation, as the two spots contribute slightly different frequencies depending on stellar latitude. We report the first few rows of the derived stellar rotation periods in Table~\ref{table: St Periods}.

\begin{figure}[!ht]
    \centering
    \label{figure: Syn_Per_p_values}
    \includegraphics[width=0.47\textwidth]{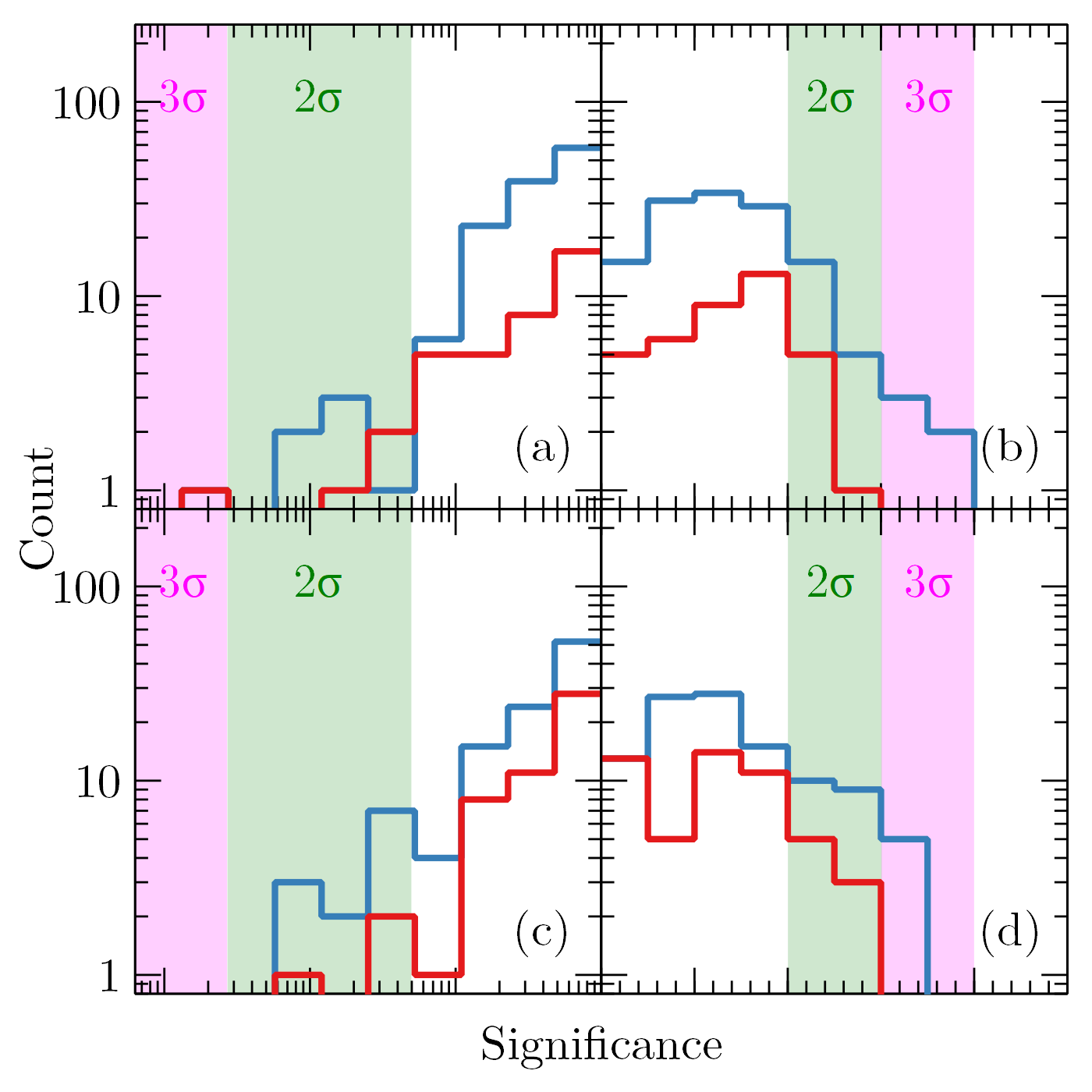}
    \caption{Histograms showing significance metrics, $p_{KU}$ and $\sqrt{TS}_{VM}$, for flare phase correlation with $P_{\star,\text{rot}}$ and $P_{\text{Syn.}}$. The green region represents detections within the interval of $3\sigma<x<2\sigma$, and the magenta region $3\sigma<x<2\sigma$ for either the p-value distribution for the Kuiper test or the $\sqrt{TS}_{VM}$ for UBL analysis. (a) Distribution of p-values from the Kuiper test. (b) Distribution of the $\sqrt{TS}_{VM}$ from the UBL analysis. (c-d) Identical to (a-b), but for $P_{\text{Syn}}$.}
\end{figure}

We investigate correlations between stellar rotational periods and flare epochs by applying both the Kuiper goodness-of-fit test and the unbinned likelihood analysis presented in Section~\ref{section: Analysis Methods}. Sixty-six exoplanet hosts and 123 TOI hosts had more than three flares identified. We only consider $P_{\star,\mathrm{rot}}<30\,\mathrm{days}$ to avoid correlations arising from insufficient sampling. We assume an arbitrary epoch to phase fold the flares, $t_{epoch}=2457000\,\mathrm{BJD}$. Where possible, we defer to the literature values over derived values. The distribution of p-values from the Kuiper test and the $\sqrt{TS}_{VM}$ for the UBL analysis for both samples is shown in Figure~\ref{figure: Syn_Per_p_values}.

\begin{deluxetable}{c|ccccc}

\tabletypesize{\footnotesize}

\tablecaption{First and last two rows of the stellar period catalog.}

\tablehead{\vspace{-0.2cm} & $P_{\star,rot}$& $B_{\mathrm{avg}}$&  & \\
\vspace{-0.2cm} Sys. ID & & & $A_{\mathrm{avg}}$&$R^{2}_{\mathrm{avg}}$& Samp.\\
& (Days) & (Lag) &&}
\startdata 
AU Mic & $4.87\pm0.02$ & 0.869 & 0.036 & 0.977  & CPH\\
CoRoT-18 & $5.41\pm0.03$ & 0.456 & 0.274 & 0.982  & CPH\\
\vdots & \vdots & \vdots & \vdots & \vdots &  \vdots\\
1007 & $1.712\pm0.003$ & 0.446 & 0.259 & 0.973 & PCH\\
1009 & $0.97\pm0.03$ & 0.467 & 0.198 & 0.993 & PCH\\
\enddata
\tablecomments{The units on $B_\mathrm{avg}$ are in units of the ACF lag. Confirmed planet host (CPH) and TOI PC hosts (PCH) are denoted in the 'Samp.' column.}
\label{table: St Periods}
\end{deluxetable}
\begin{deluxetable}{c|ccccc}

\tabletypesize{\footnotesize}

\tablecaption{Phase correlations between flare phase and stellar rotational period.}

\tablehead{Sys. ID & $N$ & $p_{\mathrm{KU}}$ & $\sqrt{TS}_{VM}$  & $\overline{\Delta\log{Z}}$ & $P_{\star,\mathrm{rot}}$ (d)}
\startdata 
TOI-1063  & 30 & 0.002 & 3.59 & 7.9   & 7.485  \\
TOI-168   & 5  & 0.043 & 3.52 & 12.2  & 6.442  \\
TOI-936   & 40 & 0.033 & 3.35 & 209.4 & 4.015  \\
TOI-4180  & 14 & 0.050 & 3.18 & 16.0  & 3.168  \\
\enddata
\tablecomments{ $\overline{\Delta\log{Z}}$  quantifies the average Bayesian evidence of the flares in the sample described in Section~\ref{subsection: Tier 3}.}
\label{table: St_Per_K}
\end{deluxetable}
We identify six hosts with a detection significance above $3\sigma$ for either detection metric. Notably, all but one targets have $N_{\mathrm{flare}}>10$. We summarize significant detections in Table~\ref{table: St_Per_K}, and discuss individual targets in Section~\ref{section: Discussion}.

\subsubsection{Flare Correlation with Stellar-Planetary Synodic Period}
\label{subsubsection: Synodic Period}

Lastly, we investigate the correlation between flares and the synodic period of close-orbiting planets and the stellar rotational period. The synodic period is defined as:
\begin{equation}
\frac{1}{P_{syn}}=\left|\frac{1}{P_1}-\frac{1}{P_{2}}\right|; \quad P_{1}<P_{2}
\end{equation}
corresponding to the period of one object with respect to the rotating frame of the other. Induced flares occurring between the orbiting planet and a structure on the stellar surface, such as a coronal streamer or hot spot, have the highest power at this period. We search for flare clustering with respect to the synodic period for exoplanet hosts and TOI planet candidate hosts with constrained stellar rotation periods and $P_{\text{syn}}<30\,\text{d}$, chosen to ensure sufficient sampling with targets with at least one TESS sector available. We compile the results into a catalog, with the first entries shown in Table~\ref{table: Syn_Per_K} and the statistical test results in Figure~\ref{figure: Syn_Per_p_values}.

\begin{deluxetable}{c|cccccc}

\tabletypesize{\footnotesize}

\tablecaption{Significant detections in phase correlations between flare phase and $P_{\text{syn}}$.}

\tablehead{Sys. ID & $P_{\text{syn}}$ (d) & $N$ & $p_{\mathrm{KU}}$ & $\sqrt{TS}_{VM}$  & $\overline{\Delta\log{Z}}$ & Sample}
\startdata 
TOI-4501.01 & 2.90 & 5 & 0.006 & 3.49 & 12.3 & PC\\
TOI-5078.01 & 2.07 & 6 & 0.011 & 3.28 & 6.4 & PC\\
TOI-4131.01 & 10.48 & 9 & 0.011 & 3.26 & 19.8 & PC\\
TOI-724.02 & 7.90 & 10 & 0.031 & 3.10 & 13.6 & PC\\
\enddata
\tablecomments{CP and PC stand for confirmed planet and planet candidate, respectively.}
\label{table: Syn_Per_K}
\end{deluxetable}

\subsection{False-Positives}
\label{subsection: False-Positives}
While \textsc{ardor} performs well in the PR space, false positives are an inevitability of any classification problem. Here, we focus entirely on the output of Tier 3, as it handles most cases that would otherwise be misclassified by Tier 2. 

The most frequent false positives encountered are those related to the ingress/egress of deep transits. These are flagged as flares for two reasons. First, the variance and median of the photometry are computed based on a local window. Deep transits will bias the median significantly such that a sharp ingress/egress can mimic the rising action of a flare, which will deviate several standard deviations above the transit-biased median. Second, the baseline model computed by \textsc{allesfitter} yields low evidence when fitting the ingress/egress returns; thus, the flare model is predisposed to return higher evidence than the baseline, even if neither model fits the light curve well. This could be solved by defining ingress/egress windows, as well as modeling a linear baseline, and similarly with mid-transit windows.  In the final flare catalog, we visually vet flares correlating strongly with predicted ingress/egress by identifying repeated flares at similar phases near ingress/egress.

Infrequently, asteroids will pass through \textsc{ardor}'s vetting process, such as the event during Sector 61 on Kepler-13 shown in Figure~\ref{figure: Asteroid}. Visual vetting identifies these false positives immediately since they are symmetric and do not decay exponentially. Additionally, they typically have high amplitudes and long durations, leading to anomalously high EDs. \textsc{ardor} does not explicitly vet for asteroids, but the procedure described in \citet{Gunther2020} could be implemented.
\begin{figure}
    \centering
    \label{figure: False_Pos_WASP_88}
    \includegraphics[width=0.47\textwidth]{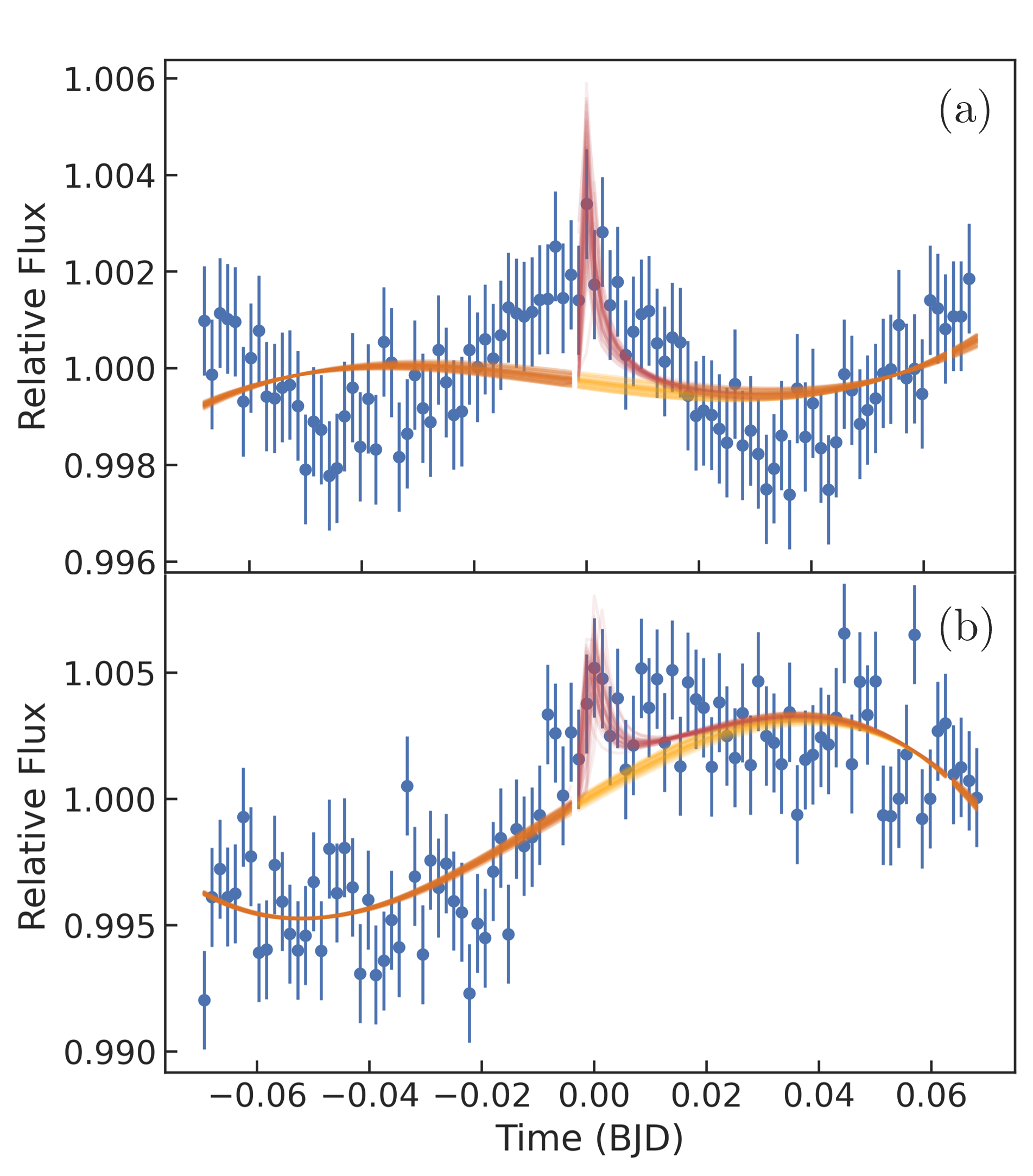}
    \caption{False-positive signals described in \ref{subsection: False-Positives}. (a) A detected 'flare' from TOI-2001. The high-frequency variability created many false positives for this target that were not properly detrended. The Bayes factor was $\Delta\log{Z}=3.95$ in favor of the flare model. (b) A detected 'flare' from WASP-88. There appears to be a minor spike during the transit egress that was erroneously identified as a flare due to the decrease in the median flux of the local window due to the deep transit. The Bayes factor was $\Delta\log{Z}=6.01$ in favor of the flare model.}
\end{figure}
\begin{figure}
    \centering
    \label{figure: Asteroid}
    \includegraphics[width=0.47\textwidth]{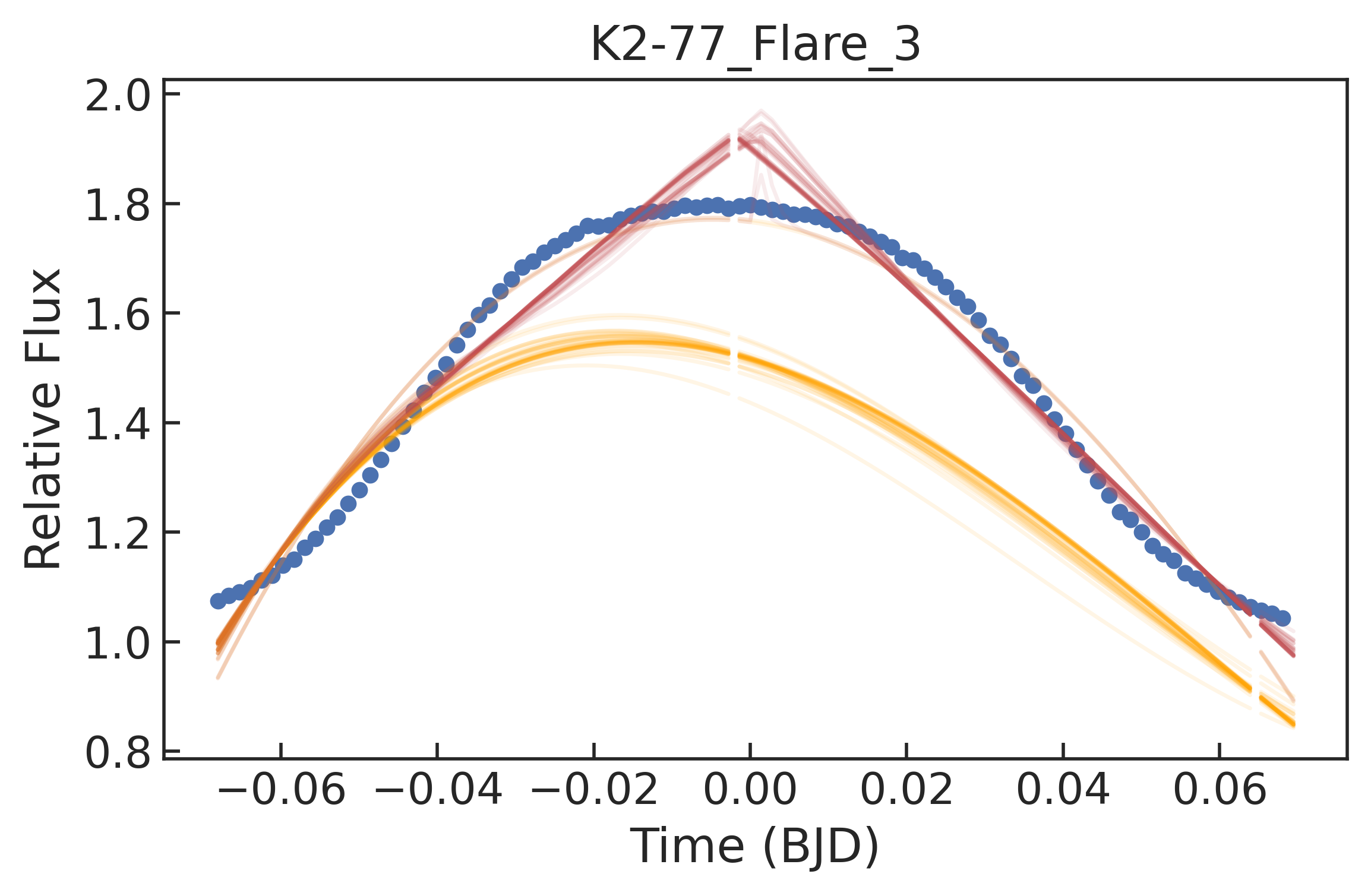}
    \caption{False-positive signals described in \ref{subsection: False-Positives}. (a) A detected 'flare' from TOI-2001. The high-frequency variability created many false positives for this target that were not properly detrended. The Bayes factor was $\Delta\log{Z}=3.95$ in favor of the flare model. (b) A detected 'flare' from WASP-88. There appears to be a minor spike during the transit egress that was erroneously identified as a flare due to the decrease in the median flux of the local window during the deep transit. The Bayes factor was $\Delta\log{Z}=6.01$ in favor of the flare model.}
\end{figure}

\section{Discussion}
\label{section: Discussion}
The systems identified in Section~\ref{section: Results} fall into three groups for flares correlated with $P_{\text{orb}}$: eccentric systems with flaring consistent with periastron-dependent SPIs ($\beta_{\text{SPI}}>0.2$); geometrically agnostic, statistically significant flare clustering; and flare clustering in TOI planet candidates. We discuss the signal identified for each target and contextualize its potential to be driven by magnetic SPIs or a different astrophysical source. Discussion is not exhaustive for each target presented in Section~\ref{table: results}, as we focus on the most promising candidates. 

We identify one high-priority target, TOI-1062\,b, exhibiting significant flare clustering at periastron and our best candidate for induced flares. Additionally, we discuss Gl 49\,b and HD 163607\,b, both exhibiting moderately significant flare clustering in addition to a large $\beta_{\text{SPI}}$. We then discuss AU Mic\,b, the target of a similar correlated flare study by \citet{Ilin2022}, and find a $3\sigma$ detection with no clustering at periastron. Additionally, we discuss significant flare correlations with $P_{\text{syn}}$ and $P_{\text{rot},\star}$, and conclude by considering salient follow-up observations as well as the role future survey missions, like ULTRASAT, serve in confirming and identifying induced flares.

\subsection{Induced Flare Candidates}
\label{subsection: Induced Flare Candidates}
This section highlights three close-in, eccentric systems, TOI-1062\,b, Gl 49\,b, and HD 163607\,b, which exhibit significant flare clustering around periastron. We discuss the plausibility of magnetic SPIs in the context of system parameters, 
\subsubsection{TOI-1062 b}
\label{subsubsection: TOI-1062 b}
TOI-1062\,b shows highly clustered flares, with detection metrics of $p_{KU}=2.2\times10^{-5}$, $\sqrt{TS}_{VM}=5.1$, and $\beta_{\text{SPI}}=0.63$, all the most significant metrics found in the sample. TOI-1062\,b is a super-Earth with $R_{P}=2.265\,R_{\oplus}$ and $M_{P}=10.15\,M_{\oplus}$, with a Sun-like G-type host star \citep{Otegi2021}. Periastron at $0.043\,\mathrm{AU}$ ($\sim9\,R_{\star}$) constitutes a high-likelihood sub-Alfvénic periastron. The flare epochs, when phase-folded to the stellar rotation and synodic period, are not significant.

\footnote[1]{$P_{\star}:\sqrt{TS}_{VM}=0.89\,,P_{syn}:\sqrt{TS}_{VM}=0$}

The flares and the orbital configuration of the system are shown in plot (a) of Figure~\ref{fig: Peri_Polar_Plots}. Six flares were detected with an average detection evidence of $\overline{\Delta\log{Z}}=5.1$. The maximum flare amplitudes recorded up to $0.03\%$, consistent with low ED flares. The flares cluster around periastron, with $\Delta\phi=|\omega-\nu|=2^{\circ}$. With an estimated Alfvén surface of $0.062^{+0.2}_{-0.05}\,\mathrm{AU}$, Monte-Carlo sampling suggests a $\sim70\%$ of sub-Alfvénic periastron approach. This makes TOI-1062\,b our strongest candidate for induced flares. The upper limit on equatorial magnetic field strengths of a $10\, M_{\oplus}$ planet by the scaling laws of \citet{Christensen2010} is around $\sim10\,\text{G}$. Using the energy and time estimation in \ref{equation: Field Energy Difference}, induced flare interactions are predicted to produce flares with $t_{dur}=80\text{ s}$ and $E_{\text{flare}}=6.7\times10^{29}\text{ erg}$. While the predicted time scale is consistent with the low ED flares observed, the energy is underestimated by four orders of magnitude when following the bolometric flare energy prescription in equation~\ref{equation: flare Lumin}. Additionally, $\omega_{p}=297^{\circ+27}_{-19}$  places the line of sight of the interaction within $20^{\circ}$ of secondary eclipse. If SPIs drive the flares, this indicates that the interaction region capable of inducing flares extends over many stellar radii. We suggest follow-up observations to search for alternative magnetic SPI mechanisms, such as phase-dependent coherent radio emissions or the observation of the calcium triplet, as indicators of phase-dependent chromospheric activity, like those observed by \citet{pineda2023} and \citet{Shkolnik2008}. We also propose using ZDI to constrain the average equatorial magnetic field strength and to characterize the large-scale stellar field, which can be used as a prior in MHD simulations of the system.

\subsubsection{Gl 49 b}
\label{subsubsection: Gl 49 b}
Gliese 49 is an M dwarf ($T_{eff}=3805\pm51\,\mathrm{K}$ with an eccentric ($e=0.36\pm0.1$), likely sub-Alfvénic hot Neptune companion with $r_{p}=0.058\,\mathrm{AU}$. Gliese 49\,b has only been detected using RV methods and has no constraint on inclination. The 14 flares detected from this host display moderate clustering, showing marginal statistical significance with $\sqrt{TS}_{VM}=2.5$ and $p_{\text{KU}}=0.044$. The flare locations prefer periastron, with $\beta_{\mathrm{SPI}}=0.43$, which can be seen in subplot (b) in Figure~\ref{fig: Beta_Polar_Plots}. Gl 49\,b does not flare frequently, and an additional $\sim650\,\text{d}$ of photometric baseline would be required to claim significance. Flaring activity, coupled with strong stellar activity and stellar class, suggests strong stellar field strengths that could resist perturbations. The planet's high eccentricity could lend credence to highly local SPIs, such as the 'bend-and-snap' mechanism that couples to coronal streamers, as the driving period would be the orbital period. Similar to TOI-1062, ZDI imaging should be used to probe the large-scale magnetic field polarity and strength. Future TESS photometry should also be analyzed for additional flares.

\subsubsection{HD 163607 b}
\label{subsubsection: HD 163607 b}
One of the most eccentric planets ($e=0.744\pm0.007)$ ) discovered, HD 163607\,b is the longest period target in our sample with $P_{orb}=75.2203\pm0.0094\,\text{d}$ \citep{Luhn2019}. With a likely sub-Alfvénic periastron approach of $r_{p}=0.09\,\text{AU}$, HD 163607\,b is a Jupiter-sized gas giant orbiting a main-sequence G-type dwarf, making it an ideal candidate to search for periastron-dependent induced flares. We identify 21 flares with non-significant metrics $\sqrt{TS}=1.62$, $p_{KU}=0.28$, but with $\beta{_\text{SPI}}0.42$ and clustering center close to periastron with $\Delta\phi=28^{\circ}$. Despite its long orbital period, the phase space is well sampled since the star has been observed over 37 TESS sectors. The required $t_{3\sigma}=1466\,\mathrm{d}$ will likely be achieved since the target is close to TESS's continuous viewing zone. In addition to continued photometric monitoring, this host is ideal for spectroscopic follow-up using stellar activity indicators, since the system architecture and host characteristics are ideal for periastron-dependent induced flares \citep{Shkolnik2008}.

\subsection{Correlated Flares with Planetary Orbital Period}
\label{subsection: Planetary Orbital Period Correlation}
Two low eccentricity targets, AU Mic\,b and TOI-1807\,b, show statistically significant flare clustering ($>3\sigma$). These targets are unlikely to be candidates for periastron-dependent induced flares. Due to interest in AU Mic\,b as an induced flare candidate, we discuss our results of AU Mic\,b within the broader context of SPI studies, and analyze how differences in methodology impact our ultimate conclusion. TOI-1807\,b has an eccentricity consistent with 0; however, its designation as an ultra-hot super Earth allows for close-in interactions of coronal structures, which we discuss briefly.

\subsubsection{AU Mic b}
\label{subsubsection: AU Mic b}

AU Mic is a young and magnetically active M-dwarf that hosts three close-in planets, with a fourth planet postulated from transit-timing variation analysis \citep{Plavchan2020, Martioli2021, Wittrock2023}. AU Mic's magnetogram has been mapped using Zeeman-Doppler imaging, revealing large field strengths of $2-3\, \mathrm{kG}$. This magnetogram has been used as a prior in magnetohydrodynamical simulations, showing AU Mic\,b residing primarily in the sub-Alfvénic regime. AU Mic\,b is a Neptune-like planet with $R_{p}=3.96\,R_{\oplus}$ and $M_{p}=20.12\,M_{\oplus}$. While not a hot Jupiter, it is reasonable that a planet with this mass could generate a strong planetary field. 

The detection metrics, at $p_{KU}=6.9\times10^{-3}$ and $\sqrt{TS}_{VM}=3.1$ ($N=153$), suggest non-uniform flaring, which can be seen in figure (b) of Figure \ref{fig: Peri_Polar_Plots}. AU Mic\,b is only slightly eccentric, with $e=0.006$, with the strongest signal originating $\sim50^{\circ}$ from periastron. The distribution of flares does not appear unimodal, either, deviating from the expectation that induced flares originate near periastron with an eccentric orbit. Additionally, AU Mic's strong magnetic field reduces its susceptibility to perturbation, assuming induced flares depend on $(B_r)^{\alpha}$, where $\alpha>0$. This observation does not provide overwhelming support for our instantaneous separation model, but does not preclude other induced flare mechanisms, such as those described in Section 2.1 in \citet{Lanza2018}. Another induced flare study conducted in \citet{Ilin2022} found a potential detection in the high energy flare sample ($ED>1\, \mathrm{s},\, N=77$) with a custom AD test statistic of $p\sim0.07$, with the differences in results stemming from methodology (discussed thoroughly in Section~\ref{subsection: Impacts Methodology}). 

The stellar rotation has been believed to be a source of phase-modulated flares. The idea that systems with active hot spots exhibit an increase in flare rate when the hot spot is in our line of sight is physically plausible. However, this flare-rotation correlation has not been identified in other photometric flare studies \citep{Doyle2018, Doyle2020}. Indeed, when our data is phase folded with respect to both the equatorial rotation period ($T_{\mathrm{eq}}=4.84\,\mathrm{days}$) and the polar rotational period ($T_{\mathrm{eq}}=5.1\,\mathrm{days}$) we find no significant detections across any metric for either period ($P_{\star}:\sqrt{TS}_{VM}=0.82\,,P_{syn}:\sqrt{TS}_{VM}=0.63$) \citep{Klein2021}.

The synodic period between an active hot spot and the stellar surface, along with the orbital period, is an additional source of phase-correlated flaring \citep{Castro2024}. A rotating hot spot modulates the light curve of AU Mic. Interactions may occur between extended coronal streamers or loops each time the planet passes the hot spot during its orbit, consistent with the synodic period. In the case of AU Mic\,b and the equatorial period of the stellar surface, the synodic period is:
\begin{equation}
\label{eq: Synodic Period}
T_{\mathrm{synodic}}=T_{eq}^{-1}-T^{-1}_{pl}=11.29\, \mathrm{Days}.
\end{equation}
Similar to the stellar rotational period, this does not result in a significant signal. 
\subsubsection{TOI-1807 b}
\label{subsubsection: TOI-1807 b}
The only non-eccentric target discussed, TOI-1807 is a young K-type ($T_{eff}=4914\,\mathrm{K}$) star slightly smaller than the Sun with $M_{\star}=0.8 \, M_{\odot}$ and $R_{\star}=0.75\,R_{\odot}$ \citep{Paegert2021}. It hosts a hot, ultra-short period super-Earth, TOI-1807\,b, with a semi-major axis of $0.01\,\mathrm{AU}$ \citep{Hedges2021}. This safely places TOI-1807\,b in the sub-Alfvénic regime. Only five flares were found in the available TESS data, but they displayed a significant detection using unbinned likelihood analysis with $\sqrt{TS}_{VM}=3.1$. 

Four of the five flares were found in the first TESS sector, displaying a short-term increase in activity. These flares had amplitudes of a few percent, with $\overline{\Delta\log{Z}}=8.4$, indicating clear flaring events. The temporal clustering of these flares could indicate issues with detrending due to the rapidly evolving, long-term trend in the photometry; however, the detrending methods described in Section~\ref{subsection: Tier 0} appear to effectively eliminate the baseline across all TESS sectors.

Given the close orbital separation of this target, interactions with large-scale stellar structures such as helmet streamers and coronal loops are expected. In this scenario, enhanced flare power would be observed at the synodic period between the stellar rotation and the planet’s orbit, as demonstrated for AU Mic\,b (\ref{subsubsection: AU Mic b}). For this system, the synodic period is $T_{\mathrm{synodic}} = 0.585,\mathrm{days}$. However, when the flare epochs are phase-folded at this period, the signal strength decreases to the level of a non-detection, consistent with a uniform distribution of flares. This indicates that the observed modulation is correlated with the orbital phase of the planet, rather than with recurring interactions associated with fixed stellar surface structures. Although the circular orbit of this planet rules out periastron-driven flare enhancement, the statistical significance of the detected modulation, combined with the extreme stellar and planetary conditions, renders this system a notable case of potential star–planet interaction.

\subsection{Detections around TOI Planet Candidates}
\label{subsection: TOI Detections}
This section outlines all TOI planet candidates with at least one metric indicating a $>3\sigma$ detection: TOI-6276.01, TOI-2323.01, TOI-1232.01, and TOI-723.01. Since these targets are not confirmed planets, there are no constraints on their arguments of periastron or eccentricities. To compensate, we sample different possible orbital configurations with eccentricities limited by the Roche limit, described in Section~\ref{subsubsection: Sampling Orbits}. The significance of the detection and the plausibility of the induced flare based on tentative parameters derived by the TESS SPOC pipeline.

\subsubsection{Sampling Orbits}
\label{subsubsection: Sampling Orbits}
The TOI reports generated by the SPOC pipeline estimate the semi-major axis but do not estimate the eccentricity \citep{Guerrero2021}. We sample the possible orbits for each target-of-interest by computing the most considerable eccentricity allowed, $e_{\mathrm{max}}$, through dynamical interactions in multi-planet systems or the Roche limit \citep{roche1849}. In the case that both are applicable, the smaller $e_{\mathrm{max}}$ is chosen. We compute the Roche limit through the following:
\begin{equation}
\label{equation: Roche Limit}
d\approx2.4R_{\star}\left( \frac{\rho_{\star}}{\rho_{P}}\right)^{1/3}.
\end{equation}
We compute $\rho_{\star}$ by assuming a uniform stellar density and estimate $M_{\star}$ using the mass-radius relationship of $R_{\star}\propto M_{\star}^{0.8}$. We compute $\rho_{P}$ assuming a uniform planetary density and estimate $M_{P}$ using the planetary mass-radius relationships derived in \citet{Muller2024}. In the case of multi-planetary systems, we assume that the orbits do not cross. The results for the targets-of-interest in Figure~\ref{fig: TOI_Polar_Plots} are summarized in Table~\ref{table: TOI Roche}.

\begin{deluxetable}{c|ccccc}

\tabletypesize{\footnotesize}

\tablecaption{$e_{max}$ estimates for TOI targets. \label{table: TOI Roche}}

\tablehead{\vspace{-0.2cm} & a &  & & $r_{max}$ &  \\
\vspace{-0.2cm} TOI PC &  & $e_{min}$ & $e_{max}$&  & Limit \\
& (AU) &  & & (AU) &  }

\startdata 
6276.01 & 0.048 & 0 & 0.25 & 0.058& Dynamic \\
2329.01 & 0.066 & 0 & 0.65 & 0.023&Roche \\
1232.01 & 0.118* & 0 & 0.82 & 0.021&Roche \\
 723.01 & 0.023 & 0 & 0.50 & 0.011&Roche \\
1520.01 & 0.069 & 0 & 0.71 & 0.019&Roche \\
1887.01 & 0.029 & 0 & 0.70 & 0.010&Roche \\
5051.01 & 0.042 & 0 & 0.30 & 0.029&Roche \\
1187.01 & 0.041 & 0 & 0.63 & 0.015 &Roche \\
\enddata

\tablecomments{*This value exceeds the $0.1\mathrm{AU}$ limit in Section~\ref{section: Data}. This is due to the difference in the derived semi-major axis from the SPOC report and the estimate from the period in equation~\ref{equation: TOI Condition}.}
\vspace{-1cm}
\end{deluxetable}

\subsubsection{TOI-6276.01}
\label{subsubsection: TOI-6276.01}
TOI-6276 is a K-type ($T_{eff}=4633.9\, \mathrm{K}$ star with $R_{\star}=0.63\, R_{\odot}$. The one planet candidate found has an orbital period of 6.2 days and a semi-major axis of $a=0.048\,\mathrm{AU}$. This places TOI-6276.01 as likely sub-Alfvénic. Five flares were detected on this target with $\overline{\Delta \log{Z}}=13.5$. The detection metrics of $\sqrt{TS}_{VM}=3.2$ and $p_{AD}=0.0098\times10^{-3}$ make it the most compelling TOI from our survey. The polar flare histogram for this target is shown in (a) of Figure \ref{fig: TOI_Polar_Plots}. 

Given the size of the planet candidate, magnetic SPIs seem unlikely; however, since the parameter space in which planetary fields reside is mainly unknown, this candidate warrants additional monitoring. Contextualizing both the possible planet and the host will be important in determining the plausibility of induced flaring.

\subsubsection{TOI-2329.01}
\label{subsubsection: TOI-2329.01}
TOI-2329 is a G-type ($T_{eff}=5424\, \mathrm{K}$) star slightly larger than the Sun at $R_{\star}=1.33\,R_{\odot}$. It hosts a hot Jupiter candidate with an estimated radius of $ R_p =0.96\, R_J$ and semi-major axis of $a=0.066\, \mathrm{AU}$. Since the host star is larger than our Sun, and the solar Alfvén surface has been seen to extend up to $0.1\, \mathrm{AU}$, it is highly plausible that this potential planet is sub-Alfvénic. Similar to TOI-6276, this host exhibited only seven flares and produced only one significant metric of $\sqrt{TS}_{VM}=3.1$. Individual flare epochs can be seen in (b) of Figure \ref{fig: TOI_Polar_Plots}. The large size of the candidate planet leads to the potential for a strong planetary field, making induced flares a strong possibility. As with TOI-6276.01, follow-up observations of this target are necessary to capture additional flare events and to understand the magnetic environment in which the host star and its potential planet reside.

\subsubsection{TOI-1232.01}
\label{subsubsection: TOI-1232.01}
TOI-1232 is a Sun-like, G star ($T_{eff}=5739\,\mathrm{K}$ with a potential warm Jupiter that orbits $\sim0.12\,\mathrm{AU}$ from its host, placing it on the border of a potential sub-Alfvénic orbit. Multiple detection metrics indicate clustered flaring, including $\sqrt{TS}_{VM}=3.1$, $p_{KU}=0.003$, and $p_{AD}=0.008$. This target exhibited 12 flares with a $\overline{\Delta \log{Z}}=6.1$, with typically low amplitude and short duration flaring (except a single $\sim30\%$ amplitude flare). Similar to TOI-2329, the candidate's large size leads to a greater capacity for induced flares. A follow-up to determine the orbital configuration of this system would help determine if the flares cluster at periastron, which would be compelling evidence for induced flaring.
 
\subsubsection{TOI-723.01}
\label{subsubsection: TOI-723.01}
\begin{figure*}[p]
\label{fig: TOI_Polar_Plots}
\centering
\gridline{\fig{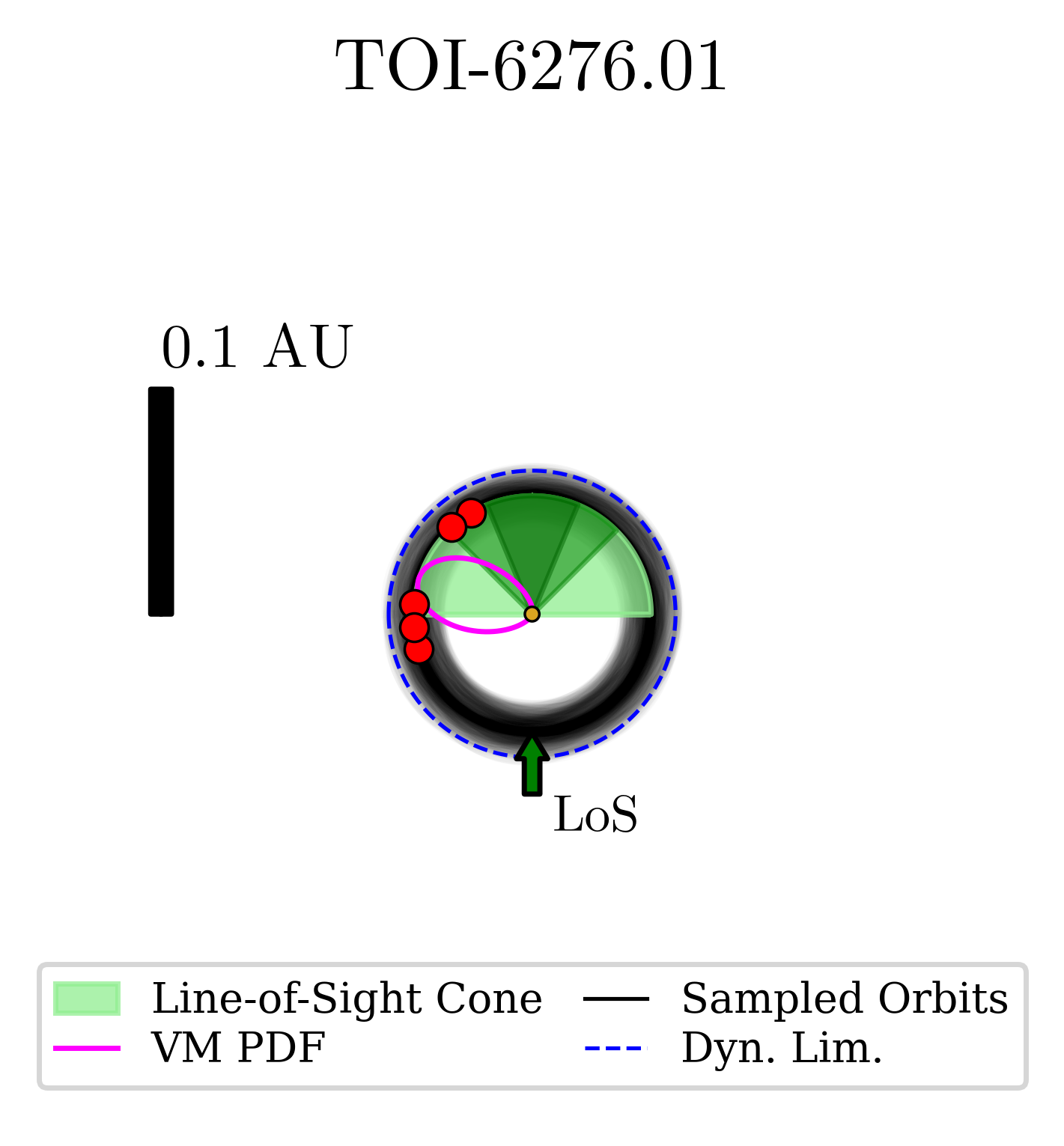}{0.3\textwidth}{(a)}
          \fig{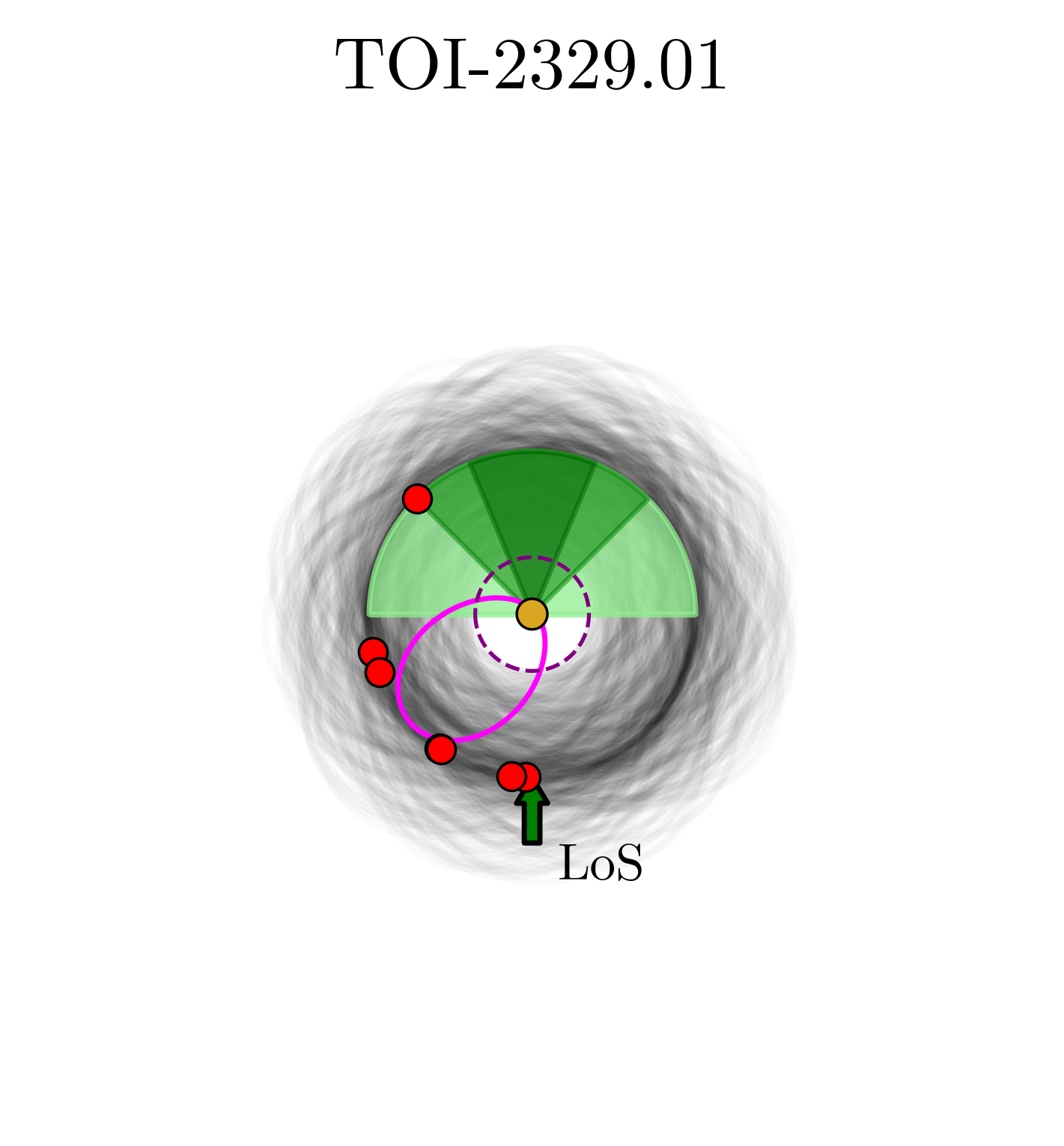}{0.3\textwidth}{(b)}
          \fig{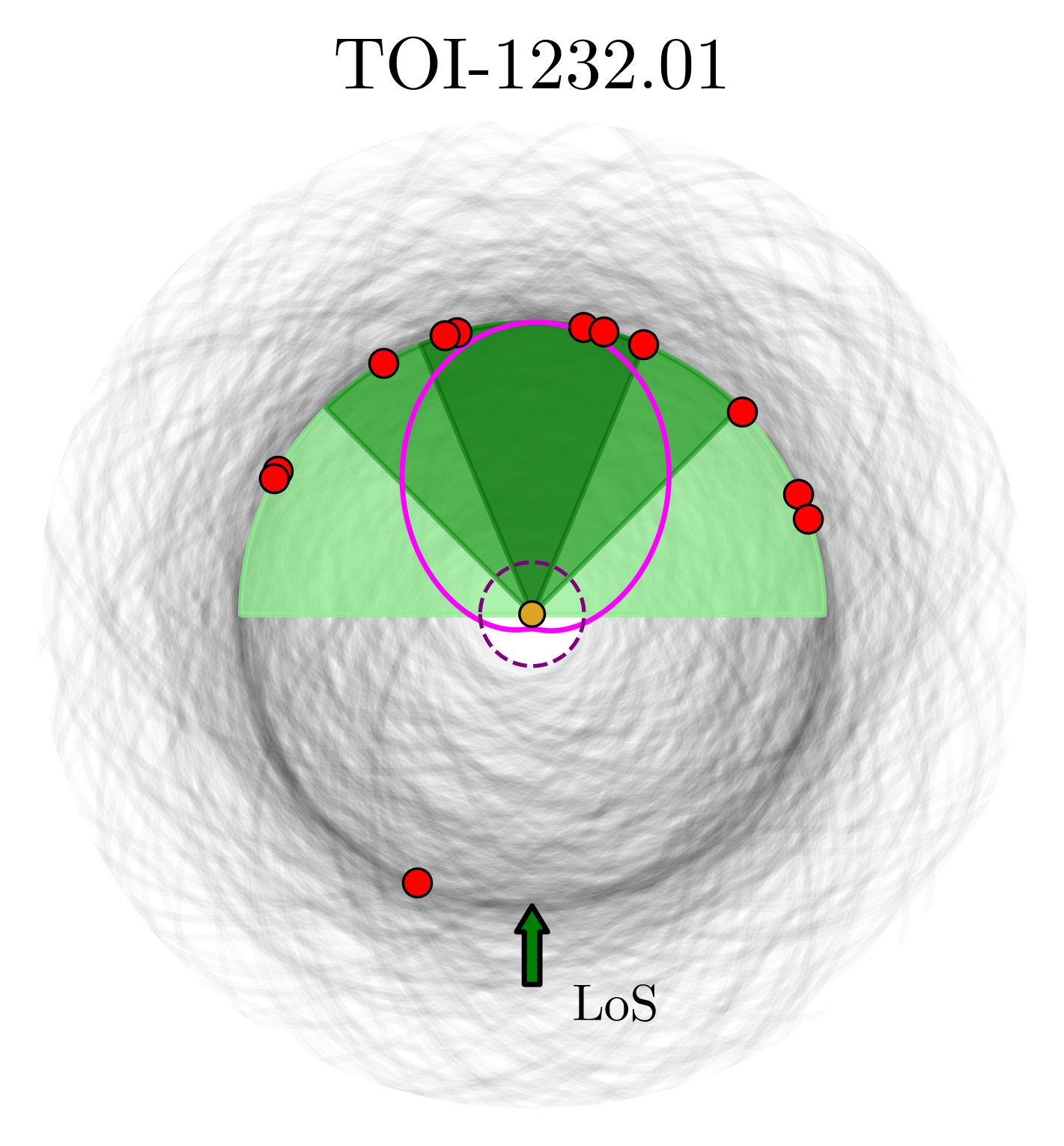}{0.3\textwidth}{(c)}}
\gridline{\fig{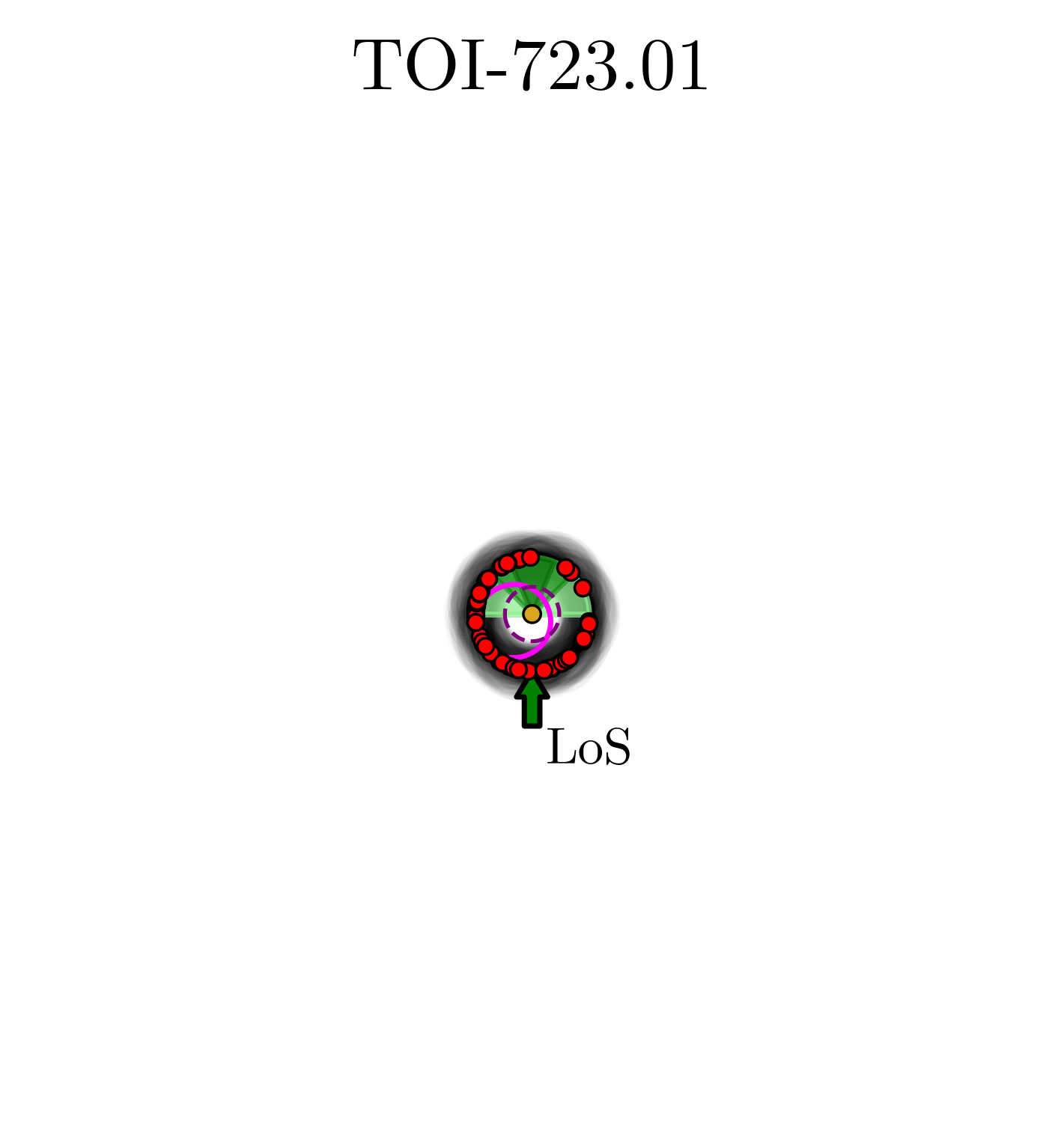}{0.3\textwidth}{(d)}
          \fig{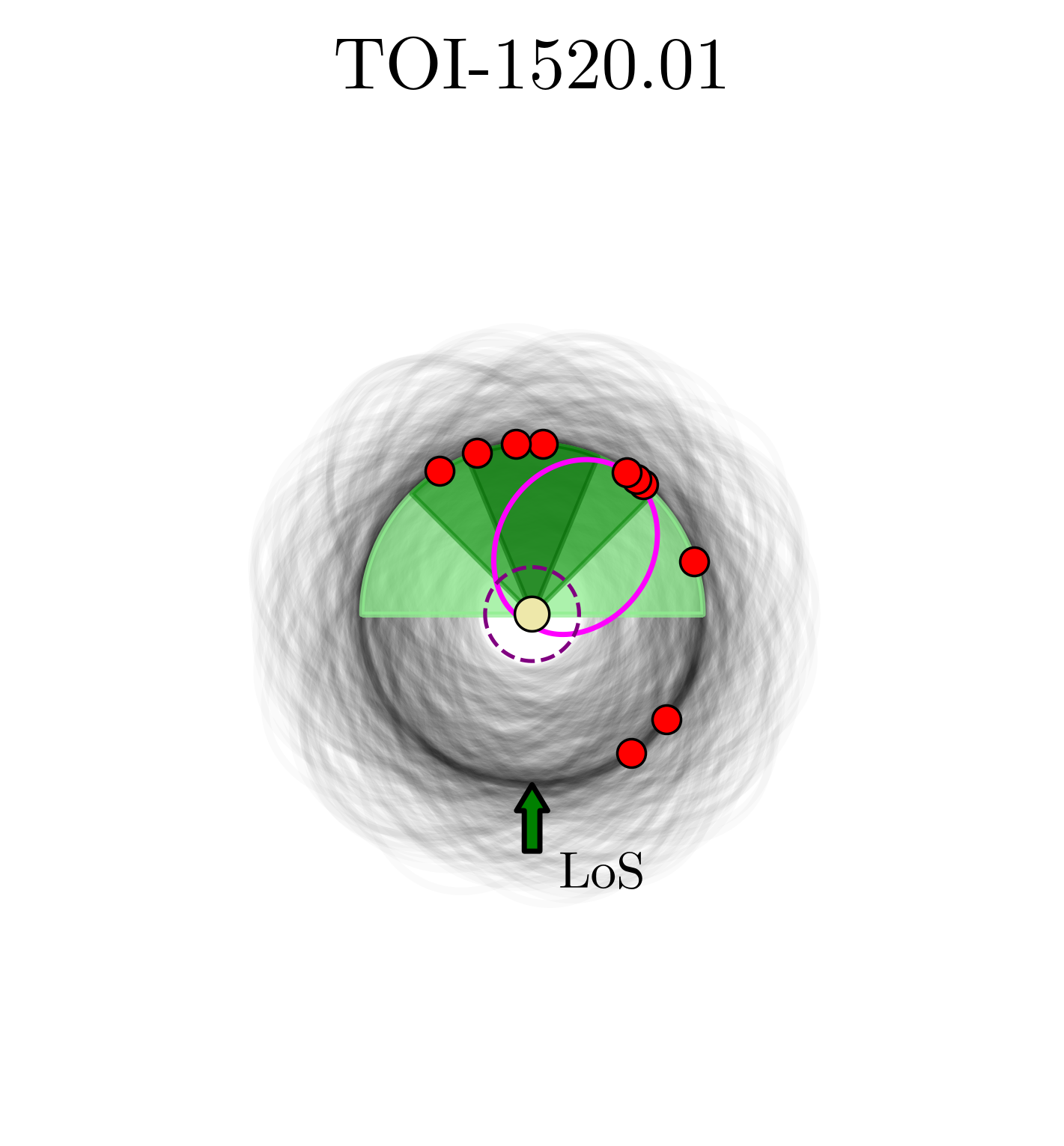}{0.3\textwidth}{(e)}}
\gridline{\fig{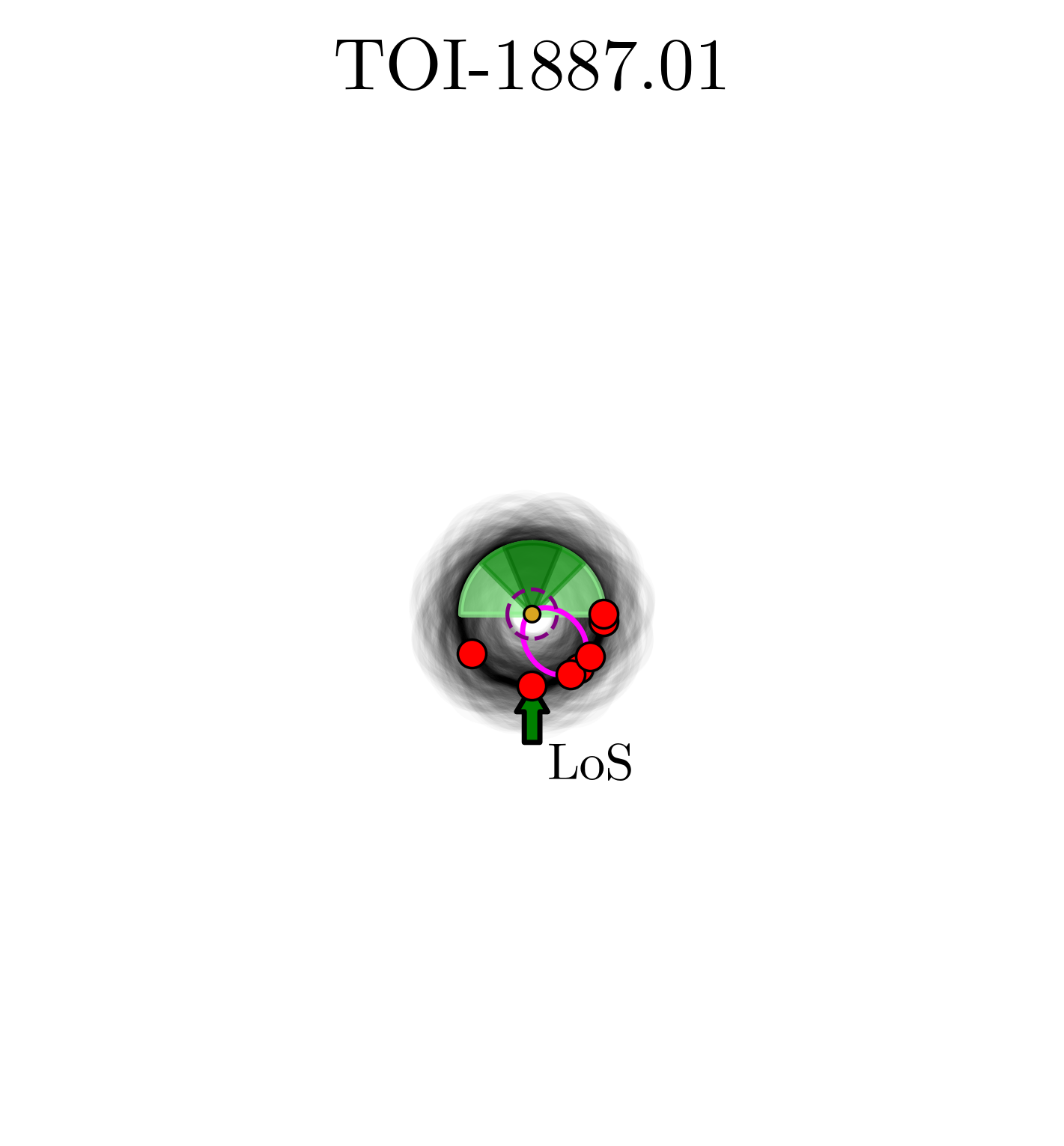}{0.3\textwidth}{(e)}
          \fig{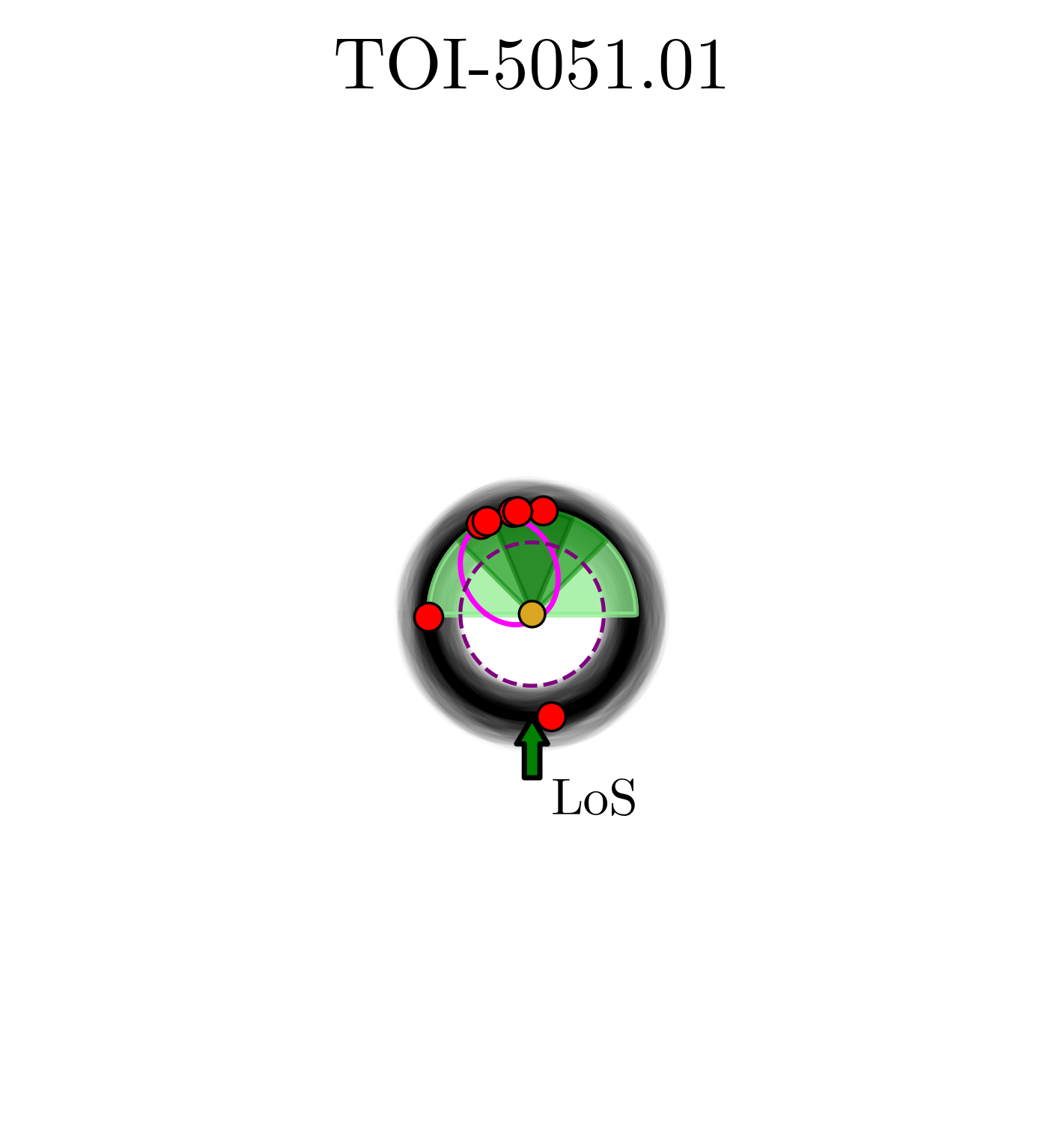}{0.3\textwidth}{(f)}
          \fig{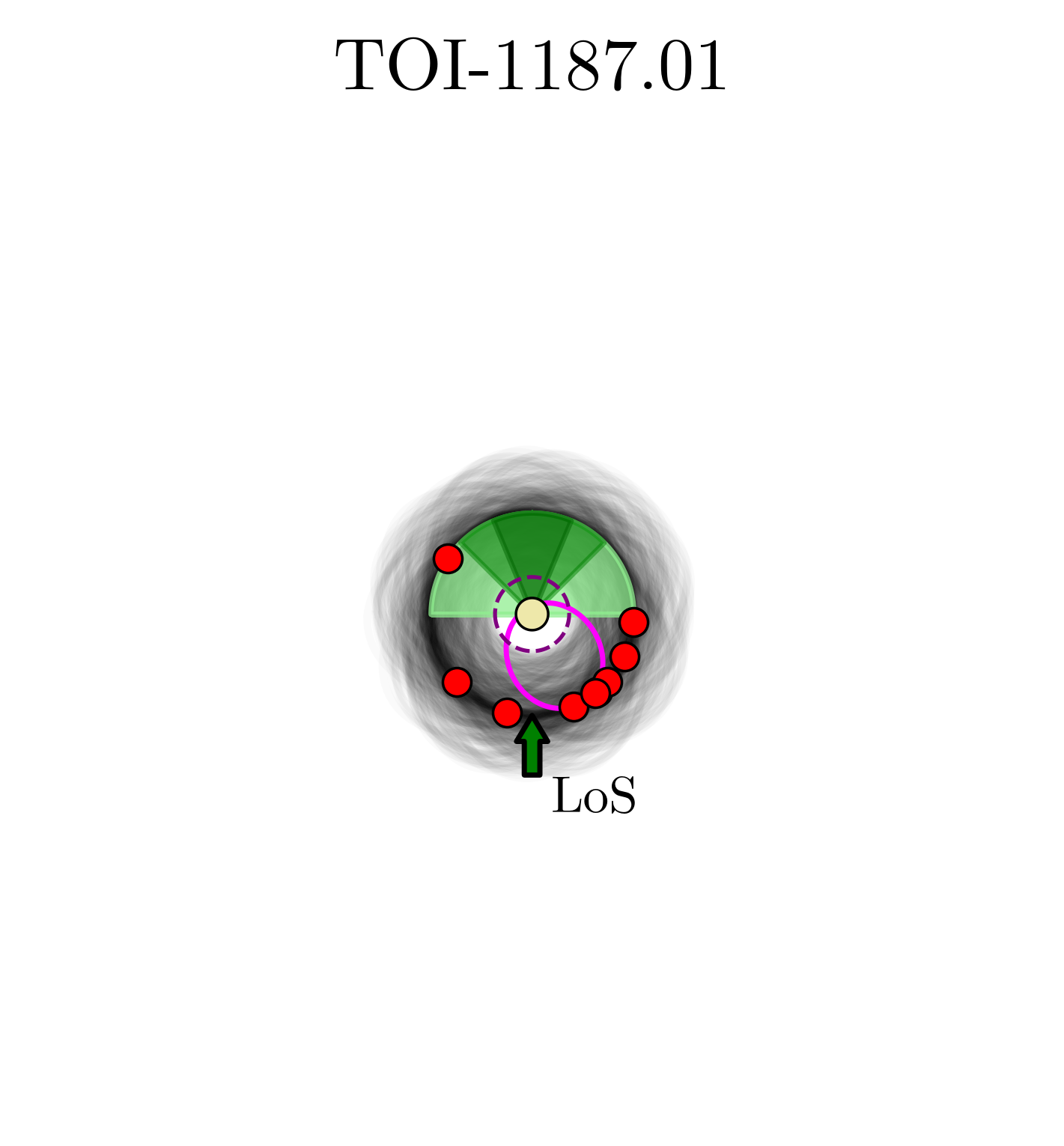}{0.3\textwidth}{(g)}}
\caption{Polar projections of all TOI planet candidate systems of interest. All figures are to scale with the marker in (a) denoting the length of $0.1\,\mathrm{AU}$. The cyan line shows the best-fit von Mises PDF projected onto the orbit. TOI PCs do not have a constraint on eccentricity, so we sample 500 possible orbital configurations (see Section~\ref{subsubsection: Sampling Orbits}). Darker regions correspond to the orbital space that is more likely to be occupied by the planet. The dashed purple lines are the individual flare epochs derived in this work. The shaded green regions represent different possible line-of-sight constraints when the planet is behind the stellar limb, showing $\pm90^{\circ}$, $45^{\circ}$, $22.5^{\circ}$ cones. The green arrow denotes the observational line-of-sight. The estimated  Alfvén surfaces, described in Section~\ref{subsubsection: Alfvén Surface}, are shown in red.}
\end{figure*}

The last target with a significant detection, TOI-723, is a K star ($T_{eff}=4907\,\mathrm{K}$ slightly smaller than our Sun and hosts a close-in Earth-sized candidate with an expected $a=0.022\,\mathrm{AU}$. This is very likely a sub-Alfvénic orbit. The host is active, with 53 flares found in the same. The detections for all metrics are greater than $ 2\sigma$, with $\sqrt{TS}_{VM}=3.0$ being the most significant. Individual flare epochs are shown in (c) of Figure \ref{fig: TOI_Polar_Plots}. Most flares are short-duration and low-amplitude, occurring consistently across all TESS sectors. We suggest follow-up observations to both validate the planet's existence and to illuminate the system's magnetic environment, eccentricity, and additional flare epochs to confirm this detection.

\subsection{Correlated Flares with Synodic Period}
\label{subsection: Synodic Period Correlation}
We discuss two TOI planet candidates, TOI-724.02 and TOI-4131.01, exhibiting significant flare clustering with the synodic period, $P_{syn}$. Similar to the targets discussed in Section~\ref{subsection: TOI Detections}, these are prime targets for validation to constrain orbital parameters and their eccentricities to determine the plausibility of SPIs.
\subsubsection{TOI-724.02}
\label{subsubsection: TOI-724.02}
TOI-724 is a Sun-like ($T_{eff}=5330\pm150\,\text{K}$, $R_{\star}=0.88\pm0.05\,R_{\odot}$) G dwarf with a close-in planet candidate. The planetary solution from the SPOC pipeline places TOI-724.02 as a super Earth with $R_{P}=2.02\,R_{\oplus}$, $a=0.045\,\text{AU}$, and $P_{orb}=3.212\,\mathrm{d}$. This system exhibits significant stellar activity, which was used to constrain a stellar rotation period of $P_{\star, rot}=5.41\,\mathrm{d}$. Ten flares were identified with an average $\overline{\Delta\log{Z}}=13.6$. The rotation period found using \texttt{SpinSpotter} yields a different solution compared to \citet{martins2020}, which finds $P_{\star, rot}=9.67\,\mathrm{d}$, approximately a two times alias. The difference likely arises from the use of the ACF compared to Fourier-like methods. Flares are not correlated with either stellar rotation period, but cluster significantly with the synodic period with $P_{syn}=(1/3.212\,\mathrm{d}-1/5.41\,\mathrm{d})^{-1}=7.90\,\mathrm{d}$ at $3.1\sigma$ significance in UBA and $>2\sigma$ significance using the Kuiper test. Validating and determining the orbital geometry will contextualize the significance of this detection.

\subsubsection{TOI-4131.01}
\label{subsubsection: TOI-4131.01}
TOI-4131 is a faint, Sun-like star ($T_{eff}=5263\pm120\,\text{K}$, $R_{\star}=0.94\pm0.05\,R_{\odot}$) identified using the FAINT pipeline. The planetary candidate solution places TOI-4131.01 as a hot Jupiter with $R_{p}=11.12\pm0.7\,R_{\oplus}$ and $a=0.033\pm0.003\,\mathrm{AU}$. Nine, long-duration and high-ED flares were identified with this target with $\overline{\Delta\log{Z}}=19.8$ and a clustering significance of $\sqrt{TS}_{VM}=3.26\sigma$. Assuming TOI-4131's magnetic environment is comparable to that of our Sun, this places TOI-4131 within the Alfvén surface and within interaction distance of large-scale structures. SPIs are likely given the companion size and distance. However, two stars have been identified in the same TESS pixel, which leads to a non-zero possibility of contamination in the light curve. Additionally, higher resolution photometry is needed to confirm the origins of the flares and to validate the planet candidate as a hot Jupiter.

\subsection{Correlated Flares with Stellar Rotation Period}
\label{subsection: Stellar Rotation Correlation}
Five planet or planet candidate hosts were found to have significant flaring ($3\sigma>$) clustered with the stellar rotation rate: HIP 67522, TOI-168.01, TOI-936.01, TOI-1063.01, and TOI-4180.01. HIP 67522, TOI-936, and TOI-1063 all have $N_{\text{flare}}\geq30$, providing significant evidence that flare epochs are strongly correlated with the stellar rotation rate, and we encourage subsequent flare studies to analyze further correlation in future flaring events. A search for correlation with stellar rotation was performed to eliminate noise sources from induced flare candidates.  None of these targets emerged as significant in either $P_{orb}$ or $P_{syn}$ signals. Consequently, further analysis of these targets is outside the scope of this paper, but they are reported for future studies.

\subsection{Detection Comparison: HIP 67522}
\label{subsection: Comparison}
\citet{Ilin2024} surveyed 1,812 targets in both TESS and Kepler photometry. Notably, they report a significant detection of HIP 67522, with $N=12$ and $p=0.009$. We replicated the analysis for HIP 67522 using the flare phases publicly available from \citet{Ilin2024}, selecting flares with $ED<1\,\mathrm{s}$. Using the same flares in this study, we obtain a similar $\sim2.3\sigma$ result of $\sim p_{\mathrm{AD}}=0.011$, with the discrepancy arising from the custom Anderson-Darling test used in their methodology. The target had a $\sqrt{TS}_{VM}=2.951$, consistent with the reported prominent clustering. When all flares ($N=16$) from the public catalog are included, the signal becomes less significant with $p_{AD}\ sim 0.03$. Additionally, three of the reported flares are degenerate with each other, having flare epochs within $\sim2\,\mathrm{min}$ of other flares. Removing degenerate flares significantly decreases the detection, as expected, with $p_{\mathrm{AD}}\sim 0.14$.

We compared the flares identified by \citet{Ilin2024} with those retrieved by \texttt{ardor}, noting discrepancies in the number of non-degenerate flares recovered ($N=12$ for the former, $N=14$ for the latter) and uniqueness. Of the $N=12$ non-degenerate flares reported in \citet{Ilin2024}'s public catalog, seven were recovered by \texttt{ardor}. We then combined all unique epochs for $N=19$, which resulted in a non-detection of $\sim p_{\mathrm{AD}}=0.33$ and $\sqrt{TS}_{VM}=1.26$. Methodological differences account for these discrepancies, which are discussed in detail in Section~\ref{subsection: Impacts Methodology}. While we recreated the significant results claimed in \citet{Ilin2024}, the flares identified in this study suggest that not all flares were recovered by either pipeline, and using all recovered flares does not yield a significant detection for this target. 

HIP 67522 has a periodic baseline whose frequency changes significantly throughout a single TESS light curve, requiring more rigorous detrending than other targets. Small amplitude flares may be lost if the detrending algorithm over-fits the data. The SG filter is effective at removing the bulk of the periodic signal; however, artifacts may still be present that could impact the detectability of flares. HIP 67522 is a critical case study illustrating how subtle methodological differences can lead to significantly different outcomes. It is worth further validation with additional flare-vetting paradigms and more rigorous detrending models, such as Gaussian kernels.

\subsection{Investigating the SPI Signal of HD 118203}
\label{subsubsection: HD 118203}

\begin{figure}[!ht]
    \centering
    \label{figure: HD 118203 ACF}
    \includegraphics[width=0.47\textwidth]{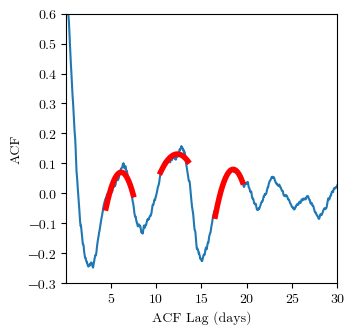}
    \caption{The ACF (blue) of HD 118203 with the best-fit parabolas (red) used to quantify $B, A, R^{2}$, with a derived stellar rotational period $P_{rot}=6.35\pm0.07\,\text{d}$.}
\end{figure}

\citet{Castro2024} identified a periodic signal in the TESS photometry of HD 118203, consistent with the orbital period of the close-in planet HD 118203\,b. The identified photometric modulation, with period $\approx6.1\,\text{d}$, approximately coincides with HD 118203's orbital period of $6.1349890\pm0.0000013\,\text{d}$ \citet{Maciejewski2024}. The analysis could not rule out stellar rotation as the cause of the modulation, deriving inconsistent periods from ELODIE RV data of $~110$ and $470 \text{ d}$. As part of the stellar rotation catalog of exoplanet hosts in Section~\ref{subsubsection: Stellar Rotation}, we used \texttt{SpinSpotter} to generate the autocorrelation function to see if the periodic signal is consistent with stellar rotation using all available TESS photometry from sectors 15, 16, 22, 49, and 76. Sector 76 was not included in the analysis by \citet{Castro2024}, which utilized data with a 20-second cadence. After applying a transit mask, the parabolic solution returned a period of $P_{rot}=6.35\pm0.07\,\text{d}$, with $B=0.566$, $A/B=0.166$, and $R^{2}=0.951$. We show the ACF in Figure~\ref{figure: HD 118203 ACF}. This solution is consistent with the selection function for stellar rotation defined in Section~\ref{subsubsection: Stellar Rotation}. We note that \citet{Holcomb2022} sets a strict selection criterion of $A/B > 0.25$ and a relaxed selection criterion of $A/B > 0.15$, where HD 118203 falls under the latter. The ACF  in Figure~\ref{figure: HD 118203 ACF} shows two overtones at the derived period, though subsequent peaks are ill-defined. 

Multiple studies have measured values of $v\sin{i}$ and $P_{rot}$ for HD 118203. The first by \citet{da2006} used ELODIE to derive a $v\sin{i_{\star}}=4.7 \text{ km/s}$ giving a stellar rotational period of $\approx22\text{ d}$, with no reported uncertainty. A second study by \citet{Luck2016} used the Sandiford Echelle spectrometer (SES) to derive a $v\sin{i}=5.32\pm0.5\,\mathrm{km/s}$, giving an upper bound $P_{rot,\star}<18.9\pm0.4\,\text{d}$. Work done by \citet{Pepper2020} and \citet{Zhang2024} use the LS periodogram method to derive stellar rotation periods of $20\pm5 \text{ d}$ and $23\pm4 \text{ d}$ respectively, with \citet{Zhang2024} also reporting $v\sin{i_{\star}}=5.1\pm1\text{ km/s}$, leading to $i_{\star}=90\pm14^{\circ}$. Lastly, a study by \citet{Knudstrup2024} uses the Fiber-fed Echelle Spectrograph (FIES), and finds a value of $v\sin{i}=4.9\pm0.4\text{ km/s}$, but also finding a derived stellar rotational value of $P_{rot}=6.1\pm0.4\text{ d}$ using the same methods in \citet{Mcquillan2014} by utilizing the AFC in Kepler photometry. Our results using \texttt{SpinSpotter} and the ACF are consistent with this result. Thus, it is clear that the Lomb-Scargle and ACF methods return intrinsically different $P_{rot}$ values using the same TESS photometry, while $v\sin{i_{\star}}$ is consistent across observations. The differing values of $P_{rot}$ have a significant impact on stellar inclination, and thus projected and true obliquity.

The TESS light curve of HD 118203 shows non-sinusoidal periodicity (see Figure 3 in \citet{Zhang2024}). If the modulation is due to star spots, spot evolution is rapid and non-uniform. In this case, the use of Fourier-based methods in non-sinusoidal photometric baselines can be inconsistent. Consequently, the sinusoidal fit to the derived period of $23\pm4 \text{ d}$ in \citet{Zhang2024} does not encapsulate the higher frequency features in the photometry, and the LS power at that period is relatively low. Conversely, the ACF is shown to be robust across non-sinusoidal baselines caused by spot evolution \citep{Mcquillan2014}. On the surface, the $P_{rot}=6.34\pm0.07\text{ d}$ and $P_{rot}=6.1\pm0.4\text{ d}$ appear to be the best solutions to the stellar rotational period, which places the bound $18.9^{\circ}<i_{\star}<21.4^{\circ}$. However, the near pole-on stellar inclination would significantly reduce the detectability of spot modulation, which would favor the edge-on solution found in \citet{Zhan2019}. However, modulation may be possible to observe given the high SNR of the target, making it unclear whether the $\sim6.1\,\text{d}$ modulation is caused by rotation or SPI. If we assume our result of $P_{\text{rot}}=6.34\pm0.04\,\text{d}$ is correct, using $N=100,000$ Monte Carlo samples assuming Gaussian error in each of the input parameters, we derive $i_{\star}\sim17.5$ and $\psi\sim89^{\circ}$ using the $\lambda=-23^{\circ+25}_{-38}$ solution from \citet{Knudstrup2024}. We avoid ascribing error due to the complex distributions from which each parameter is drawn, and report the maximum likelihood estimate solutions to contextualize our $P_{\text{rot}}$ value in terms of its effect on $i_{\star}$ and $\lambda$ values. As \citet{Pepper2020} suggests, HD 118203 is a prime target for astereoseismology, which will provide an independent probe into $i_{\star}$. Given the difference in the competing solutions, any constraint will provide clarity on the origin of the signal and on the true obliquity of the system.

\subsection{Impacts of Methodology}
\label{subsection: Impacts Methodology}
When comparing the results of this work with those of other flare studies, discrepancies emerge in the number and uniqueness of flares retrieved. This poses critical methodological questions and challenges the approach to the problem of induced flares. In this section, we outline the limitations of previous methods used in detecting flares and how we address them. We also discuss our perspective on maximizing the SNR of induced flares.

\subsubsection{Maximizing Short-Duration Flare Recovery}
\label{subsubsection: Cadence}
One paradigm adopted by flare surveys, such as \citet{Chang2015, Ilin2022}, requires three consecutive data points to be above $3\sigma$ from the median relative flux. Motivated by the search for superflares, this method is reliable for identifying large flares. However, this approach limits the adequate parameter space of flaring events. It does not maximize sensitivity to the photometric limit, excluding small amplitude and short-duration flares, which are regimes where induced flares may have the highest power. Using the predicted flare energy budget proposed in Section \ref{section: SPMI Models} and shown in Figure \ref{Fig: Estimated Results} for main-sequence stars, a significant fraction of expected induced flare amplitudes are below TESS's photometric limit.

As outlined in (b) of Figure \ref{fig: IR Param Figure}, a clear $40\%$ improvement in short amplitude flares across all FWHM (particularly with $FWHM<2\, \mathrm{min}$) occurs when using \texttt{ardor} over this previous detection threshold. While it may lead to an increase of false positives through Tier 2 of \textsc{ardor}, the Bayesian model comparison eliminates most of these false positives. Motivated by the energy estimates by \citet{Lanza2018}, it is reasonable to assume induced flares may reside in the low-energy flare limit. This is supported by the historical lack of detections of correlated super-flares with close-in planets \citep{Rubenstein2000}. 

\subsubsection{Model Agnostic Induced Flare Detection}
\label{subsubsection: Model agnostic induced flare detection}
Our results do not place priors on the energy, amplitudes, or durations of the flares produced via induced flaring. The processes that lead to induced flaring are complex, time and spatially dependent, and are poorly constrained from observation. While well-motivated theoretical models contextualize the plausibility of induced flaring, many simplifying assumptions do not reflect the reality of each system. To reconcile with this, we only assume that over a sufficiently long observational baseline, on average, induced flares will cause a clustering of observed flares with respect to the orbital period. This avoids constraining searches for flares based on a particular model that does not accurately reflect the complex physics occurring in the system.

\subsection{UV Photometry with ULTRASAT}
\label{subsubsection: UV Photometry}
Although TESS and Kepler provide long-baseline photometry in the infrared and optical, stellar flare spectra peak in the UV/X-ray range, the commonly adopted $9000\,\mathrm{K}$ blackbody assumption in determining the bolometric energy of flares has a peak emission at $\sim0.32\, \mathrm{\mu m}$, well into the ultraviolet band. Co-observation of visible and UV photometry with Kepler and GALEX has shown that the $9000\,\mathrm{K}$ assumption underestimates the UV flare energy budget by up to a factor of six \citep{Berger2024}. The ULTRASAT mission, slated for launch in 2027, will provide long-baseline UV photometry in tandem with TESS \citep{Shvartzvald2024}. It will give an order-of-magnitude SNR increase for most stellar targets, enabling comparative statistics to assess the completeness of the white-light flare surveys conducted by TESS.

\section{Conclusion}
\label{section: Conclusion}
Exoplanetary magnetic fields remain an elusive yet important parameter in exoplanetary studies, and their role in planetary formation, atmospheric loss, and habitability is significant. Direct detection of exoplanet magnetic fields via cyclotron maser instability or synchrotron emission is an exciting frontier with null results so far. Consequently, indirect observation through phase-correlated magnetic star-planet interactions, such as planet-induced stellar flaring, is an equally promising probe. 

We presented a photometric flare detection pipeline, \texttt{ardor}, which detects flares in three increasingly strict tiers. We presented computational, injection-recovery, and precision-recall metrics supporting its efficacy as an effective flare detection tool in 2-minute and 20-second TESS cadence photometry. We presented two toy models that describe induced flaring: the von Mises PDF and the dipole-dipole energy model. We performed flare injection simulations to quantify the detectability of induced flares between G and M-type stars, while varying the induced flare strength to identify the SPI strength at which induced flares would be consistently detectable with typical TESS data volumes.

Using \texttt{ardor}, we conducted an expansive photometric flare survey using archival TESS photometry, targeting exoplanets with highly sub-Alfvénic periastrons as well as all TOI planet candidate hosts. This survey yielded three promising candidates for induced flaring: TOI-1062\,b, Gl 49\,b  and HD 163607\,b. TOI-1062\,b shows the greatest promise for induced flaring due to its moderate eccentricity ($ e = 0.177$) and the high likelihood of a sub-Alfvénic approach. The flare excess occurs relatively close to the periastron within its orbit, which supports dependence on induced flaring on instantaneous separation. We strongly suggest this target for additional SPI signals, such as observation using HST and LOFAR.

In-depth comparisons with \citet{Ilin2024} demonstrate important nuances in methodological choices. The strongest induced-flare candidate in that study, HIP 67522\,b, was found to be a non-detection with the flares found using \texttt{ardor}. In both cases, flares found by one pipeline were missed by the other, and vice versa. The signal disappeared when all unique flares from both pipelines were considered in tandem. While \citet{Ilin2024} places an $ED>1\,\mathrm{s}$ prior on considered flares, we argue induced flares likely exist in a low energy, low amplitude regime consistent with the findings in Section\ref{section: SPMI Models}.

The SPI signal of HD 118203\,b remains a compelling result. The difference in stellar rotational solutions obtained using the ACF and Fourier-based methods is an important degeneracy to break in order to determine the true stellar inclination. We propose expedited analysis using astereoseismology to find an independent constraint on $i_{\star}$. With an independent constraint, the observed photometric modulation identified by \citet{Castro2024} can be validated as either spot activity due to stellar rotation or, more excitingly, an SPMI.

Significant work can still be done in improving \texttt{ardor}'s performance in avoiding the false-positives outlined in Section\ref{subsection: False-Positives}, as well as meta-studies between its performance and the multitude of similar white-light flare studies conducted on TESS data.

\section{Acknowledgments}
\label{section: Acknowledgements}
We acknowledge support from the McDonnell Center for the Space Sciences at Washington University in St. Louis.


\software{allesfitter, ardor, Veusz, astropy, matplotlib, scipy, astropy, astroquery, numpy, pandas, aflare, LightKurve, SpinSpotter}
\facility {Exoplanet Archive, MAST}


\bibliography{references}

\end{document}